\documentclass[12pt,preprint]{aastex}
\usepackage{amssymb}

\slugcomment{Submitted to ApJ}
\shortauthors{Ardila, Basri}
\shorttitle{GHRS observations}
\begin{document}

\newcommand \kms{km s$^{-1}$}
\newcommand \funit{$\rm{\ ergs\ sec^{-1}\ cm^{-2}\ \AA}$}
\newcommand \vsini{$v$sin$i$}
\newcommand \caii{\ion{Ca}{2}}
\newcommand \hei{\ion{He}{1}}
\newcommand \heii{\ion{He}{2}}
\newcommand \feii{\ion{Fe}{2}}
\newcommand \hal{H$\alpha$}
\newcommand \hbeta{H$\beta$}
\newcommand \rsun{R_\odot}
\newcommand \msun{M_\odot}
\newcommand \lsun{L_\odot}
\newcommand \degs{$^\circ$}
\newcommand \htwo{H$_2$}
\newcommand \lya{H$_{Ly\alpha}$}
\newcommand \mgii{\ion{Mg}{2}}
\newcommand \sii{\ion{Si}{1}}
\newcommand \siiv{\ion{Si}{4}}
\newcommand \civ{\ion{C}{4}}
\newcommand \nad{NaD}
\newcommand \siiii{\ion{Si}{3}]}
\newcommand \oiv{\ion{O}{4}]}
\newcommand \oi{\ion{O}{1}]}
\newcommand \cii{\ion{C}{2}}
\newcommand \ciii{\ion{C}{3}]}

\title{Observations of T-Tauri Stars using HST-GHRS: I. Far Ultraviolet Emission Lines}

\author{David R. Ardila\altaffilmark{1}, Gibor Basri\altaffilmark{2}, Frederick M. Walter\altaffilmark{3}, Jeff A. Valenti\altaffilmark{4}, Christopher M. Johns-Krull\altaffilmark{5} }
\altaffiltext{1}{Astronomy Dept., Univ. of California, Berkeley, CA 94720, ardila@garavito.berkeley.edu}
\altaffiltext{2}{Astronomy Dept., Univ. of California, Berkeley, CA 94720, basri@soleil.berkeley.edu}
\altaffiltext{3}{Department of Physics and Astronomy, SUNY, Stony Brook NY 11794-3800, fwalter@astro.sunysb.edu}
\altaffiltext{4}{Space Telescope Science Institute, Baltimore, MD 21218, valenti@stsci.edu}
\altaffiltext{5}{Space Sciences Laboratory, Berkeley, CA 94720, cmj@ssl.berkeley.edu}

\begin{abstract}
We have analyzed GHRS data of eight CTTS and one WTTS. The GHRS data consists of spectral ranges 40 \AA\ wide centered on 1345, 1400, 1497, 1550, and 1900 \AA. These UV spectra show strong \siiv, and \civ\ emission, and large quantities of sharp ($\sim40$ \kms) \htwo\ lines. All the \htwo\ lines belong to the Lyman band and all the observed lines are single peaked and optically thin. The averages of all the \htwo\ lines centroids for each star are negative which may indicate that they come from an outflow. We interpret the emission in \htwo\ as being due to fluorescence, mostly by \lya\, and identify seven excitation routes within 4 \AA\ of that line. We obtain column densities ($10^{12}$ to $10^{15}\rm{cm^{-2}}$) and optical depths ($\sim1$ or less) for each exciting transition. We conclude that the populations are far from being in thermal equilibrium. We do not observe any lines excited from the far blue wing of \lya, which implies that the molecular features are excited by an absorbed profile. \siiv\ and \civ\ (corrected for \htwo\ emission) have widths of $\sim200$ \kms, and an array of centroids (blueshifted lines, centered, redshifted). These characteristics are difficult to understand in the context of current models of the accretion shock. For DR Tau we observe transient strong blueshifted emission, perhaps the a result of reconnection events in the magnetosphere. We also see evidence of multiple emission regions for the hot lines. While \civ\ is optically thin in most stars in our sample, \siiv\ is not. However, \civ\ is a good predictor of \siiv\ and \htwo\ emission. We conclude that most of the flux in the hot lines may be due to accretion processes, but the line profiles can have multiple and variable components. 
\end{abstract}

\section{Introduction}

The peculiar spectral characteristics of Classical T Tauri stars (CTTSs) are usually interpreted as the result of a circumstellar disk accreting onto a magnetized young star. Considerable development of this idea has occurred during the last decade. The current paradigm asserts that the inner region of the disk is truncated by the stellar magnetic field and partially ionized material slides down to the stellar surface. Close to the surface the gas produces a strong shock, in which the kinetic energy of the infalling stream is transformed into random thermal motions. It is thought that radiation from the shock reprocessed by the stellar surface is responsible for the UV and optical continuum excess (`veiling'). This process has been modeled in detail by \citet{1998ApJ...509..802C} and \citet*{2000ApJ...544..927G}, who use the strength of the ultraviolet continuum to obtain accretion rates and spot sizes for CTTS.

Given the high temperature of the gas in the inner stellar magnetosphere, wavelengths shorter than the Balmer jump contain important spectral diagnostics. For material infalling at $\sim300$ \kms, the strong shock conditions imply a shock temperature of $\sim10^6$K, which decreases monotonically as the gas settles onto the star \citep{1998ApJ...509..802C}. With densities ranging from $10^{10} \rm{cm^{-3}}$ to $10^{14} \rm{cm^{-3}}$ (depending on the accretion rate, area of the accretion column and the specific model), the region is similar to the chromosphere, transition region, and corona of the underlying star. Even though the physics involved is very different, one expects to observe transition region diagnostics, like \civ\ at $1550$\AA\  (which, when it is collisionally dominated in the low density limit, has maximum population at $10^5$K, see \citealt{1992ApJ...398..394A}) and \siiv\ at $1400 $\AA\  (formed at $63000$K in the same limit), and chromospheric diagnostics such as the \mgii\ resonance doublet at 2800 \AA.

Analyses of low-resolution observations by the International Ultraviolet Explorer (IUE) suggest that a large fraction of the \civ\ emission does indeed come from processes related to accretion (as opposed to intrinsic transition region emission). \citet*{2000ApJ...539..815J} have shown that the flux in the \civ\ resonance doublet is up to two orders of magnitude larger than in main-sequence and RS CVn stars with the same dynamo number (a measure of magnetic activity). Furthermore, such emission is well correlated with the estimates of accretion rates obtained from optical data \citep*{1995ApJ...452..736H}. On the other hand, as \citet{2000ApJ...539..815J} argue, due to the different internal structure of T-Tauri Stars (TTSs) compared to these other stars, it is not clear what fraction of the line is due to accretion as opposed to other possible sources. Even though the evidence that supports the connection between accretion and large transition region line emission is strong, it is not conclusive. 

If due to accretion, the line shapes of the transition lines should reflect the conditions in the accretion shock. \citet*{1994ARep...38..540G} have studied the \civ\ and \siiv\ lines using high resolution observations of CTTS by IUE. They argue that the profiles of these lines should have at least two distinct peaks: one due to emission coming from the pre-shock zone and another from emission coming from material immediately after the shock. Another emission peak may appear, due to shocked material close to the star. Such prediction has been formalized in the models developed by \citet{1998ARep...42..322L}. Given the low signal-to-noise of the IUE observations, it is not clear to what extent these two and three-peaked profiles are actually observed. 

Semiforbidden ultraviolet line ratios have been used \citep{2000ApJ...539..815J,1999MNRAS.304L..41G,2000ARep...44..323L} to estimate the density of the shocked material. \citet{2000ApJ...539..815J} show that the density of the shock, as measured by these ratios, is 3 to 4 orders of magnitude less than the densities assumed in the models by \citet{1998ApJ...509..802C} but coincide with the densities predicted by \citet{1998ARep...42..322L}. Theoretically, the shock density depends on parameters that are not well determined, such as the accretion rate, the reddening of the star, and the covering fraction of the shock. 

Comparing observations of \civ\ and \siiv\ lines to models is complicated by the fact that the UV spectra of TTSs show strong \htwo\ emission lines, and some of those are blended with the doublets. The \htwo\ lines were first identified by \citet{1981Natur.290...34B} and \citet*{1984MNRAS.207..831B} in T Tau. Based on IUE low-resolution spectra, they suggest that the lines are the result of resonant fluorescence of molecular hydrogen (excitation from $X^1\Sigma^+_g$ to $B^1\Sigma^+_u$, the Lyman band) by \lya.  The \htwo\ line R(3)1-8 is at $-166$ \kms\  from  the red member of the \civ\ doublet and therefore it alters its profile.  The same is true for \siiv\ and the \htwo\ lines R(0)0-5 (at $-8$ \kms), R(1)0-5 (at $+44$ \kms), and P(3)0-5 (at $-25$ \kms). In the Sun, \siiv\ and \civ\ have themselves been identified as pumping mechanisms for \htwo\ \citep{1978ApJ...226..687J,1979MNRAS.187..463B}.

\htwo\ ultraviolet emission has also been identified in the large sample of TTSs presented by \citet*{2000ApJS..129..399V}. At least in the case of T Tau, the emission is extended more than 5 arcsecs (\citealt{1981Natur.290...34B}; \citealt*{1997AJ....114..744H}), and has a very complex structure which includes filaments and knots \citep{1999A&A...348..877V}.

The analysis of \htwo\ observations in CTTSs has centered around the heating mechanism for the molecule, and the origin of the emission (either the disk or nebular material surrounding CTTSs). As \htwo\ is supposed to be the most abundant species in a TTS system, analysis of its physical conditions has important implications for a wide range of topics, from the gas-disk dissipation timescale to the characteristics of the outflow. While a lot of work has been done about the observational characteristics of the infrared \htwo\ emission (e.g. \citealt{1999ApJ...521L..63T}; \citealt{1997Ap&SS.255...77V}; \citealt*{1997AJ....114..744H}), relatively little work has been published regarding the ultraviolet fluorescent lines in the context of CTTS (\citealt*{LamzinonDF}; \citealt{1987ApJ...322..412B}; \citealt{1998sigh.conf..360W}), perhaps because of the scarcity of high-resolution observations of the fluorescent lines. ISO observations of the far IR \htwo\ lines of T Tauri \citep{1999A&A...348..877V} reveal two different components of different temperatures (500 and 1500 K) contributing to the emission. Furthermore, while the bulk of the heating is due to shocks (either from the outflow against the ISM or from infalling material into the disk of T Tau S), the signature of fluorescence processes is observed, even in the IR lines from low-lying states. In the context of the UV lines, \citet{1998sigh.conf..360W} have shown, using low-resolution GHRS observations, that one can generally detect \htwo\ emission from CTTSs, while none is detected in Weak T Tauri Stars (WTTSs). By assuming that the observed \htwo\ comes from the disk, they conclude that the gas-disk dissipation timescale is similar to the dust-disk dissipation timescale.

Most of the ultraviolet observations of TTSs have been performed with IUE. In spite of the prodigious amount of data this spacecraft produced, observations of TTSs suffered from its small aperture and the low sensitivity (by today's standards) of its Vidicom cameras. This situation improved with the launch of the Goddard High Resolution Spectrograph (GHRS), which flew on board the Hubble Space Telescope from 1990 to 1997; however, only a handful of TTSs were observed with this instrument.

The objective of the present work is to describe the main features of the ultraviolet range (mainly in two windows 40 \AA\  wide centered on 1400 and 1550) for a sample of eight CTTSs and one WTTS. The data were obtained with the GHRS instrument in the early 1990s. A portion of these data has been independently analyzed (\citealt*{2000A&A...357..951E}; \citealt{2000AstL...26..225L,2000AstL...26..589L,2000ARep...44..323L,LamzinonDF}). It has also been described in conference reports \citep{1996csss....9..419C,1993AAS...183.4007V}. The main features in these spectra are the \siiv and \civ\ resonance doublets, and strong \htwo\ lines. In a companion paper we analyze the \mgii\ resonance doublet, part of this same set of data.

The remainder of this paper is organized as follows: We start with a description of the observations. In section 3 we comment on the general characteristics of the data. In section 4.1 we analyze the characteristics of the \htwo\ lines present in the spectra. As mentioned above, some of these lines are blended with \civ\ and \siiv. In section 4.2 we indicate how to obtain unblended profiles for the resonance doublets. Section 4.3 contains an analysis of the unblended profile. Our conclusions are in section 5.

\clearpage
%\documentclass{aastex}
%\begin{document}

\begin{deluxetable}{cccccccccc}
\tabletypesize{\scriptsize}
\tabcolsep 1pt
\tablewidth{0pt}
%\newcommand \msun{M_\odot}
%\newcommand \rsun{R_\odot}

%\tablenum{1}
\tablecaption{\label{par}Parameters of Observed Stars}
\tablehead{
\colhead{ } &
\colhead{Spectral} &
\colhead{ } &
\colhead{$R_{\ast}$} & 
\colhead{$M_{\ast}$} &
 \colhead{$A_V$\tablenotemark{a}}&
\colhead{${\dot M}$\tablenotemark{b}}& 
\colhead{$\rm{V_{r}}$\tablenotemark{c}} &
\colhead{Inclination\tablenotemark{d}} &
\colhead{ } \\
\colhead{Name} & 
\colhead{Type} & 
\colhead{Class} & 
\colhead{($\rsun$)} & 
\colhead{($\msun$)} & 
\colhead{(mag)} &
\colhead{($10^{-8} \msun/yr$)} & \colhead{(km/s)}  & \colhead{($\deg$)} &\colhead{Multiplicity \tablenotemark{e}}
}
\startdata

BP Tauri & K7 & CTTS & 2.0 & 0.5 & 0.51 & 2.88 &  
15.8 \tablenotemark{f} & $<$ 50  & S \tablenotemark{g}\\
T Tauri NS & K0 & CTTS & 3.4 & 2.0 & 1.7 & 4.00 &
 19.1 \tablenotemark{f} &  20 & M (0.7'') \tablenotemark{g}\\
DF Tauri AB& M0 & CTTS & 3.4 & 0.3 &0.45 & 17.7 & 
15.8 \tablenotemark{h} & 80 & M (0.09'') \tablenotemark{i}\\
RW Aurigae ABC & K4 & CTTS & 1.4\tablenotemark{j} & 1.1\tablenotemark{j} & 1.14 & 16 &
 16 \tablenotemark{k} &  40 & M (1.5'')\tablenotemark{g} \\
DG Tauri & K5 & CTTS & 2.3 & 0.7 &1.6 & 50 &
 15.9 \tablenotemark{h} &  40\tablenotemark{l} & S \tablenotemark{i}\\
DR Tauri & K7 & CTTS & 2.7 & 0.4 & 1.2 & 30 &
 27.6 \tablenotemark{m}&  $<$40& S  \tablenotemark{i} \\
RY Tauri & K1 & CTTS & 2.4 & 1.6 & 0.29 & 0.25 &
 16.4 \tablenotemark{f} &  90 & S \tablenotemark{i} \\
RU Lupi & K7 & CTTS & 3.2 & 0.3 & 1.28\tablenotemark{n} & 77\tablenotemark{o} &
  -0.5  \tablenotemark{f} & 0 \tablenotemark{p} & S \tablenotemark{p} \\
HBC 388 & K1\tablenotemark{q} & WTTS & 1.5 & 1.3  & 0.1 & \nodata & 15.4 & 45 \tablenotemark{r}& S \tablenotemark{i} \\ 
 
\enddata

\tablecomments{The data for this table were taken from \citet{1999ApJ...510L..41J}, Table 1, unless indicated otherwise.}
\tablenotetext{a}{From GHBC for all except RW Aur and RY Tau. For these two, the values are from \citet{1999ApJ...510L..41J}}
\tablenotetext{b}{From GHBC for all except RW Aur and RY Tau. For these two, we have taken the HEG95 values and divided by 10.}
\tablenotetext{c}{Heliocentric velocity}
\tablenotetext{d}{Unless otherwise indicated, it is calculated from $v \sin i$ and $P_{rot}$ values quoted by \citet{2000ApJ...539..815J}. Rounded to one significant figure.}
\tablenotetext{e}{S=single, M=Multiple, the number indicates the separation}
\tablenotetext{f}{\citet{1988cels.book.....H}}
\tablenotetext{g}{\citet{1993A&A...278..129L}}
\tablenotetext{h}{\citet{1991AJ....101.1050H}}
\tablenotetext{i}{\citet*{1993AJ....106.2005G}}
\tablenotetext{j}{\citet{2001A&A...369..993P}}
\tablenotetext{k}{\citet{1999A&A...352L..95G}}
\tablenotetext{l}{\citet{2000ApJ...537L..49B}}
\tablenotetext{m}{\citet{2000AJ....119.1881A}}
\tablenotetext{n}{\citet{1994AJ....108.1071H}}
\tablenotetext{o}{\citet{2000ApJ...539..815J}, Table 4. Obtained via a correlation between accretion rate and CIV.}
\tablenotetext{p}{\citet{1996A&A...306..877L}}
\tablenotetext{q}{Unless otherwise indicated the data for this line come from \citet{1988AJ.....96..297W}}
\tablenotetext{r}{$P_{rot}=2.74 d$ from \citet{1997A&A...322..825P}. $v \sin i=20$ from \citet{1998A&A...334..592S}}

\end{deluxetable}
\tabletypesize{\normalsize}

%\end{document}

\section{Observations}
The characteristics of the observed targets are summarized in Table \ref{par}. All (except RU Lupi) belong to the Taurus star formation region, and we assume that all are at a mean distance of 140 pc from the Sun. The target stars have a range of accretion rates and include a weak TTS (WTTS), HBC 388 (other names are V1072 Tau and NTTS 042417+1744). This star does not show IR excess that could be attributed to a circumstellar disk \citep{1996AJ....111.2066W} and its \hal\ line is very weak \citep{1988AJ.....96..297W}. Therefore, spectral diagnostics related to the accretion process are expected to be negligible, and its emission can be assumed to be due only to magnetic-related phenomena (chromosphere, transition region, and corona). 

As indicated in Table \ref{par}, there are three multiple systems in the sample. In each case, all members of the system are within the slit. For DF Tau, both components have similar brightness. Speckle observations of the system suggest that it is composed of a CTTS and a WTTS. The orbital inclination from these observations is estimated to be $i\sim65^o$ \citep{1995A&A...304L..17T}, consistent with the $\sim80^o$ derived from analysis of variability. The companion of T Tau N is a very embedded young object, with a circumstellar disk. Reports of a third member of the system remain unconfirmed \citep{1998ApJ...508..736S}. RW Aur ABC is a triple system, with a factor of 10 difference in luminosity between the primary (A) and the secondary (BC) \citep*{1997ApJ...490..353G}. We unpublished high resolution spectra that suggest that RW Aur B is possibly a very weak K7 CTTS. 

In Table \ref{par} we use mostly the reddenings and accretion rates from \citet{1998ApJ...492..323G}, \citet{1998ApJ...509..802C}, and \citet{2000ApJ...544..927G} (GCHB). RW Aur, RY Tau, and RU Lup do not have these values measured in those works. For  RW Aur and RY Tau we use the extinction values determined by \citet{2000ApJ...539..815J}. In general, accretion rates determined by GCHB are one order of magnitude less than those determined by \citet{1995ApJ...452..736H} (HEG95). Therefore, for RY Tau and RW Aur we have divided the accretion rate from HEG95 by 10, to obtain the values quoted in the Table. Note that the value for the reddening of RU Lup is taken from \citet{1994AJ....108.1071H} and the accretion rate is taken from \citet{2000ApJ...539..815J}. The accretion rates divide the CTTS sample in three subsets: stars with low (RY Tau), `average' (BP Tau, T Tau), and high accretion rate.

\subsection{UV Observations}

We obtained these data using the Goddard High Resolution Spectrograph (GHRS; see \citealt{1994PASP..106..890B}), one of the first generation instruments aboard the Hubble Space Telescope. 

The GHRS is a modified Czerney-Turner spectrograph with Digicon detectors. The in-orbit performance of the GHRS is described by \citet{1995PASP..107..871H}. These data were obtained in GO programs executed in
cycles 3 and 5, as well in GTO observations in cycles 2 and 3.
Table~\ref{tbl-log1} presents the log of observations.
%\documentclass{aastex}
%\begin{document}

\begin{deluxetable}{lrrccl} 
\tabletypesize{\small}
\tablecolumns{6} 
\tablewidth{0pc} 
\tablecaption{Log of Observations\label{tbl-log1}}
\tablehead{
\colhead{Target} &\colhead{$\lambda_{cen}$} & 
   \colhead{UT start} & \colhead{Exposure (sec.)} & 
   \colhead{Readouts} & \colhead{Root} \\
\colhead{Date} &\colhead{(\AA)} & 
   \colhead{(UT)} & \colhead{(sec.)} & 
   \colhead{} & \colhead{}}
\startdata 
\cutinhead{Pre-COSTAR}
RU Lup &   1400.6  &  3:18:59 &  768 &  3 & Z10T0104 \\
8/24/92 &  1549.7  &  3:34:38 & 1280 &  5 & Z10T0105 \\
        &  1640.3  &  4:43:59 & 1280 &  5 & Z10T0107 \\
        &  1899.5  &  5:10: 8 & 1408 &  5 & Z10T0109 \\
        &  2324.7  &  6:26:50 &  512 &  2 & Z10T010C \\
        & \\
BP Tau  &  1400.8  & 17:54:11 & 1689 &  6 & Z18E0107 \\ %s-w=0.31px (1.2 km/s)
7/30/93 &  1550.0  & 19:12:58 & 1408 &  5 & Z18E0108 \\
        & \\
RW Aur  & 1400.8  & 22:13:41 &  844  &  3 & Z18E0407\\  %s-w=0.34px (1.3 km/s)
8/10/93 & 1550.0  & 23:20:25 &  844  &  3 & Z18E0408\\
        & \\
DR Tau &  1400.8 &  0:53:29  & 1689 &  6 & Z18E0307 \\ %s-w=0.41px (1.6 km/s)
8/5/93  &  1550.0 &  2:15:25  & 1689 &  6 & Z18E0308 \\
        &  1345.8 &  3:44:19  & 1408 &  5 & Z18E0309 \\
        &  1650.4 &  4:12: 7  & 1126 &  4 & Z18E030B \\
        & \\
DF Tau  &  1400.8  & 2:50:41  & 1126 &  4 & Z18E0207 \\%s-w=0.52px (2.0 km/s)
8/8/93  &  1550.0  & 4:10:34  & 1126 &  4 & Z18E0208 \\
        &  1345.8  & 5:37: 4  & 1408 &  5 & Z18E0209 \\
& \\
RY Tau  &  1400.6  &  4:25:51 & 1024 &  4 & Z1E70104 \\
12/31/93 & 1549.8  &  5:52:21 & 1280 &  5 & Z1E70105 \\
        &  1640.4  &  7:16:40 & 1280 &  5 & Z1E70106 \\
        &  1899.6  &  7:42:36 & 1408 &  5 & Z1E70108 \\
\cutinhead{Post-COSTAR}
DR Tau  &  1552.9 & 18: 2:53  & 5632 & 20 & Z2WB0207 \\ %s-w=-1.09px (-4.1 km/s)
9/7/95  &  1343.0 & 22:11:23  & 1408 &  5 & Z2WB0209 \\
        & \\
HBC388 &  1552.9  & 11:25:49 &  5068 & 18 & Z2WB0404\\
9/9/95 &  1401.0  & 15:59:47 &  5632 & 20 & Z2WB0407\\ %s-w=0.09px (0.4 km/s)
       & \\
T Tau  &  1342.9  &  2: 9:34 & 1280 &  5 & Z2WB0306 \\
9/11/95 &  1497.3  &  3:21:58 & 1280 &  5 & Z2WB0307 \\
        &  1552.9  &  3:46:36 & 1126 &  4 & Z2WB0308 \\
        &  1401.0  &  4:57:32 & 2304 &  9 & Z2WB030A \\ %s-w=-1.25px (-4.8 km/s)
        & \\
DG Tau &  1552.6  &  2: 7:40 & 5068  & 18 & Z2WB0106 \\
2/8/96 &  1400.7  &  5:23:35 &10418  & 37 & Z2WB0109,B \\%s-w=1.96px (7.4 km/s)
       & 1400.7   &  8:35:22 & 5068  & 18 & Z2WB010B \\

\enddata
\end{deluxetable}
%\end{document}

All observations before 1994 were made using the aberrated optics; later observations used the COSTAR corrective optics\setcounter{footnote}{0}\footnote{Although the RY Tau observations on 1993 December 31 occurred after the installation of COSTAR, they occurred before the GHRS corrective optics were deployed.}. The pre-COSTAR data are not seriously affected by the mirror aberrations, because of the combination of broad lines and (generally) low S/N per pixel, so we made no effort to deconvolve the aberrated line profile. 

Target acquisitions occurred in the N2 mirror, using a 3$\times$3 spiral search pattern. For the cycle 5 observations we used the BRIGHT=RETURN option. Dwell times ranged from 0.2 to 2.0 seconds; count rates ranged from 1200 to 8000 counts~s$^{-1}$. All targets were successfully acquired. Exposure times in Table \ref {tbl-log1} reflect the time observing the target, excluding the time spent exposing background.

All observations were obtained through the large science aperture,
which projects to a 2~arcsec square on the sky pre-COSTAR, and to 1.74~arcsec post-COSTAR. We used substep pattern 5 (four on-target observations offset by $\frac{1}{4}$~diode, and two background observations with the detector array deflected above and below
the spectrum) for all observations. The four on-target observations were summed using comb-addition (see Heap et al.\/ 1995). Each pixel corresponds to $\frac{1}{4}$~diode width; the instrumental resolution is about 4.4~pixels. The large science aperture subtends 8 diodes, or 32 pixels, in the dispersion direction. The instrumental resolution is constant with wavelength and is the same for each individual observation and the co-added frame. The resolving power is 20000 at 1500 \AA.

All observations utilized the side 2 (long wavelengths) optics with the CsTe Digicon detector and the medium wavelength gratings (G160M, G200M, G270M). To minimize the effect of GIMP-induced image motions, all observations were read out about every 5 minutes. No significant drifts were seen.

\subsubsection{GHRS Data Reduction}

We converted the spectra from GEIS format to GHRS format data files
using the HRS\_ACQUIRE software \citep{blackwell1993}. The data were extracted and calibrated using version 2.20 of the IDL-based CALHRS software \citep{1993AJ....105..831B} with the default calibration files appropriate for the dates of the observations.

Because our targets are not bright and most of the flux in the ultraviolet is in emission lines, proper background subtraction is crucial. The GHRS background is primarily due to passage of cosmic rays through the Digicon window. The mean background rates vary with geomagnetic latitude, and exhibit non-Poisson fluctuations. Cosmic ray events are characterized by enhanced
counts affecting a number of contiguous diodes. Because
the integration time on the background is significantly less than that on the target, the background rates are much more sensitive to these bursts of counts from cosmic rays. 

We examined all pairs of background bins for evidence of non-Poissonian fluctuations. In those cases where the background count rates differed by more than 5$\sigma$, we replaced the affected substep bin with the rate from the other background bin. Retention of the high background bins had resulted in negative continua in some cases. We then subtracted a uniform mean background rate (this is the same as fitting the background with a zero-order polynomial). 

\subsubsection{Wavelength Calibration}

The default dispersion solution, based on the carrousel position,
is expected to be accurate to $\pm$0.1~diodes. To verify accuracy of the wavelength scale, we obtained WAVECAL observations as part of eight of the observations. No WAVECAL exposures accompany the RU~Lup or RY~Tau observations. In the second DR~Tau observation the WAVECAL accompanied the C~IV observation, in all other cases it preceeded the Si~IV observation. For all observations not preceeded by a WAVECAL observation, we used the preceeding SPYBAL\footnote{The SPYBAL (spectral Y-balance) is a WAVECAL observation at a default carrousel position which is performed whenever the grating is changed in order to center the spectrum on the Digicon array. WAVECAL observations are made at the same carrousel position as the following observation.} observation to determine zero-point of the wavelength scales. 

To verify the internal accuracy of the wavelength scales, we also reduced those observations preceeded with WAVECAL observations with the preceeding SPYBAL observation and compared the wavelength scales. The mean offsets are 0.3$\pm$1.0~diodes (1.0$\pm$3~km~s$^{-1}$), where the uncertainty is the RMS deviation in the measurements. This can be taken as a representative statistical uncertainty on the wavelength scale. There may be systematic offsets of up to $\pm$6~pixels ($\pm$20~km~s$^{-1}$) due to the target centering process, primarily for the pre-COSTAR observations. Further offsets may result from non-uniform filling of the aperture in the case of extended emission.

All readouts from a particular observation were then interpolated to a common linear wavelength scale and summed. An effect of the interpolation is to smooth the data, and we scaled the error vectors accordingly.

\section{General Characteristics}

\begin{figure}
\plotone{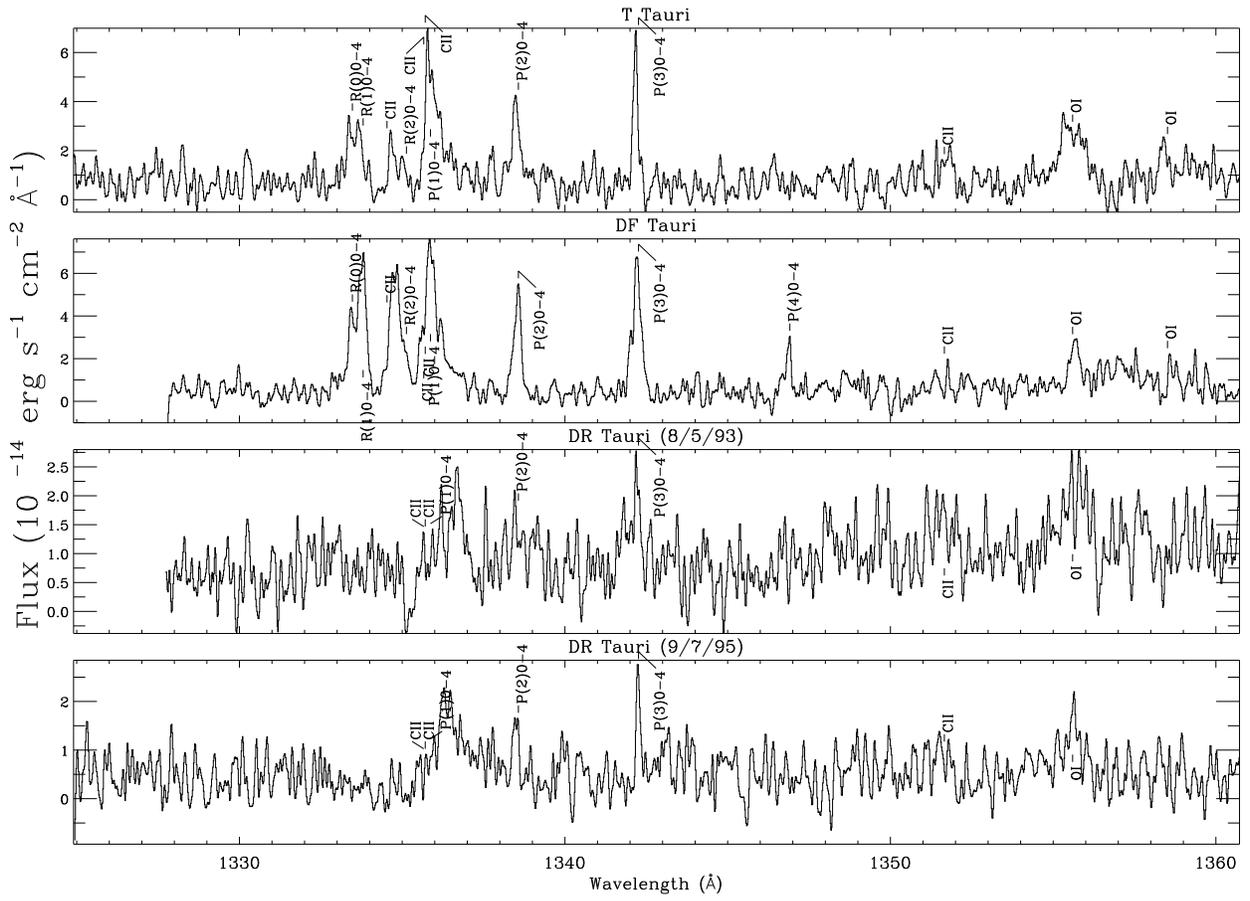}
\caption{\label{bluest_spec} Smoothed spectra centered around 1342 \AA.}
\end{figure}

\begin{figure}
\plotone{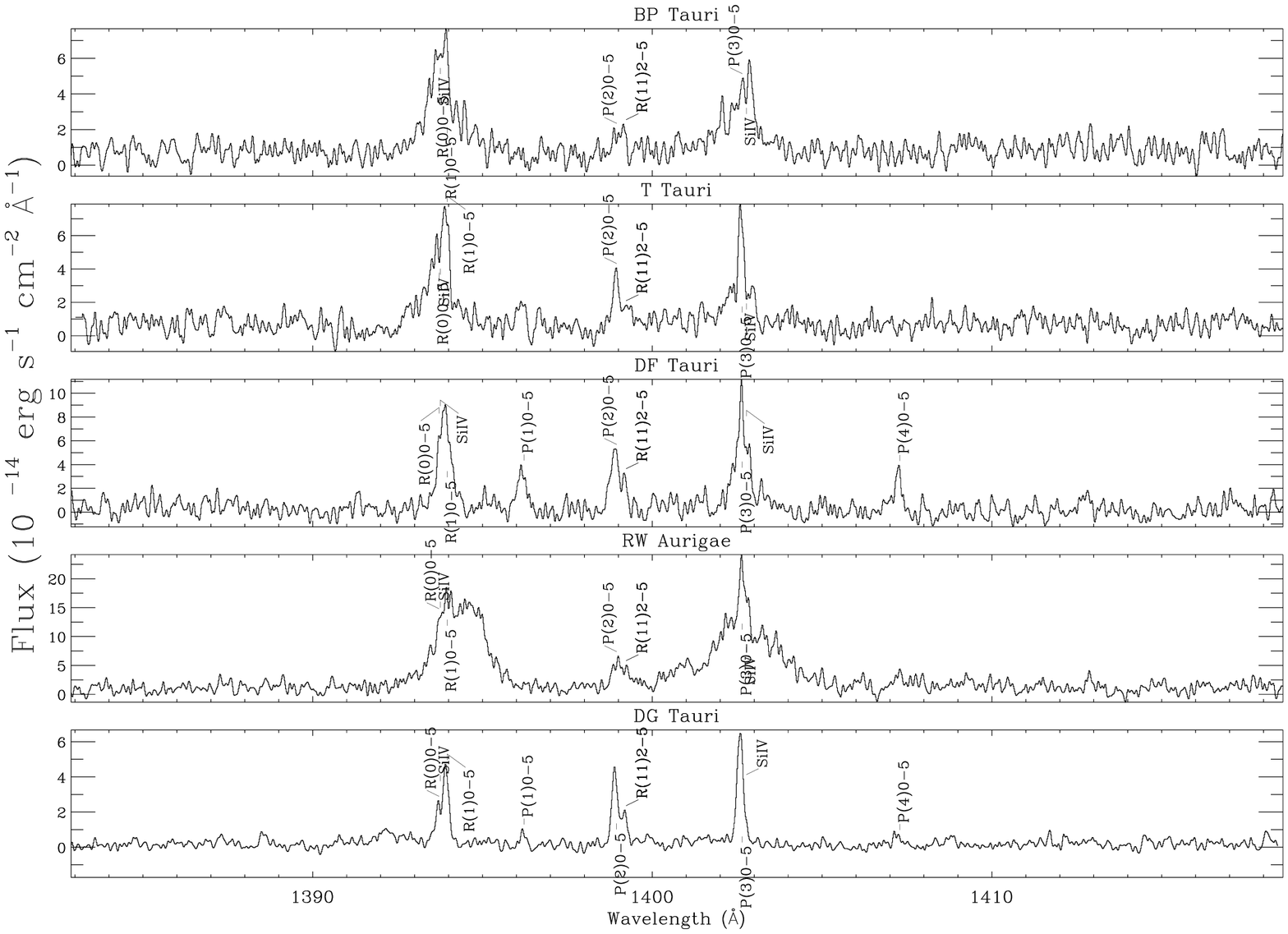}
\caption{Smoothed spectra centered around 1400 \AA. This region contains the \siiv\ doublet.\label{siiv_spec_1}}
\end{figure}

\begin{figure}
\plotone{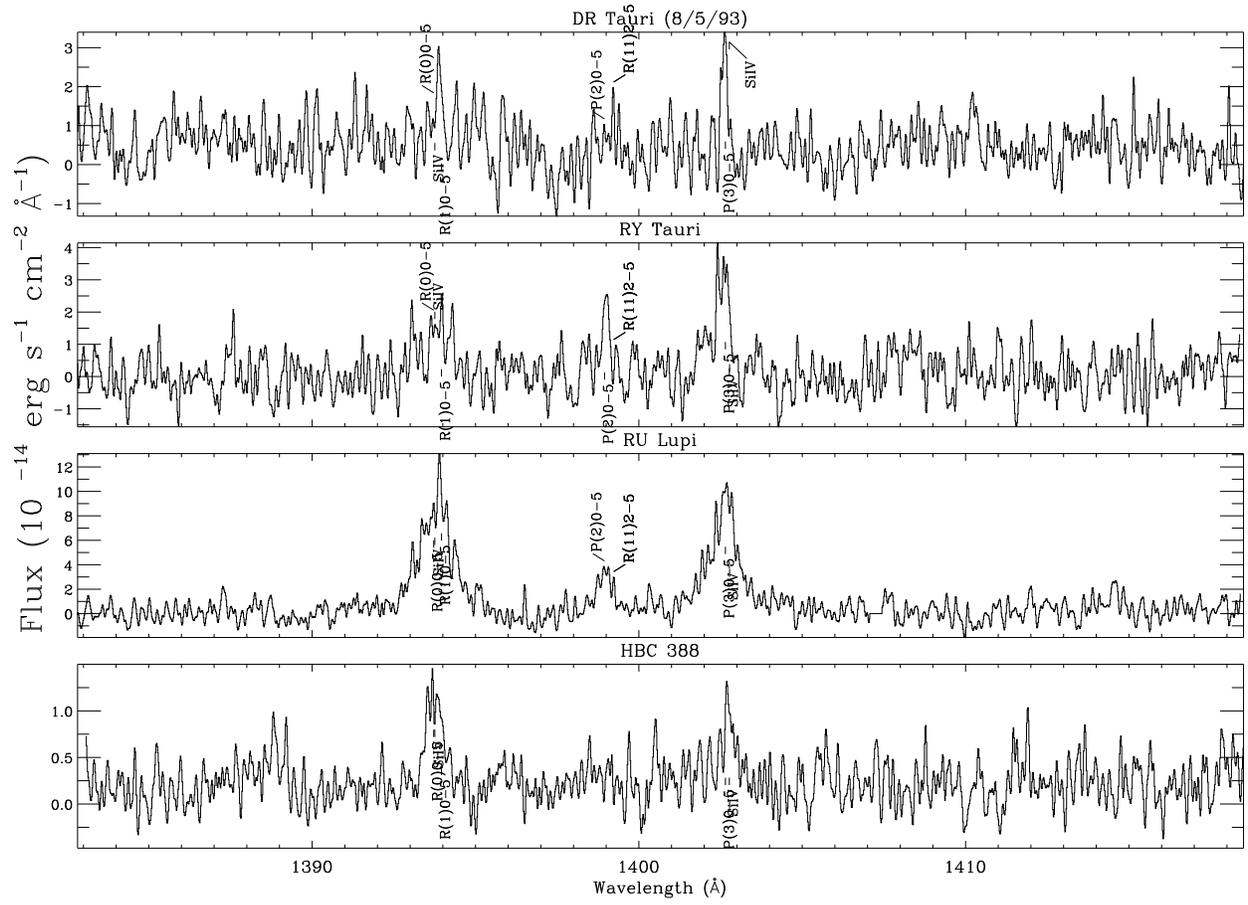}
\caption{Smoothed spectra centered around 1400 \AA, cont.\label{siiv_spec_2}}
\end{figure}

\begin{figure}
\plotone{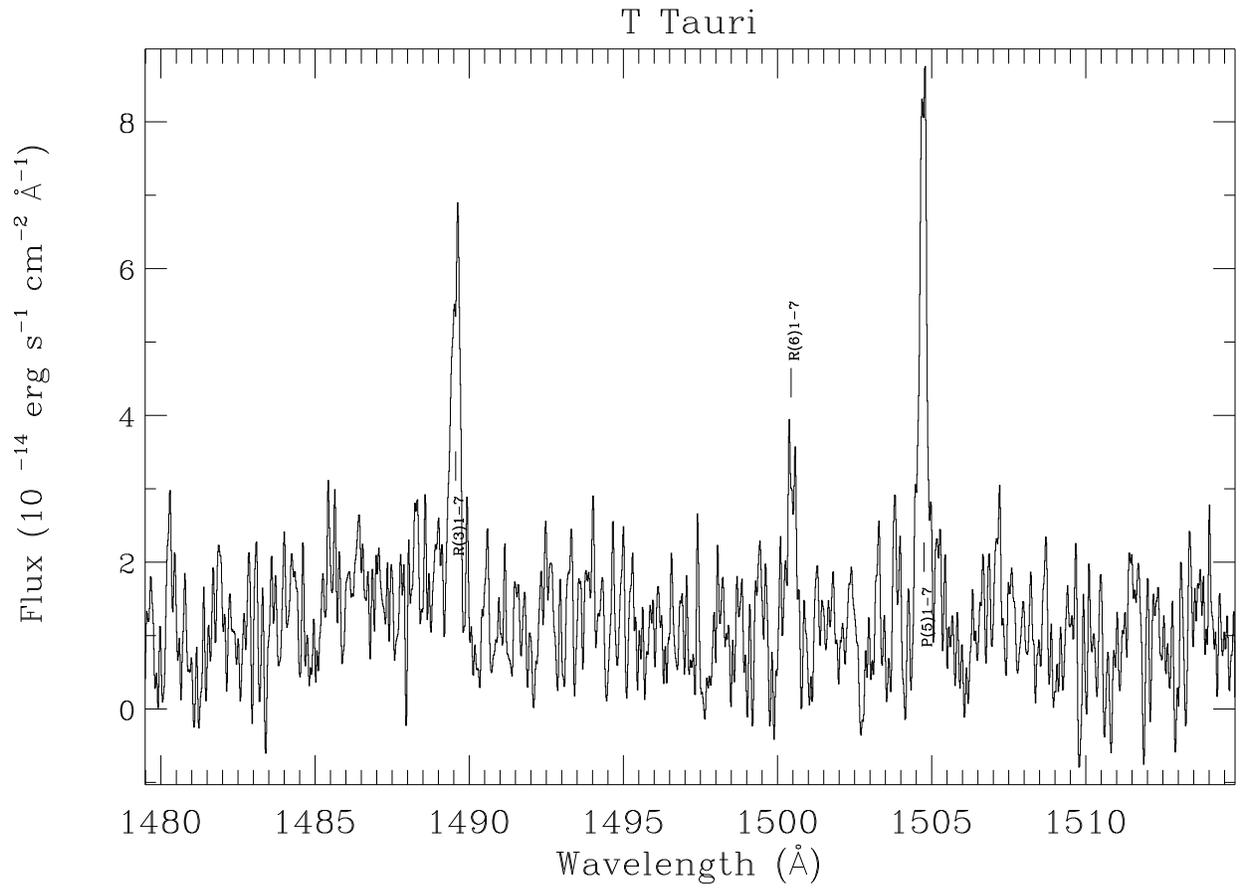}
\caption{We only have one spectrum centered around 1500\AA. \label{t_1500_spec}}
\end{figure}

\begin{figure}
\plotone{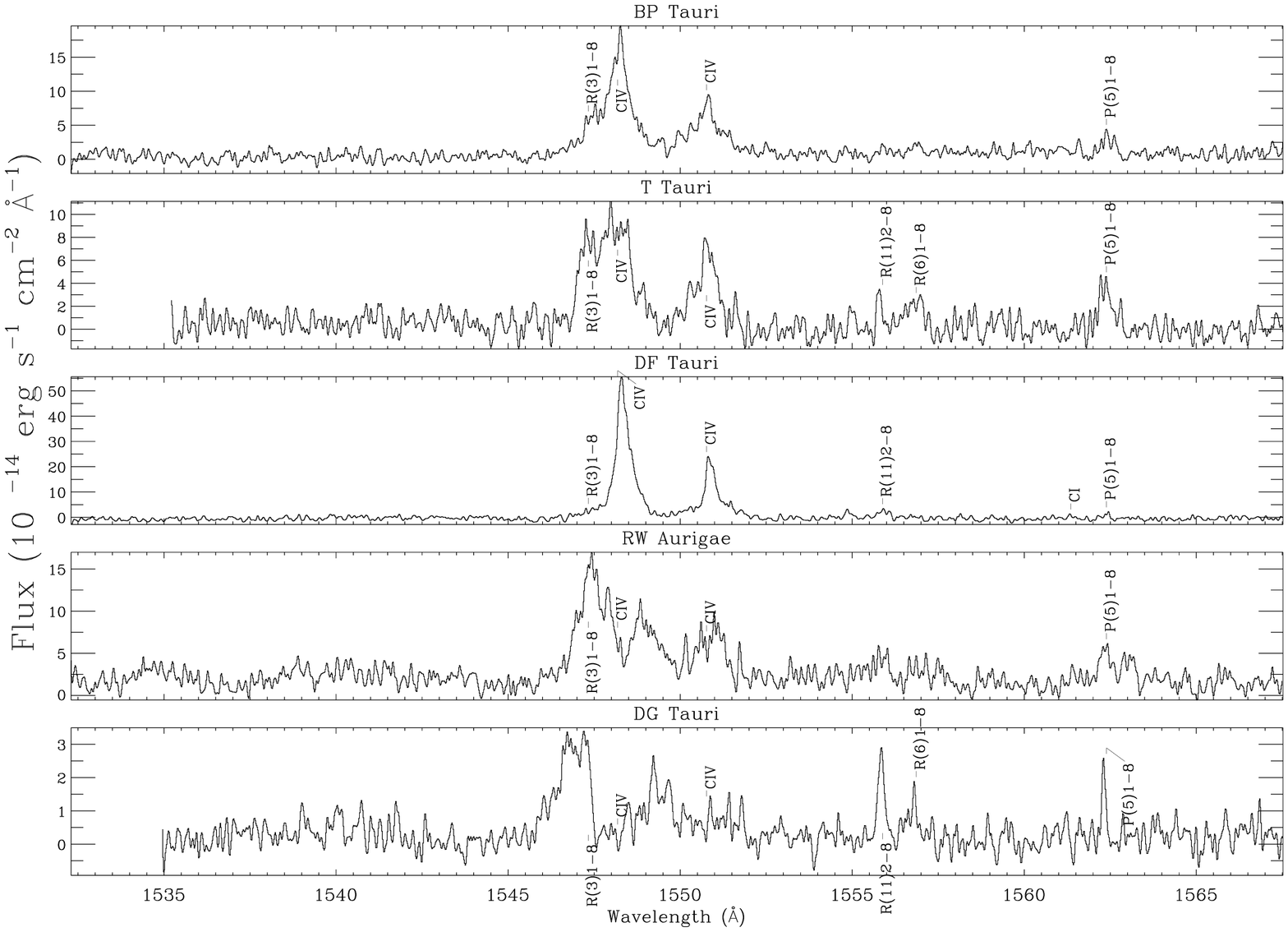}
\caption{Smoothed spectra centered around 1550 \AA. This region contains the \civ\ doublet.\label{civ_spec_1}}
\end{figure}

\begin{figure}
\plotone{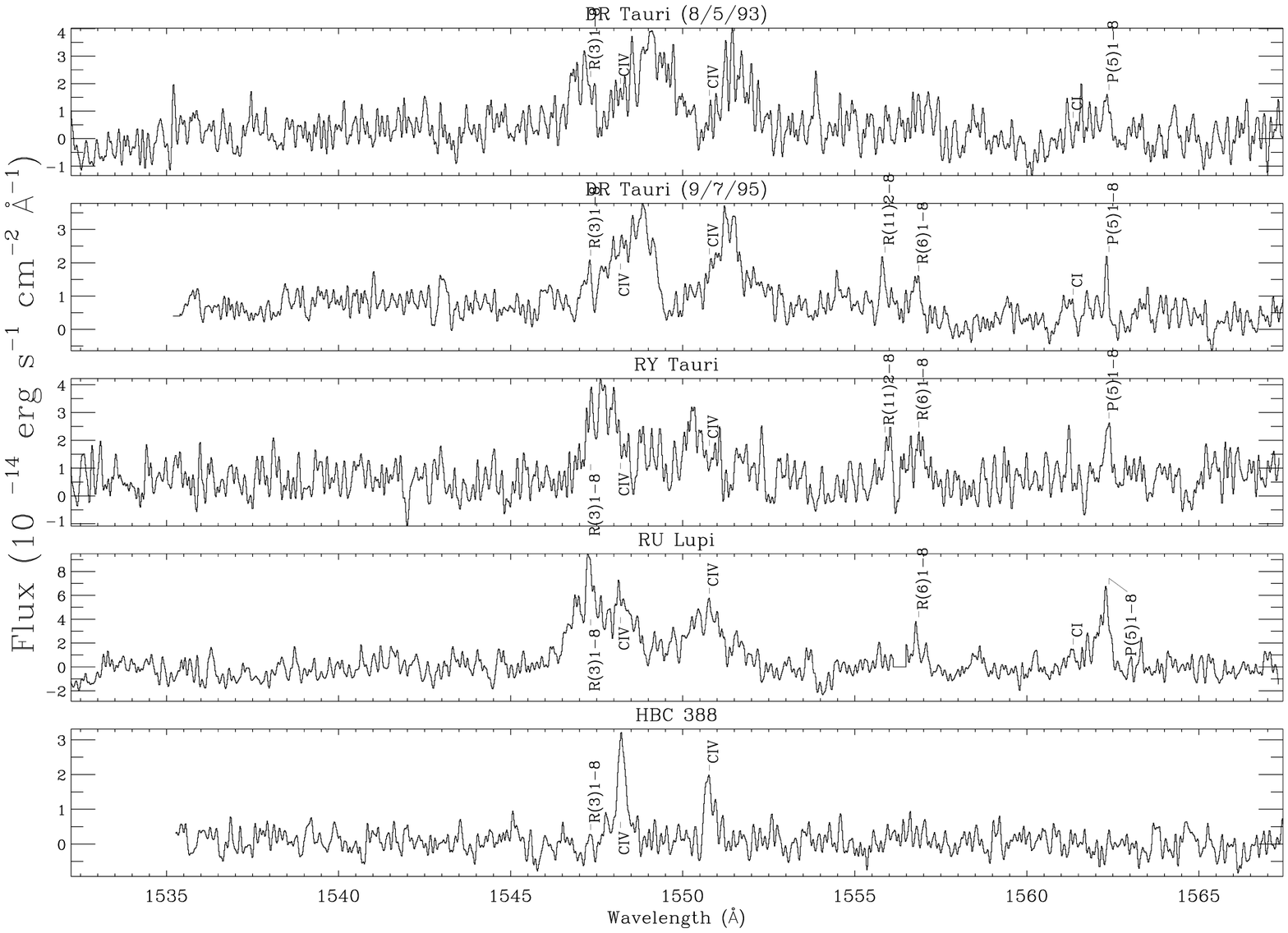}
\caption{Smoothed spectra centered around 1550 \AA, cont.\label{civ_spec_2}}
\end{figure}

\begin{figure}
\plotone{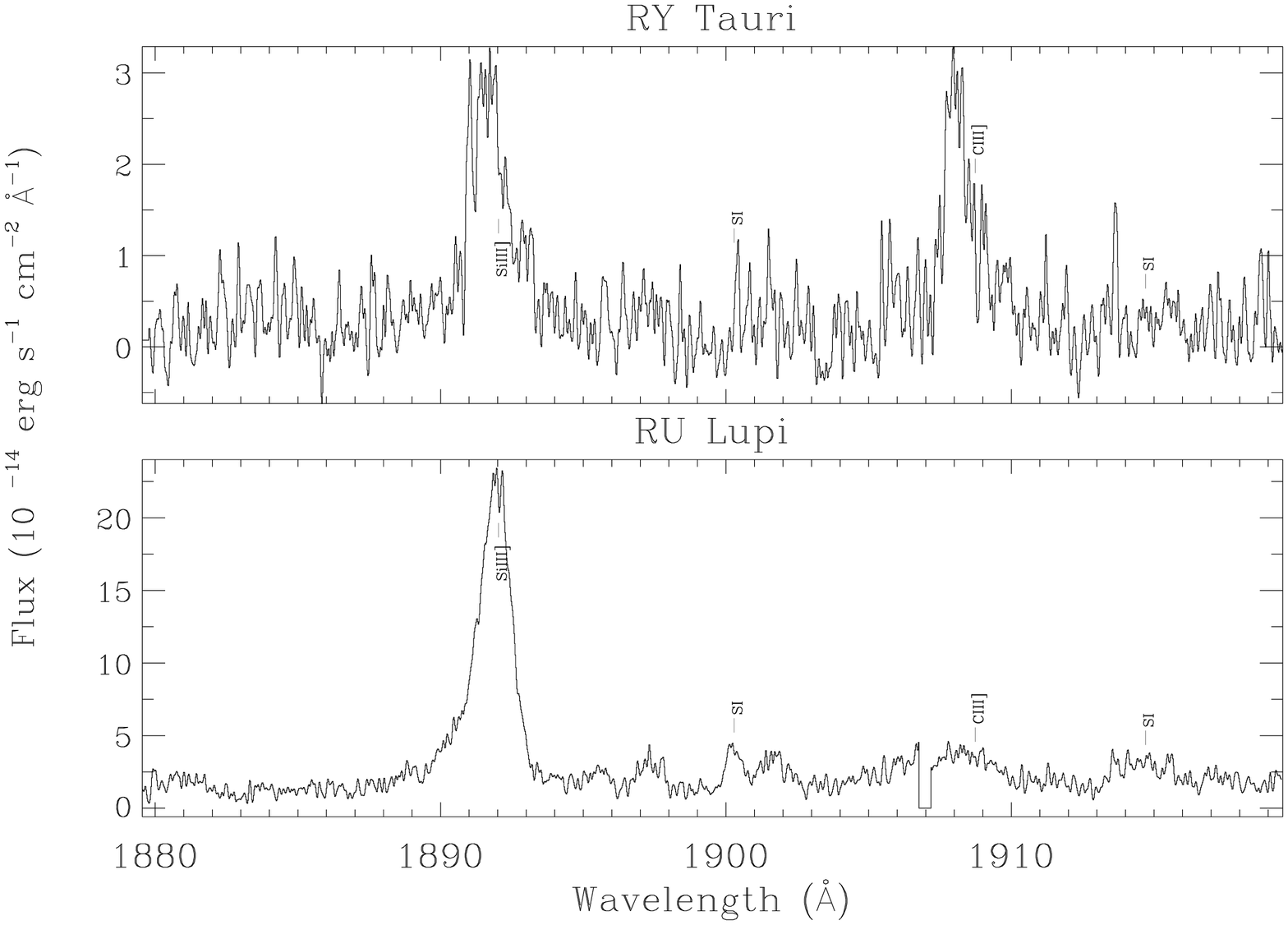}
\caption{Smoothed spectra centered around 1900 \AA. \label{siii_spec}}
\end{figure}

Figures \ref{bluest_spec} to \ref{siii_spec} show our ultraviolet data. All the spectra plotted are in the stellar rest frame. Spectra centered around the same wavelength are grouped together. To facilitate line identifications, the spectra have been convolved with a Gaussian having FWHM=0.072 \AA\ (four pixels). For each spectrum we mark the lines for which we measure a flux. Note that in the spectra of RU Lup one bad diode
         was not properly accounted for at the time these data were
         reduced. We have zeroed the affected points.


The main features of the UV spectra are the \siiv\ (1393.755, 1402.770 \AA) and \civ\ (1548.187, 1550.772 \AA) resonance doublets. In both doublets the blue member has an f-value that is twice that of the red member. All doublet lines are fairly wide ($\sim150$ to $\sim300$ \kms) and show a wide array of shapes and fluxes, from the very symmetric broad peaks of BP Tau to the narrow peaks (in \siiv) and blueshifted features (in \civ) for DG Tau.

Atomic lines from other species (\oi\ and \cii) are also present in emission for some of the spectra. They indicate the presence of an extended low temperature plasma. Although \cii\ is blended with a group of \htwo\ lines which make its analysis difficult, it is clear that these low temperature lines show blueshifted absorption, probably due to the stellar wind. The data do not have enough signal-to-noise to characterize these lines very precisely. 

Most of the spectra show conspicuous \htwo\ features, which blend with the \siiv\ and \civ\ lines, as indicated in the introduction. The exception is HBC 388, for which only the peaks of the \siiv\ and \civ\ lines are significant above the noise. All the detected \htwo\ transitions correspond to the Lyman Band. The lines are generally sharp (FWHM$\sim40$ \kms).

\section{Analysis}
As indicated below, we have measured the parameters of \htwo, \civ\ and \siiv. The results are shown in Table \ref{all_lines}. For each measurement we have used the noise vector provided with the data to obtain an error estimate. The numbers in Table \ref{all_lines} without error estimates are deduced from other parameters. 

In general, we have measured fluxes, full widths at half-maxima, and velocity shifts, by fitting a Gaussian or a combination of Gaussians to the profiles, using a Marquardt routine. The fits have been done on the unsmoothed spectra. For some stars we perform line decomposition in terms of a broad component (BC) and a narrow component (NC). For DF Tau, we decompose each member of the \civ\ doublet in three components labeled C1, C2, C3. The \htwo\ lines are grouped lines that originate in the same upper v', J' level. We have measured fluxes or upper limits in all the \htwo\ lines that are detected in at least one star. If a given line is not explicitly indicated it means that it was not in the spectral range observed for a given star. For those lines that are not detected we obtain 2-$\sigma$ upper limits. This limits are obtained assuming a width equal to the average of all the other \htwo\ lines in the star, integrating a smoothed noise vector over the width, and multiplying by two. Such procedure produces upper-limits which coincide with those produced by more formal Montecarlo simulations. These limits are indicated in Table \ref{all_lines} without centroid velocity. Some of \htwo\ line fluxes were obtained by estimating their intensity from other lines in the same fluorescent route (Section \ref{molhyd}). They are quoted in the Table without errors. In quoting the parameters for \civ\ and \siiv\ in Table \ref{all_lines}, the \htwo\ emission has been subtracted from the profile, as discussed in section \ref{si_c_no_h2}. 

\clearpage
%\documentclass{aastex}
%\begin{document}
%\newcommand \heI{\ion{He}{1}}
%\newcommand \hal{H$\alpha$}
%\newcommand \hbeta{H$\beta$}
%\newcommand \hgamn{H$\gamma$}
%\newcommand \hdeln{H$\delta$}
%\newcommand \kms{km s$^{-1}$}
%\newcommand \nosp{$\!\!\!\!$}
%\newcommand \htwo{H$_2$~}
%\newcommand \mgii{\ion{Mg}{2}~}

\begin{deluxetable}{ccccc} 
\tabletypesize{\scriptsize}
\tablecolumns{5} 
\tablewidth{0pc} 
\tablecaption{\label{all_lines}Line Measurements} 
\tablehead{\colhead{Line} & 
\colhead{Center\tablenotemark{a}}& 
\colhead{FWHM\tablenotemark{b}} &
\colhead{Flux \tablenotemark{c}}& 
\colhead{Unred. Flux \tablenotemark{c}}  }
\startdata 
\cutinhead{BP Tauri}

R(0)0-5 (1393.72 \AA)&\nodata & $60$ & $<0.6$& $<2 $ \\
P(2)0-5 (1398.95 \AA)& $-4\pm7$&$50\pm10$&$0.3\pm0.1$&$1.2\pm0.4$ \\
&\\
R(1)0-5 (1393.96 \AA)&\nodata & $60$ & $<0.6$& $<2$ \\
P(3)0-5 (1402.65 \AA)&\nodata & $60$ & $<0.6$& $<2$ \\
&\\
P(1)0-5 (1396.22 \AA)&\nodata & $60$ & $<0.6$& $<2$ \\
&\\
P(4)0-5 (1407.29 \AA)&\nodata & $60$ & $<0.5$& $<2$ \\
&\\
R(3)1-8 (1547.33 \AA)& $13\pm9$ & $60\pm20$ & $0.5\pm0.2$ & $1.7\pm0.8$ \\
P(5)1-8 (1562.39 \AA)& $6\pm6$ & $70\pm10$ & $0.9\pm0.2$ & $3.1\pm0.7$ \\
&\\
R(6)1-8 (1556.86 \AA)&\nodata & $60$ & $<1$& $<4$ \\
&\\
R(11)2-5 (1399.23 \AA)& $-11\pm7$ & $30\pm20$ & $0.3\pm0.2$ & $1.0\pm0.6$\\
R(11)2-8 (1555.88 \AA)&\nodata & $60$  & $<1$& $<4$ \\
&\\
SiIV (1393.755 \AA) NC& $-1.0\pm5$ & $100\pm20$ & $3.0\pm0.5$ & $11\pm2$ \\
BC & $5\pm20$ & $400\pm100$ & $3\pm1$ & $11\pm5$ \\
SiIV (1402.770 \AA) NC& $5\pm8$ & $100\pm30$ & $1.3\pm0.4$ & $5\pm2$ \\
BC & $10\pm30$ & $340\pm60$ & $2.9\pm0.9$ & $11\pm4$ \\
&\\
&\\
CIV (1548.187 \AA)\ NC& $11\pm2$ & $81\pm4$ & $4.3\pm0.3$ & $15\pm1$ \\
BC & $-10\pm5$ & $300\pm10$ & $11.7\pm0.6$ & $40\pm2$ \\
CIV (1550.772 \AA)\ NC &$6\pm3$ & $80\pm6$ & $2.1\pm0.3$ & $7.1\pm0.9$ \\
BC & $-0\pm6$ & $300\pm10$ & $6.2\pm0.4$ & $21\pm1$ \\
\cutinhead{T Tauri}

R(0)0-4 (1333.47 \AA)&$-22\pm7$&$30\pm20$&$0.5\pm0.2$&$50\pm30$ \\
P(2)0-4 (1338.57 \AA)&$-20\pm5$&$50\pm10$&$1.0\pm0.3$&$110\pm30$\\
R(0)0-5 (1393.72 \AA)&$-5$&$40$&$0.36$&$32.4$ \\
P(2)0-5 (1398.95 \AA)& $-5\pm1$&$40\pm2$&$0.73\pm0.04$&$63\pm3$\\
&\\
R(1)0-4 (1333.80 \AA)&$-33\pm4$&$50\pm20$&$0.5\pm0.2$&$60\pm20$\\
P(3)0-4 (1342.26 \AA)&$-21\pm2$ & $37\pm5$ & $1.0\pm0.2$ & $110\pm20$ \\
R(1)0-5 (1393.96 \AA)&$-8$&$40$&$0.68$&$58.7$\\
P(3)0-5 (1402.65 \AA)&$-8$ &$40$&$0.96$ & $82.2$ \\
&\\
P(1)0-5 (1396.22 \AA)&\nodata & $50$ & $<0.5$& $<50   $ \\
&\\
P(4)0-4 (1346.91 \AA)&\nodata & $50$ & $<0.5$& $<50   $ \\
P(4)0-5 (1407.29 \AA)&\nodata & $50$ & $<0.5$& $<40   $ \\
&\\
R(3)1-7 (1489.56 \AA)&$-5\pm4$&$70\pm10$&$2.2\pm0.5$&$140\pm30$ \\
P(5)1-7 (1504.75 \AA)&$-9\pm3$&$63\pm7 $&$2.4\pm0.3$&$160\pm20$\\
R(3)1-8 (1547.33 \AA)&$-12$ & $91$ &$2.1$&$120$ \\
P(5)1-8 (1562.39 \AA)& $-3\pm2$&$90\pm3$&$2.1\pm0.1$&$120\pm6$ \\
&\\
R(6)1-7 (1500.44 \AA)&$-0.4\pm7$&$80\pm20$&$1.3\pm0.4$&$80\pm30$ \\
R(6)1-8 (1556.86 \AA)&$-14\pm9$ & $130\pm20$ & $1.8\pm0.3$ & $110\pm20$ \\
&\\
P(13)2-3 (1325.34 \AA)&\nodata & $50$ & $<0.5$& $<60   $ \\
R(11)2-5 (1399.23 \AA)&\nodata & $50$ & $<0.6$& $<50   $ \\
R(11)2-8 (1555.88 \AA)&$-18\pm2$&$24\pm5$&$0.6\pm0.2$&$36\pm9$ \\
&\\
SiIV (1393.755 \AA)&$-15\pm7$ & $220\pm20$ & $3.6\pm0.4$ & $320\pm30$ \\
SiIV (1402.770 \AA)&$-50\pm10$ & $220\pm30$ & $2.2\pm0.4$ & $190\pm30$ \\
&\\
CIV (1548.187 \AA)&$-25\pm6$ & $250\pm10$ & $11.8\pm0.9$ & $700\pm50$ \\
CIV (1550.772 \AA)&$0\pm6$ & $150\pm10$ & $4.8\pm0.6$ & $280\pm40$ \\

\cutinhead{DF Tauri}
R(0)0-4 (1333.47 \AA)&$-7\pm8$ & $60\pm20$ & $1.0\pm0.3$ & $3\pm1$ \\
P(2)0-4 (1338.57 \AA)&$-6\pm2$&$58\pm4$&$1.3\pm0.1$&$4.6\pm0.4$\\
R(0)0-5 (1393.72 \AA)&\nodata & $30$ & $<0.4$& $<1$ \\
P(2)0-5 (1398.95 \AA)& $-12\pm4$ & $70\pm10$ & $1.7\pm0.3$ & $6\pm1$ \\
&\\
R(1)0-4 (1333.80 \AA)&$-5\pm3$ & $60\pm10$ & $1.8\pm0.4$ & $6\pm1$ \\
P(3)0-4 (1342.26 \AA)&$-8\pm2$ & $62\pm4$ & $1.8\pm0.1$ & $6.1\pm0.4$ \\
R(1)0-5 (1393.96 \AA)&$-18\pm3$ & $94\pm6$ & $3.8\pm0.3$ & $12\pm1$ \\
P(3)0-5 (1402.65 \AA)&$-7\pm3$ & $89\pm7$ & $3.4\pm0.3$ & $11\pm1$ \\
&\\
P(1)0-5 (1396.22 \AA)&$-14\pm5$ & $80\pm10$ & $1.4\pm0.3$ & $4\pm1$ \\
&\\
P(4)0-4 (1346.91 \AA)&$-6\pm5$&$30\pm10$&$0.4\pm0.2$&$1.3\pm0.6$ \\
P(4)0-5 (1407.29 \AA)&$-8\pm6$ & $40\pm10$ & $0.8\pm0.3$ & $2.7\pm1$ \\
&\\
R(3)1-8 (1547.33 \AA)&\nodata & $30$ & $<0.5$& $<2 $ \\
P(5)1-8 (1562.39 \AA)& $5\pm4$&$24\pm8$&$0.4\pm0.1$&$1.1\pm0.4$ \\
&\\
R(6)1-8 (1556.86 \AA)&\nodata & $30$ & $<0.4$& $<1 $ \\
&\\
R(11)2-5 (1399.23 \AA)& $-7\pm5$ & $30\pm10$ & $0.4\pm0.2$ & $1.2\pm0.6$ \\
R(11)2-8 (1555.88 \AA)&\nodata & $30$  & $<0.4$& $<1 $ \\
&\\
SiIV (1393.755 \AA) & \nodata & $300$ & $<4 $& $<14$ \\
SiIV (1402.770 \AA) &  \nodata & $300$ & $<4$ & $<14$\\
&\\
CIV (1548.187 \AA)\ C1& $10\pm10$ & $350\pm30$ &$10\pm1$ & $28\pm4$  \\
C2 & $19.5\pm0.9$ & $41\pm5$ & $6.0\pm0.5$ & $18\pm1$ \\
C3 & $42\pm2$ & $122\pm4$ & $17.7\pm0.9$ & $52\pm3$ \\
CIV (1550.772 \AA)\ C1& $10\pm10$ & $320\pm20$ & $9\pm1$ & $25\pm3$ \\
C2 & $18\pm1$ & $52\pm3$ & $5.3\pm0.4$ & $15\pm1$ \\
C3 & $60\pm20$ & $130\pm30$ & $1.8\pm0.6$ & $5\pm2$ \\

\cutinhead{RW Auriga}

R(0)0-5 (1338.57 \AA)&\nodata & $50$ & $<1$ & $<23$ \\
P(2)0-5 (1398.95 \AA)&$4\pm3$ & $80\pm20$ & $1.8\pm0.7$ & $40\pm10$ \\
&\\ 
R(1)0-5 (1393.96 \AA)&$-20$&$60$&$1.4$&$28$ \\
P(3)0-5 (1402.65 \AA)&$-0\pm2$&$49\pm5$&$2.4\pm0.3$&$47\pm6$ \\
&\\
P(1)0-5 (1396.22 \AA)&\nodata & $50$ & $<0.8$& $<16  $ \\
&\\
P(4)0-5 (1407.29 \AA)&\nodata & $50$ & $<0.8$& $<16   $ \\
&\\
R(3)1-8 (1547.33 \AA)&$15.4$&$60$&$1.5$&$23$ \\
P(5)1-8 (1562.39 \AA)& $-6\pm5$&$70\pm10$&$1.4\pm0.3$&$21\pm5$ \\
&\\
R(6)1-8 (1556.86 \AA)&\nodata & $50$ & $<1$& $<15   $ \\
&\\
R(11)2-5 (1399.23 \AA)& $5\pm6$&$20\pm20$&$<0.6$&$<12$ \\
R(11)2-8 (1555.88 \AA)&\nodata &$50$  & $<1$& $<15   $ \\
&\\
SiIV (1393.755 \AA)& $160\pm20$ & $290\pm30$ & $19\pm2$ & $380\pm40$ \\
SiIV (1402.770 \AA)& $-20\pm8$ & $500\pm20$ & $25\pm2$ & $490\pm40$ \\
&\\
CIV (1548.187 \AA)\tablenotemark{d}& \nodata & \nodata&$14\pm5$ & $220\pm80$ \\
CIV (1550.772 \AA)\tablenotemark{d}&\nodata & \nodata& $9\pm4$ & $140\pm60$\\
\cutinhead{DG Tauri}

R(0)0-5 (1393.72 \AA)&$-4\pm6$&$40\pm10$&$0.4\pm0.1$&$27\pm9$ \\
P(2)0-5 (1398.95 \AA)&$-9\pm2$&$39\pm5$&$0.9\pm0.1$&$59\pm8$ \\
&\\ 
R(1)0-5 (1393.96 \AA)&$-4\pm3$&$38\pm6$&$0.9\pm0.2$&$60\pm10.$ \\
P(3)0-5 (1402.65 \AA)&$-10\pm1$&$47\pm3$&$1.5\pm0.1$&$97\pm9$ \\
&\\
P(1)0-5 (1396.22 \AA)&$-3\pm5$ & $60\pm20$ & $0.3\pm0.1$ & $19\pm7$ \\
&\\
P(4)0-5 (1407.29 \AA)&$-20\pm10.$&$40\pm30$&$<0.4$&$<40$ \\
&\\
R(3)1-8 (1547.33 \AA)&$0.0$& $27.3$& $0.32$&$15.0$ \\
P(5)1-8 (1562.39 \AA)&$-12\pm2$ & $24\pm6$ & $0.3\pm0.1$ & $14\pm5$ \\
&\\
R(6)1-8 (1556.86 \AA)&$-7\pm6$ & $30\pm10$ & $0.2\pm0.1$ & $10\pm7$ \\
&\\
R(11)2-5 (1399.23 \AA)&$-9\pm5$&$40\pm10$&$0.4\pm0.1$&$27\pm9$ \\
R(11)2-8 (1555.88 \AA)&$-4\pm3$ & $42\pm7$ & $0.6\pm0.1$ & $28\pm6$ \\
&\\
SiIV (1393.755 \AA)& \nodata & $200$& $<1.3$& $<80$  \\
SiIV (1402.770 \AA)& \nodata & $200$& $<1.3$& $<80$  \\
&\\
CIV (1548.187 \AA)&$-260\pm6$ & $170\pm10$ & $2.7\pm0.3$ & $140\pm20$ \\
CIV (1550.772 \AA)& $-270\pm10$ & $230\pm30$ & $1.7\pm0.3$ & $90\pm20$ \\

\cutinhead{DR Tauri (8/5/93)}

R(0)0-4 (1333.47 \AA) &\nodata &$40$  & $<0.4$& $<10 $ \\
P(2)0-4 (1338.57 \AA) &\nodata & $40$ & $<0.3$& $<7 $ \\
R(0)0-5 (1393.72 \AA) &\nodata &$40$  &$<0.4$& $<10$ \\
P(2)0-5 (1398.95 \AA)  &\nodata & $40$ & $<0.3$& $<8$ \\
&\\
R(1)0-4 (1333.80 \AA) &\nodata & $40$ & $<0.4$& $<10$ \\
P(3)0-4 (1342.26 \AA) &$-11\pm3$ & $45\pm7$ & $0.36\pm0.07$ & $10\pm2$ \\
R(1)0-5 (1393.96 \AA) &$-15\pm6$&$50\pm20$&$0.6\pm0.3$&$14\pm7$\\
P(3)0-5 (1402.65 \AA) &$-4\pm2$&$30\pm5$&$0.6\pm0.1$&$13\pm3$ \\
&\\
P(1)0-5 (1396.22 \AA) &\nodata & $40$ & $<0.4$& $<10$ \\
&\\
P(4)0-4 (1346.91 \AA) &\nodata & $40$ & $<0.3$& $<8$ \\
P(4)0-5 (1407.29 \AA) &\nodata & $40$ & $<0.4$& $<8$ \\
&\\
R(3)1-8 (1547.33 \AA) &$-15$&$40$&$0.3$&$5.3$ \\
P(5)1-8 (1562.39 \AA) & $-15\pm7$ & $50\pm20$ & $0.4\pm0.2$ & $8\pm4$ \\
&\\
R(6)1-8 (1556.86 \AA) &\nodata & $40$ & $<0.5$& $<9$ \\
&\\
R(11)2-5 (1399.23 \AA)  &\nodata & $40$ & $<0.4$& $<9$ \\
R(11)2-8 (1555.88 \AA) &\nodata & $40$ & $<0.5$& $<8$ \\
&\\
SiIV (1393.755 \AA)& \nodata &$150$   & $<1.5$& $<40$ \\
SiIV (1402.770 \AA)& \nodata & $150$  & $<1.5$& $<40$ \\
&\\
CIV (1548.187 \AA)  C1& $-250\pm10$ & $90\pm20$ & $0.9\pm0.3$ & $16\pm5$ \\
C2&$130\pm10$ & $150\pm20$ & $2.0\pm.2$ & $40\pm4$ \\

CIV (1550.772 \AA) C1 &$-260\pm20$ & $90\pm20$ & $1\pm1$ & $20\pm20$ \\
C2&$147\pm9$ & $150\pm20$ & $1.8\pm0.4$ & $31\pm6$ \\

\cutinhead{DR Tauri (9/7/95)}

R(0)0-4 (1333.47 \AA) &\nodata & $20$ & $<0.1 $& $<3$ \\
P(2)0-4 (1338.57 \AA) &$-19\pm5$ & $50\pm10$ & $0.3\pm0.1$ & $9\pm3$ \\
&\\
R(1)0-4 (1333.80 \AA) &\nodata & $20$ & $<0.1$& $<2$ \\
P(3)0-4 (1342.26 \AA) &$-3\pm3$&$20\pm2$&$0.25\pm0.07$&$7\pm2$ \\
&\\
P(4)0-4 (1346.91 \AA) &\nodata & $20$ & $<0.1$& $<3$ \\
&\\
R(3)1-8 (1547.33 \AA) &$-14$&$20$&$0.25$&$0.43$ \\
P(5)1-8 (1562.39 \AA) & $-14\pm2$&$14\pm4$&$0.21\pm0.05$&$3.7\pm0.9$ \\
&\\
R(6)1-8 (1556.86 \AA) &$-5\pm4$&$29\pm8$&$0.19\pm0.07$&$3\pm1$ \\
&\\
R(11)2-8 (1555.88 \AA) &$-13\pm2$ & $26\pm5$ & $0.21\pm0.05$ & $3.7\pm0.9$ \\

&\\
CIV (1548.187 \AA) & $75\pm6$ & $240\pm20$ & $3.3\pm0.3$ & $59\pm5$ \\
CIV (1550.772 \AA)&$110\pm6$ & $190\pm10$ & $2.4\pm0.2$ & $42\pm4$ \\

\cutinhead{RY Tauri}

R(0)0-5 (1393.72 \AA)&\nodata & $50$ &$<0.6$ &$<1.3$   \\
P(2)0-5 (1398.95 \AA)& $9\pm4$&$48\pm8$&$0.7\pm0.2$&$1.5\pm0.4$ \\
&\\
R(1)0-5 (1393.96 \AA)&$2\pm5$ & $20\pm20$ & $<0.6$ & $<1.3$ \\
&\\
P(3)0-5 (1402.65 \AA)&$-2\pm6$ & $60\pm20$ & $1.0\pm0.3$ & $2.1\pm0.6$ \\
&\\
P(4)0-5 (1407.29 \AA)&\nodata & $50$ & $<0.6$& $<1.3$ \\
&\\
P(1)0-5 (1396.22 \AA)&\nodata & $50$ & $<0.6$& $<1.3$ \\
&\\
R(3)1-8 (1547.33 \AA)&$-4$&$50$&$0.47$&$2.95$ \\
P(5)1-8 (1562.39 \AA)& $-8\pm4$ & $44\pm8$ & $0.5\pm0.1$ & $1.0\pm0.2$ \\
&\\
R(6)1-8 (1556.86 \AA)&\nodata & $50$   & $<0.6$& $<1.2$ \\
&\\
R(11)2-5 (1399.23 \AA)&\nodata &  $50$  & $<0.6$& $<1.3$ \\
R(11)2-8 (1555.88 \AA)&\nodata &  $50$  & $<0.6$& $<1.2$ \\
&\\
SiIV (1393.755 \AA)& \nodata & $150$& $<2$& $<4.3$ \\
SiIV (1402.770 \AA)& \nodata &$150$ & $<2$& $<4.3$  \\
&\\
CIV (1548.187 \AA)& $78\pm8$ & $160\pm20$ & $2.7\pm0.4$ & $5.4\pm0.8$ \\
CIV (1550.772 \AA)&$80\pm10$ & $140\pm20$ & $1.5\pm0.3$ & $3.0\pm0.6$ \\

\cutinhead{RU Lupi}
R(0)0-5 (1393.72 \AA)&\nodata&$40$&$<0.7$& $<20$ \\
P(2)0-5 (1398.95 \AA)&  $-4\pm7$ & $30\pm20$ & $0.4\pm0.4$ & $10\pm10$  \\
&\\
R(1)0-5 (1393.96 \AA)&$-12$ &$30$& $0.37$ &$10.7$\\ 
P(3)0-5 (1402.65 \AA)&$-12$&$30$&$0.51$&$14.7$ \\
&\\
P(1)0-5 (1396.22 \AA)&\nodata & $40$ & $<0.5 $& $<20$ \\
&\\
P(4)0-5 (1407.29 \AA)&\nodata &$40$  & $<0.6$& $<20$ \\
&\\
R(3)1-8 (1547.33 \AA)&$-17$&$40$&$2.0$&$42$\\
P(5)1-8 (1562.39 \AA)&$-28\pm4$ & $78\pm8$ & $2.3\pm0.3$ & $50\pm6$ \\
&\\
R(6)1-8 (1556.86 \AA)&$-16\pm3$&$16\pm6$&$0.3\pm0.1$&$7\pm3$ \\
&\\
R(11)2-5 (1399.23 \AA)&\nodata &$40$  & $<0.5$& $<10$ \\
R(11)2-8 (1555.88 \AA)&\nodata &$40$  & $<0.8$& $<20$ \\
&\\
SiIV (1393.755 \AA)&$5\pm5$ & $270\pm10$ & $10.8\pm0.7$ & $310\pm20$ \\
SiIV (1402.770 \AA)&$-31\pm6$ & $250\pm10$ & $8.5\pm0.6$ & $240\pm20$ \\
&\\
CIV (1548.187 \AA)&$-20\pm20$ & $340\pm40$ & $8\pm1$ & $170\pm20$ \\
CIV (1550.772 \AA)&$-10\pm10$ & $300\pm30$ & $6.6\pm0.8$ & $140\pm20$ \\

\cutinhead{HBC 388}
R(0)0-5 (1393.72 \AA)&\nodata & $40$ & $<0.2$& $<0.2$ \\
P(2)0-5 (1398.95 \AA)&\nodata & $40$ & $<0.2$& $<0.2$ \\
&\\
R(1)0-5 (1393.96 \AA)&\nodata & $40$ & $<0.2$& $<0.2$ \\
P(3)0-5 (1402.65 \AA)&\nodata & $40$ & $<0.2$& $<0.2$ \\
&\\
P(1)0-5 (1396.22 \AA)&\nodata & $40$ & $<0.2$& $<0.2$ \\
&\\
P(4)0-5 (1407.29 \AA)&\nodata & $40$ & $<0.2$& $<0.2$ \\
&\\
R(3)1-8 (1547.33 \AA)&\nodata & $40$ & $<0.3$& $<0.3$ \\
P(5)1-8 (1562.39 \AA)& \nodata & $40$ & $<0.3$& $<0.3$ \\
&\\
R(6)1-8 (1556.86 \AA)&\nodata & $40$ & $<0.3$& $<0.3$ \\
&\\
R(11)2-5 (1399.23 \AA)&\nodata & $40$ & $<0.2$& $<0.2$ \\
R(11)2-8 (1555.88 \AA)&\nodata & $40$ & $<0.2$& $<0.3$ \\
&\\
SiIV (1393.755 \AA)& $-1\pm7$ & $130\pm20$ & $0.7\pm0.1$ & $0.9\pm0.2$ \\
SiIV (1402.770 \AA)&$0\pm10$ & $110\pm30$ & $0.4\pm0.1$ & $0.5\pm0.1$ \\
&\\
CIV (1548.187 \AA)& $8\pm3$&$61\pm7$&$0.9\pm0.2$&$1.2\pm0.2$  \\
CIV (1550.772 \AA)& $-0\pm4$ & $52\pm9$ & $0.5\pm0.1$ & $0.6\pm0.1$ \\

\enddata
\tablecomments{All wavelengths are in the stellar rest frame. The errors in each measurement are indicated. For measurements without errors, the parameters of the fit have been assumed from neighboring lines. The upper limits (2-$\sigma$) of the \htwo\ lines have been obtained by integrating a smooth noise vector (multiplied by 2) over the indicated FWHM, which corresponds to the average of other \htwo\ lines. ``NC'' stands for narrow component, ``BC'' is broad component. In DF Tau, we decompose CIV in three components labeled C1, C2, and C3. }

\tablenotetext{a}{Units are \kms.}
\tablenotetext{b}{Units are \kms.}
\tablenotetext{c}{Units are $10^{-14}\rm{\ ergs\ sec^{-1}\ cm^{-2}}$. }
\tablenotetext{d}{Measured directly over the spectrum (i.e., without a Gaussian fit).}
\end{deluxetable}
%\end{document}

\subsection{Molecular hydrogen\label{molhyd}}

The spectra show a large number of \htwo\ lines. These correspond to transitions in the Lyman Band of molecular hydrogen,  $B^1\Sigma^+_u$ to $X^1\Sigma^+_g$. In this work we use the \htwo\ parameters derived by \citet{1993A&AS..101..273A}. Table \ref{all_lines} shows the measured parameters of all the lines present in at least one spectrum. Each transition is indicated in standard notation: the letter tells the change in rotational level from the upper to the lower level (J''-J'=1 is P, J''-J'=-1 is R), the number in parenthesis indicates the J'' value (the rotational level of the lower energy state), and the two final numbers indicate the vibrational levels v', v''  of the upper and lower levels, respectively, for the transition. Table \ref{all_lines} also shows the upper limits of the lines that are not observed in a given star. We have made no attempt to obtain unblended fluxes for the \htwo\ lines near \cii\ (1335 \AA). 

%\documentclass{aastex}
%\begin{document}
%\newcommand \siiv{\ion{Si}{4}~}
%\newcommand \civ{\ion{C}{4}~}

\begin{deluxetable}{cccccccccc} 
\rotate
\tabletypesize{\tiny}
\tablecolumns{10} 
\tablewidth{0pc} 
\tablecaption{\label{h2_lines}\htwo\ Lines Observed} 
\tablehead{\colhead{Line}&
\colhead{BP Tau}&\colhead{T Tau}&
\colhead{DF Tau}&\colhead{RW Aur}&\colhead{DG Tau}&\colhead{DR Tau}&
\colhead{DR Tau}&
\colhead{RY Tau}&\colhead{RU Lup}\\
\colhead{}&
\colhead{}&\colhead{}&
\colhead{}&\colhead{}&\colhead{}&\colhead{(8/5/93)}&
\colhead{(9/7/95)}&
\colhead{}&\colhead{} }
\startdata 
Average shift (\kms)& $0\pm3$ & $-9.7\pm0.7$ &$-2.2\pm0.5$ & $0.5\pm2$ & $-9.2\pm0.9$ &$-7\pm2$&$-11\pm1$ & $0\pm2$ & $-18\pm2$ \\
Average FWHM (\kms) & $57\pm8$ & $52\pm2$ &
$27.4\pm0.8$ & $52\pm4$ & $41\pm2$ &$36\pm4$&$20\pm2$ & $46\pm5$ & $37\pm5$  \\
R(0)0-4 (1333.47 \AA)& \nodata &   y  & y & \nodata & \nodata & u& u   & \nodata & \nodata \\
P(2)0-4 (1338.57 \AA)& \nodata &   y  & y & \nodata & \nodata & u& y   & \nodata & \nodata \\
R(0)0-5 (1393.72 \AA)& u   &   s  & u & s   & y   & u& u   & u   & u   \\ 
P(2)0-5 (1398.95 \AA)& y   &   y  & y & y   & y   & u& \nodata & y   & y   \\
&\\
R(1)0-4 (1333.80 \AA)& \nodata &   y  & y & \nodata & \nodata & u& u   & \nodata & \nodata \\
P(3)0-4 (1342.26 \AA)& \nodata &   y  & y & \nodata & \nodata & y& y   & \nodata & \nodata \\
R(1)0-5 (1393.96 \AA)& u   &   s  & y & s   & y   & y& \nodata & y   & s   \\
P(3)0-5 (1402.65 \AA)& u   &   s  & y & y   & y   & y& \nodata & y   & s   \\
&\\
P(1)0-5 (1396.22 \AA)& u   &   u  & y & u   & y   & u& \nodata & u   & u   \\
&\\
P(4)0-4 (1346.91 \AA)& \nodata &   u  & y & \nodata & \nodata & u& u   & \nodata & \nodata \\
P(4)0-5 (1407.29 \AA)& u   & u    & y & u   & u   & u&u    & u   & u \\
&\\
R(3)1-7 (1489.56 \AA)& \nodata & y    & \nodata  & \nodata  & \nodata & \nodata& \nodata & \nodata & \nodata \\
P(5)1-7 (1504.75 \AA)& \nodata & y     & \nodata  & \nodata  & \nodata & \nodata& \nodata & \nodata & \nodata \\
R(3)1-8 (1547.33 \AA)& y   & s & u    & s    & s   & s&s      & s   & s \\
P(5)1-8 (1562.39 \AA)& y   & y & y    & y    & y   & y&y      & y   & y \\
&\\
R(6)1-7 (1500.44 \AA)& \nodata & y & \nodata  & \nodata  & \nodata & \nodata& \nodata & \nodata & \nodata \\
R(6)1-8 (1556.86 \AA)& u   & u & u    & u    & y   & u&y      &u    & y \\ 
&\\
R(11)2-5 (1399.23 \AA)& y  & u & y    & y    & y   & u& \nodata   & u   &u \\
R(11)2-8 (1555.88 \AA)& u  & y & u    & u    & y   & u&y      & u   &u \\
&\\

\enddata
\tablecomments{The letters mean the following: u: upper limit (This is a 2-$\sigma$ limit), y: parameters measured (by fitting a Gaussian to the line), s:parameters estimated (by using other lines from the same fluorescent route). Three dots indicate that the spectral range in which the line is present is not available.}
\end{deluxetable} 

%\end{document}

Table \ref{h2_lines} summarizes the \htwo\ lines detected in each star. For optically thin lines the ratio of the flux values of two lines with the same upper level should be given by the ratio of the A-values to the wavelength of the transition. For almost every star (except DF Tau, see below) that shows at least two lines with the same upper level, this is the case. This leads us to conclude that \htwo\ levels with $v'' \geq 4$ are optically thin. 

The validity of this conclusion depends on the assumption that reddenings from Table \ref{par} are valid, and that the extinction is the same for the ultraviolet \htwo\ emission lines and for the optical spectra (from which A$_V$ is determined). Given the S/N of the spectra, the flux ratios used here are not very sensitive to errors in the absolute (and poorly determined) extinction values, as observed lines of the same route are generally less than a 1000 \AA\ from each other. For this same reason, it is difficult to estimate whether or not there is differential extinction between the star and the region responsible for the \htwo\ emission.

By using other lines in the same fluorescent route we can obtain estimates of the intensities of the lines blended with \civ\ (R(3)1-8) and \siiv\ (R(0)0-5, R(1)0-5, and P(3)0-5). The procedure is outlined in detail for some stars in section \ref{corr}. Also indicated in Table \ref{h2_lines} is the average FWHM and the average shift in velocity for all the \htwo\ lines in each star. 

In DF tau, the ratios within the fluxes of the four lines belonging to the upper level v'=2, J'=0, do not follow the values expected for optically thin emission. R(0)0-5 and P(3)0-5 coincide with the positions of the \siiv\ resonance doublet, and we have obtained their fluxes assuming that all the observed emission is \htwo. From the appearance of the spectra, this is a reasonable assumption, but the fact that we do not obtain ratios consistent with optically thin emission may be telling us that there is indeed some underlying \siiv\ emission. Roughly half of the flux in  R(0)0-5 and P(3)0-5 would have to be \siiv\ to make their fluxes have a value consistent with optically thin emission.  The two lines also have some of the largest FWHM measured for \htwo. These two facts suggest that the core of a broad \siiv\ line is affecting the measurement of \htwo. Given the quality of our spectra, we choose not to explore this issue further (but see section \ref{si_c_no_h2}).

We do not detect any significant systematic change in the FWHM of the lines with wavelength. There are no correlations between FWHM and velocity shift (either average quantities or individual lines), or between FWHM, velocity shift, and the inclination of the star. Notice, however, that the average of the line shifts is always less or equal to zero. Here we should remember that systematic errors in the velocity (up to 20 \kms) may appear due to inaccurate centering of the star in the slit. This means that velocity shifts from star to star cannot be compared. However, random centering errors should produce random positive and negative shifts. The fact that the \htwo\ lines are blueshifted (in average) may indicate that they are formed in material moving towards the observer.

The resolving power of the GHRS instrument at 1500 \AA\ is $\sim$15~\kms. The widths of the \htwo\ emission are larger than this, due either to turbulence or to a partially filled aperture. As noted in the introduction, extended \htwo\ emission is indeed detected in T Tauri stars. A uniformly filled aperture would produce a flat-topped profile with a base width larger than 100 \kms. Non-uniformly filled apertures will produce sharper, narrower lines. We do not observe flat-topped profiles, and all the detected lines have a single clear peak. \citet{2000A&A...357..951E} claim (using these same data) that a double peaked profile is present in the P(5)1-8 line (1562.39 \AA) of RW Aur. However, this is a region with very poor signal-to-noise, which leads us to conclude that the double peaked profile is not significant. The similarity of the average widths of the \htwo\ lines for all stars suggests that the emission is extended in all, although its angular scale is likely to be less than 2 arcsecs.

These \htwo\ lines are thought to be due to fluorescence, excited by strong atomic lines. In the Sun, \lya, \siiv, and \civ, have been identified, among others, as the sources of the excitation \citep{1978ApJ...226..687J,1979MNRAS.187..463B}. This conclusion is reached by using the observed strength of the \lya\ line and predicting what is the strength of the \htwo\ excitation produced. Insight into the nature of the exciting process can be obtained by calculating the populations of the levels from which the fluorescent routes are excited. 

Observations of the \lya\ line for TTS have been performed by the IUE satellite. As reported by \citet*{1993A&A...268..624B}, four observations of the \lya\ region are available in the IUE database. These correspond to TW Hya, RU Lup, DR Tau, and T Tau (only upper limits are derived for the last two). The spectra show a redshifted emission peak in the location of the \lya\ line.  \citet{1995ARep...39..322K} argue that the blue wing of the line should be strongly affected by ISM absorption. In addition, \lya\ is expected to show a strong wind signature that will absorb emission in the blue wing. This is the case in the STIS spectrum of TW Hya analyzed by \citet{TWspec}.

If we assume that the \lya\ line can be described by a Gaussian, then for RU Lup, $F_p=2\times 10^{-13}$\funit, $\sigma =2.01$\AA, where $F_p$ is the peak flux, uncorrected for extinction, and $\sigma$ is the Gaussian line width). For the reasons mentioned above, the total flux (corrected by ISM and wind absorption) is probably 3-4 times larger. As Table \ref{all_lines} show, these fluxes are larger than the fluxes in the \siiv\ and \civ\ lines, which are of the order of $10^{-14}$\funit\ (without taking the extinction into account). In the particular case of RU Lup, the \lya\ flux is 34 times larger than the flux in the blue member of the \civ\ doublet. In what follows we therefore assume that fluorescence by other atomic species is negligible compared to the fluorescence by \lya\ (and perhaps other lines of the Lyman series). 

\begin{figure}
\plotone{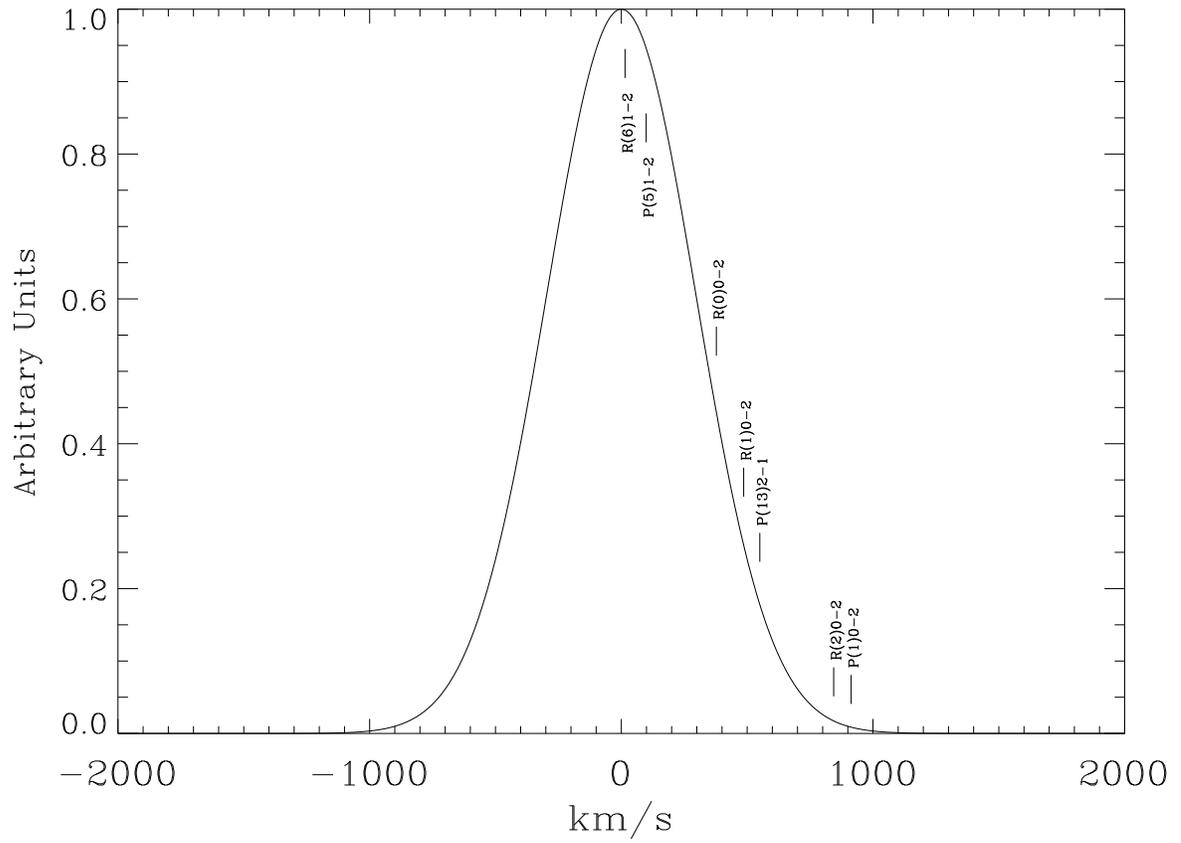}
\caption{Example of a Gaussian \lya\ line with FWHM=$680$ \kms\ (or $\sigma$=1.2 \AA). Indicated are the transitions from which fluorescence is observed in at least one star in our spectra. The line P(1)0-2 is at 1219.37, 912.4 \kms\ to the red of the \lya\ peak.\label{lyman}}
\end{figure}

Another possible pumping source for \htwo\ is the strong UV continuum of TTS. If the UV continuum were an important source of photons for the excitation of \htwo\ we would see fluorescent lines coming from a wide range of wavelengths. However, as Figure \ref{lyman} shows, all the observed lines are excited within a few \AA ngstroms of \lya. Furthermore, for TW Hya \citet{TWspec} show that the continuum is three orders of magnitude weaker than the \lya\ peak flux. Therefore, \lya\ dominates the fluorescence process. All of the fluorescent routes we observe have a strong line within 4 \AA\ ($\sim 1000$ \kms) of the center of \lya\ (Figure \ref{lyman}).

To understand the physical conditions of the region producing the emission, we need to know about the strength and angular coverage of the exciting agent (assumed here to be \lya) at the exciting wavelength in the spatial position where the excitation takes place, and the population of the state that will be excited. However, the fluorescent process produces only one observed quantity: the population of the upper level from which fluorescence takes place. Many lines are observed from a given fluorescent route, but, if optically thin,  they all measure the same quantity: the population of the upper level from which they come. Even if we knew exactly where the \htwo\ is in relation to the star, what fraction of the star emits the \lya\ radiation responsible for the excitation, and what fraction is intercepted by the \htwo, we would still have a mathematical problem with n equations (the number of fluorescent routes observed) and 2n unknowns (for each fluorescent route, the population of the exciting state and the strength of \lya\ at the excitation wavelength).

To solve this degeneracy, assumptions must be made, either about the population distribution of \htwo\ or the shape of the \lya\ line. In what follows, we assume that we know the shape and strength of the \lya\ line, and obtain the column densities of the exciting \htwo\ states. We will show that such a procedure requires non-thermal populations. An alternative procedure would be to assume a certain distribution of the levels of \htwo\ and from the observed lines reconstruct the \lya\ profile. We do not have enough observed routes to carry on this procedure (the maximum number of routes observed in a given star is six).

\subsubsection{\label{gaussian_model}A Gaussian Shape for \lya}
The flux in an \htwo\ line is given by:
\begin{equation}
{F_{ul}}={A_{sp} \over 4\pi d^2}  \\
{hc \over \lambda_{ul}} N_u A_{ul}
\end{equation}

where $F_{ul}$ is the integrated flux of the observed line (corrected for extinction), $A_{sp}$ is the fraction of the spectrograph aperture from which the emission comes, $d$ is the distance to the star, $h$ and $c$ are the Planck constant and the speed of light, respectively, $N_u$ is the column density of \htwo\ in the upper level, and $A_{ul}$ is the Einstein A-value of the transition. In the statistical equilibrium equation

\begin{equation}
N_u {\sum_{\sigma}A_{u\sigma}} = {\sum_{s}N_s B_{su} J_{\nu}}
\end{equation}

we neglect stimulated emission and collisions among upper levels. The term on the right represents the absorption of \lya\ photons ($B_{su}$ is the Einstein B-coefficient, $J_{\nu}$ is the mean specific intensity in the position where the absorption occurs). From this we obtain \citep*{1999MNRAS.302...48M}:

\begin{equation}
{F_{ul}}=\beta\ {1 \over {4\pi}}  \\
\sum_{s} {g_{u} \over g_{s}}  {\lambda^3_{us} \over \lambda_{ul}}  \\
{A_{ul}A_{us} \over {\sum_{\sigma}A_{u\sigma}}}  \\
{n_s}  {\lambda_{Ly\alpha} \over \nu_{Ly\alpha}}  F_{\lambda us} \\
 \rm{\ ergs \ cm^{-2} \ s^{-1}}
\end{equation}

where $g_{u}$ is the multiplicity of the upper level of the observed line, $\sum_{\sigma}A_{u\sigma}$ is the sum over the A-values of all the transitions from the same upper level. The sum on the left is over all the exciting transitions which contribute to a given upper level:  $g_{s}$ is the multiplicity of the level from which the line is excited, $\lambda_{us}$ is the wavelength of the exciting transition, $A_{us}$ is the A-value of the exciting transition, $N_s$ is the column density of the \htwo\ molecules in the state from which fluorescence takes place, and $F_{\lambda us}$ is the flux (in units of $\rm{ergs\ sec^{-1}\ cm^{-2}\ \AA}$) from the exciting transition. This is the \lya\ flux integrated over the \htwo\ line profile, which we assume is very sharp compared to the width of \lya\ and therefore it can be approximated by a delta function. $\lambda_{Ly\alpha} \over \nu_{Ly\alpha}$ converts the flux in per \AA\ to flux per Hz. Additionally, $\beta={A_{sp}\over {8\pi d^2_{*}}}$, where $d_*$ is the distance from the source of \lya\ to the position where the excitation takes place. In what follows, we assume $\beta=1$. In the absence of detailed information about the spatial distribution of \htwo\ this is a reasonable assumption.

From Equation 3 (with $\beta=1$ and assuming the high temperature ortho/para ratio of 3:1 for \htwo) we obtain the column density of the levels from which the exciting transition takes place (Table \ref{pop}, in order of excitation energy from the ground state). For this Table we have assumed that the integrated flux in \lya\ is 30 times larger than the blue \civ\ line flux (for RU Lup is 34 times larger), and that the \lya\ emission is Gaussian with $\sigma \sim 1.2$ \AA\ (the value measured for TW Hya). To calculate the population of a given level we use the best determined fluorescent line (highest signal-to-noise) from the route excited from that level. Furthermore, we assume that the excitation is due to the energy of the exciting line closest to \lya. For the values of $\sigma$ we are considering here this is an excellent assumption. The errors in the populations indicated in the Table have been obtained by propagating only the errors in the measured \htwo\ lines. They do not include errors in the \lya\ flux and in the width of the line, which are impossible to determine with the IUE data and should dominate this kind of calculation. Errors of up to half a magnitude in the extinction estimates will produce variations in the densities consistent with the calculated random error.

The populations are weakly correlated between the stars. This is an indication that the characteristics of the \lya\ emission dominate over differences in the temperature and density in the \htwo\ gas between different stars (Section \ref{rel_acc}). 

The order of magnitude of the column densities indicates that the populations of the levels should not be thermal. The critical density for quadrupole transitions between the v'' levels is $\sim10^6 \rm{cm^{-3}}$ (using the collisional cross sections from \citealp{1995ApJ...455L..89M}). For an emission region 1 $\rsun$ thick, a $10^{14} \rm{cm^{-2}}$ column density corresponds to a density of $10^{3} \rm{cm^{-3}}$. Therefore, the density is very subcritical and radiative effects should be considered when calculating the equilibrium populations.

%\documentclass{aastex}
%\begin{document}
%\newcommand \siiv{\ion{Si}{4}~}
%\newcommand \civ{\ion{C}{4}~}
%\newcommand \htwo{H$_2$}
%\newcommand \kms{km s$^{-1}$}

\begin{deluxetable}{cccccccccc} 
%\rotate
\tabletypesize{\tiny}
\tablecolumns{10} 
\tablewidth{0pc}
\setlength{\tabcolsep}{0.02in} 
\tablecaption{\label{pop}Derived Parameters for Pumped \htwo\ Levels} 
\tablehead{ 
\colhead{Line} & 
\colhead{BP Tau} & \colhead{T Tau} &
\colhead{DF Tau} & \colhead{RW Aur} &
\colhead{DG Tau} & \colhead{DR Tau} &
\colhead{DR Tau} & \colhead{RY Tau} &
\colhead{RU Lup} \\
\colhead{(v'', J'', g'')} & 
\colhead{} & \colhead{} &
\colhead{} & \colhead{} &
\colhead{} & \colhead{(8/5/93)} &
\colhead{(9/7/95)} & \colhead{} &
\colhead{}}

\startdata 
v''=2, J''=0, g''=1 &\\
v=378.4 \kms &\\
$\tau$&$0.06\pm0.02$& $0.30\pm0.01$&$0.25\pm0.02$&$0.5\pm0.1$ &$1.7\pm0.2$ &
\nodata &$1.1\pm0.3$ &$0.8\pm0.2$ &$0.3\pm0.2$ \\
N ($10^{14}\rm{cm^{-2}}$) & $0.09\pm0.03$ &
$0.35\pm0.01$ &  $0.18\pm0.02$ &$0.7\pm0.2$& $1.7\pm0.2$& \nodata & $0.5\pm0.2$ &  $1.0\pm0.3$ &$0.3\pm0.2$ \\
ratio & $1$ & $1$&$1$& $1$ & $1$&\nodata &$1$ &$1$ &$1$ \\
&\\
v''=2, J''=1, g''=9 &\\
v=486.5 \kms &\\
$\tau$&\nodata &$0.8\pm0.1$ &$0.52\pm0.03$ &$1.2\pm0.2$ &$5.4\pm0.5$ &$1.9\pm0.4$ & $1.5\pm0.4$& $2.2\pm0.6$ &\nodata \\
N ($10^{14}\rm{cm^{-2}}$) &\nodata & $13\pm2$ &$5.2\pm0.3$ &$22\pm3$ &$73\pm7$ &$25\pm6$ & $10\pm3$ &$40\pm10$ &\nodata\\
ratio &\nodata & $38\pm7$ &$28\pm3$ &$30\pm10$ & $43\pm7$ & $1$\tablenotemark{a} & $19\pm8$ &$40\pm10$ &\nodata\\
&\\
v''=2, J''=2, g''=5 &\\
v=843.1 \kms &\\
$\tau$&\nodata &\nodata &$4\pm2$ &\nodata &\nodata &\nodata &\nodata &\nodata &\nodata \\
N ($10^{14}\rm{cm^{-2}}$) &\nodata &\nodata &$5\pm2$ &\nodata &\nodata &\nodata &\nodata &\nodata &\nodata \\
ratio &\nodata &\nodata &$30\pm10$ &\nodata &\nodata &\nodata &\nodata &\nodata &\nodata \\
&\\
v''=2, J''=5, g''=33 &\\
v=98.5 \kms &\\
$\tau$&$0.16\pm0.04$&$0.50\pm0.02$& $0.06\pm0.02$ &$0.30\pm0.07$ &$0.6\pm0.3$& $0.4\pm0.2$ &$0.5\pm0.1$ &$0.6\pm0.1$ &$1.3\pm0.2$ \\
N ($10^{14}\rm{cm^{-2}}$) &$3.2\pm0.7$& $9.0\pm0.3$& $0.7\pm0.2$ &$6\pm1$ &$6\pm2$ &$8\pm4$ &$3.6\pm0.9$ &$9\pm2$ &$17\pm2$ \\
ratio &$40\pm1$& $27\pm2$& $3\pm1$ &$7\pm3$ &$3\pm1$ &$0.3\pm0.2$ &$7\pm3$ &$9\pm3$ &$50\pm40$ \\
&\\
v''=2, J''=6, g''=13 &\\
v=13.9 \kms &\\
$\tau$&\nodata &$0.44\pm0.06$&\nodata& \nodata &$0.3\pm0.2$&\nodata& $0.4\pm0.2$ &\nodata &$0.17\pm0.07$\\
N ($10^{14}\rm{cm^{-2}}$) &\nodata &$0.7\pm0.1$ &\nodata& \nodata & $0.4\pm0.2$ &\nodata& $0.3\pm0.1$ &\nodata &$0.21\pm0.09$\\
ratio &\nodata& $2.0\pm0.4$  &\nodata& \nodata& $0.2\pm0.1$&\nodata & $0.5\pm0.2$ &\nodata &$0.7\pm0.6$ \\
&\\
v''=1, J''=13, g''=81 &\\
v=550.9 \kms &\\
$\tau$&$0.3\pm0.2$& $0.8\pm0.2$&$0.4\pm0.2$ &\nodata &$4\pm1$ &\nodata &$2.6\pm0.6$ &\nodata & \nodata  \\
N ($10^{14}\rm{cm^{-2}}$) &$10\pm6$& $25\pm4$& $7\pm3$ &\nodata & $80\pm20$ &\nodata &$26\pm6$ &\nodata & \nodata \\
ratio &$110\pm80$& $60\pm20$ &$40\pm20$ &\nodata &$50\pm10$ &\nodata &$50\pm20$ &\nodata & \nodata \\

\enddata
\tablecomments{For each star we indicate the values derived for the exciting transition. For each exciting transition we give v'' and J'', the vibrational and rotational numbers of the level from which the transition is excited. We also give the velocity of the transition with respect to the center of the \lya\ line, the optical depth ($\tau$) of the exciting line, the column density in units of $10^{14}\rm{cm^{-2}}$, and the ratio of the column density to that of level N(v''=2,J''=0).}
\tablenotetext{a}{As we do not have a measured column density of N(v''=2,J''=0) for DR Tau (8/5/93) the ratios are given with respect to N(v''=2,J''=1)}
\end{deluxetable} 

%\end{document}

\begin{figure}
\plotone{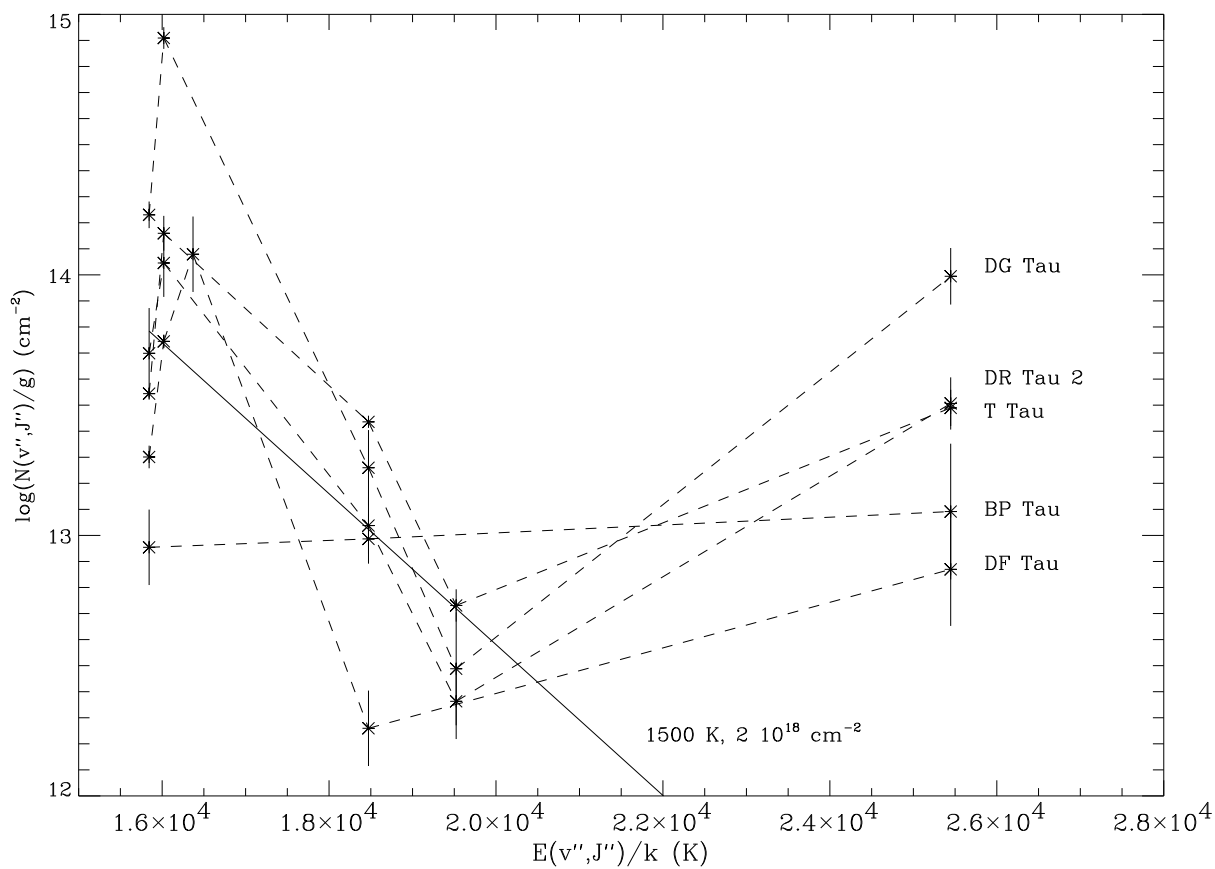}
\caption{\label{temp} Distribution of level populations for some of the stars in our sample (dashed lines). The ordinate is the logarithm of the column density of \htwo\ molecules (in a given v'', J'' level), divided by the g'' of the level (where we assume an ortho/para ratio of 3). The abscissa is the energy of the level in degrees Kevin. The solid line indicate the expected locus from a thermal distribution at 1500 K and total \htwo\ column density of $2\times10^{18}\rm{cm^{-2}}$. }
\end{figure}

This point is made clear when we take the ratios of the populations in Table \ref{pop}. The ratios are better determined than the populations themselves, as they do not suffer from the uncertainty in the \lya\ intensity. In the Table we have indicated the value of the multiplicities (g'') of each level. This is expected to be the ratio of the populations at very high temperatures and very large $\sigma$s. In any other circumstance, the ratio of the population of a state with higher excitation energy to a state with lower excitation energy will be less than this value, if the populations are in LTE. As the Table shows, population invertions are present in most stars. This means that excitation temperatures derived from the ultraviolet \htwo\ lines do not correspond to the kinetic temperature of the emitting region, and the excitation cannot be explained by any process that produces a thermal distribution (like a shock). This point is made explicitly in Figure \ref{temp}, in which we plot the level populations divided by the multiplicity of the level, versus the energy of the level, for some of our stars. If the populations were thermal, all the points for each star should lie on a straight line (whose intercept with the ordinate axis is the total \htwo\ density), and they do not. Also indicated are the lines produced by a 1500 K distribution, with a total \htwo\ column density of $2\times10^{18}\rm{cm^{-2}}$. These are the parameters found by \citet{1999A&A...348..877V} for T Tau. While the solid line occupies the same general area as the observed distributions, it does does not reproduce the observations. Errors in the assumed peak of the \lya\ line make the mean value of the populations uncertain but do not change their relative values.

Based on the densities and the A-values, we calculate the optical depth of the pumped transition. Notice that the optical depths are less than one for most stars. The effect of a high optical depth is to effectively remove the line from the fluorescent cascade, as photons re-emitted in this wavelength will be absorbed very rapidly. This tends to equalize the flux in other lines from the same route. The ratios of the observed lines indicate that they are optically thin, which confirms our optical depth estimates.

The calculations in which Table \ref{pop} is based assume that the \lya\ profile can be correctly modeled by a Gaussian. If the line shows self-reversals in the core, such assumption is false, and the populations of N(v''=2,J''=6) and perhaps N(v''=2,J''=5) are incorrect. The populations of N(v''=2,J''=6) are consistently smaller than the populations of N(v''=2,J''=5), in spite of the small energy difference between them. A NLTE self-reversal in \lya\ that affected  N(v''=2,J''=6) more than N(v''=2,J''=5) may make the populations more alike.  It is unlikely that the core self-reversal is as large as to affect the other populations, which are excited at velocities larger than 300 \kms. More fundamentally, the \lya\ line is more likely to resemble a Voigt profile. This will also change the core/wing ratios. We may be seeing some evidence of non-Gaussian wings. Two routes are excited from the v''=2, J''=1 level: one to level v'=0, J'=0 (excited at 900 \kms from the center of \lya) and the other to level v'=0, J'=2 (excited at 500 \kms). We observe the first route only in DF Tau and DG Tau, as the line P(1)0-5 at 1396 \AA. The first route predicts much larger densities (a factor of 30 and 7 for each star respectively) than the second, for the v''=2, J''=1 level. Assuming a Voigt profile with an {\it a} parameter of 0.2 will solve the discrepancy for DG Tau. For DF Tau a fairly large parameter is needed. However, while the densities of individual states change upon using a Voigt function, they are still not in thermal equilibrium. In Table \ref{pop} we use R(1)0-2, 500 \kms\ from the \lya\ center, to calculate populations for N(v''=2, J''=1), on the assumption that lines closer to the core are less affected by the non-Gaussian wings.

No lines coming from levels v''=2, J''=3 or J''=4 are detected: pumping lines for these levels are not present in the \lya\ profile. As Figure \ref{lyman} makes evident, no lines originating from the far blue side of the \lya\ line are observed either. We have searched for strong lines with exciting transitions to the blue of \lya, originating from low molecular levels, that should be observable in our spectra. We find three candidates: (v'',J'')=1, 15 (R(15)3-1 at -300 \kms), (1, 11) (P(11)1-1 at -800 \kms), and (1, 12) (R(12)1-1 at -780 \kms). There are other possible lines which would produce \htwo\ emission in our spectral ranges, but they originate in states with higher excitation energies. The line closest to the center, if we assume a population similar to that of v''=1, J''=13 should produce fluoresced lines stronger than those produced from P(13)2-1, as its A-value is $1\times 10^8 sec^{-1}$. The fact that we do not see such lines may be an indication that excitation of \htwo\ by \lya\ occurs far enough from the star such that the blue wing of \lya\ has already been absorbed by the wind. This picture can be tested with data having better signal-to-noise and wider spectral range, which will allow to decide whether or not lines coming from the blue side of \lya\ are excited by an absorbed profile. At this point, we can say that the asymmetry in the excitation routes is tantalizing but needs more study.

As argued above, the result that the populations of \htwo\ are not in thermal equilibrium does not depend strongly on the exact shape of \lya, as long as the flux decreases to the red, from the rest wavelength (this is implicit in the assumption that \lya\ is a Gaussian at 1215.67 \AA). From the \citet{1993A&A...268..624B} observations, it is not clear that this is so. The observed \lya\ line in the very noisy, underexposed IUE spectra is to the red of the rest wavelength. If the \lya\ profile absorbed by \htwo\ were redshifted (either because its blue wing is absorbed by a wind, or it is emitted in a moving region), the conclusion that the populations are not in thermal equilibrium would need to be revised, as exciting lines close to the nominal line center would see less \lya\ flux, which would translate in larger measured populations. This will tend to erase the minima observed Figure \ref{temp}, which are due to levels very close to the nominal center of \lya. On the other hand, it seems difficult to alter the very flat appearance of the trace of BP Tau in Figure \ref{temp} by changing the shape of \lya.

What can these observations tell us about the excitation mechanism and the origin of the \htwo\ emission? Two kinds of heating mechanisms are generally proposed: heating by absorption of part of the stellar and accretion luminosity (including X-rays see \citealt*{2000ApJ...541..767W}), and heating by dynamical processes, including shocks, and turbulent decay. Very little has been published about the expected populations of \htwo\ in CTTS. A preliminary model by \citet{1987ApJ...322..412B} shows that highly excited states of \htwo\ can be produced (by UV radiation) assuming that the populations in $v''\leq2$ are in thermal equilibrium at 2000 K. ISO observations of IR lines in T Tau by \citet{1999A&A...348..877V} are consistent with a C-shock that rises the gas temperature to 1500 K. The same ISO observations indicate the presence of a 440 K \htwo\ gas and conclude that yet another very low temperature component must be present to explain the total gas mass derived from CO observations. Figure \ref{temp} shows that, while a temperature in the range of 1000 to 2000 K may be appropriate, the assumption of LTE is not true.  

The fact that the average velocities of the measured \htwo\ lines are all blueshifted suggests that the emission comes from an outflow, as opposed to the disk surface layers or its inner edge. This conclusion is reinforced by the fact
 that the detected profiles are single peaked. Given the instrumental resolution, we should detect double peaked profiles if the emission comes from material within $\sim10(M/0.5\msun)\sin^2 i$ AU, where $M$ is the mass of the central star and $i$ is the inclination angle. We do not see such profiles, which implies that if the emission comes from the disk, is either originating beyond this limit or comes from a wide range of disk radii. In spite of the model uncertainties, it is clear that a large temperature, of the order of 1000 K, is necessary to produce the observed excited states from which \htwo\ will be pumped. At these distances in the disk, the temperature due to reprocessed stellar radiation should be much less. 

Analysis of forbidden(HEG95) wind diagnostics in CTTS indicate that wind temperatures are $\sim$8000K and densities are $\sim10^5$cm$^{-3}$. This values suggest that if the \htwo\ emission comes from this same flow, it should be originated in the outer, cooler and less dense parts of it (assuming a length scale of the order of one stellar radius). Therefore, the wind origin of the emission is plausible, but more observations (for example, long-slit spectroscopy) are needed before we can really understand its distribution.

\subsection{Blending of \siiv\ and \civ\ with \htwo\ \label{corr}}
As mentioned in the introduction, the \civ\ and \siiv\ doublets are affected by overlying lines of \htwo. Such lines will affect any low-resolution measurement of the doublets. In this section we will separate the emission due to \htwo\ from that due to \civ\ and \siiv. Some of the results from section \ref{molhyd} are based on these extractions.

\begin{figure}
\vskip -0.2in
\hskip 0.3in
\centering \includegraphics[angle=90,width=5in]{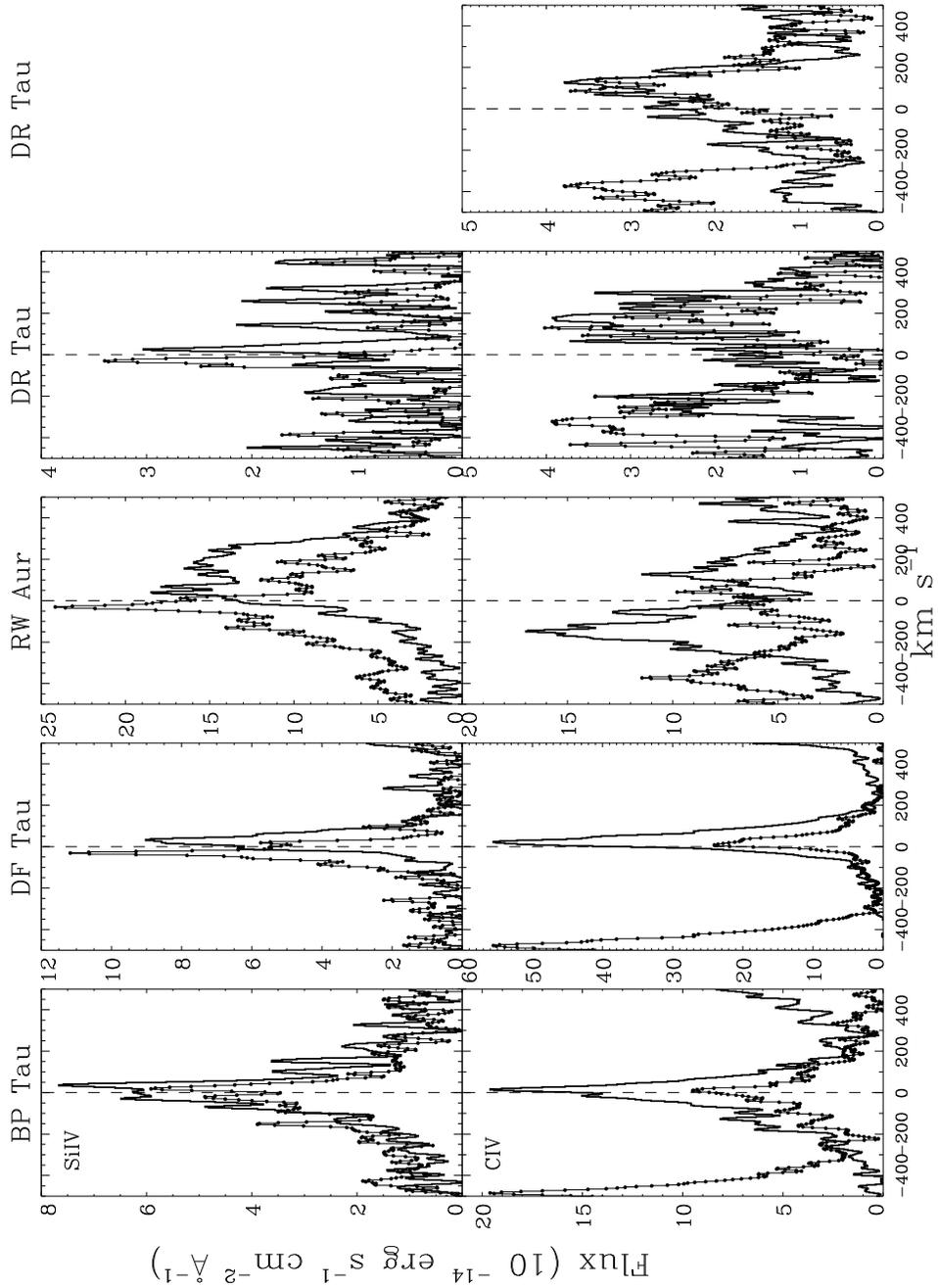}
\vskip 0.1in
\caption{\label{si_c_1}Doublet plots (\siiv\ and \civ). All the spectra have been smoothed. The thick (thin dotted) line indicates the optically thicker (thinner) member of the doublet. For these doublets, the optically thicker line is always the blue one. Zero velocity is indicated with a dashed line for all lines.}
\end{figure}

\begin{figure}
\vskip -0.2in
\hskip 0.3in
\centering \includegraphics[angle=90,width=5in]{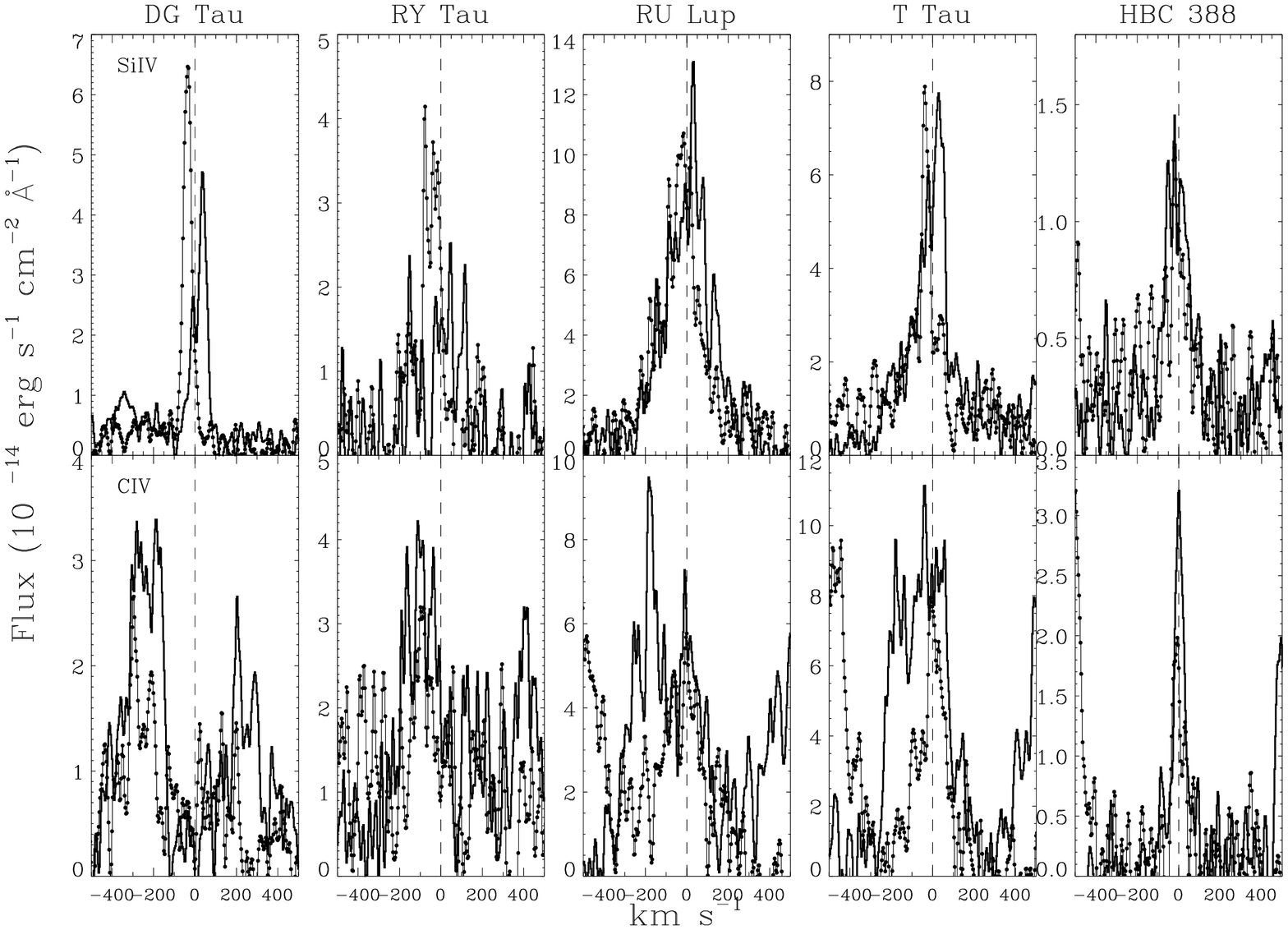}
\vskip 0.1in
\caption{\label{si_c_2}Doublet plots (\siiv\ and \civ), cont.}
\end{figure}

Figures \ref{si_c_1} and \ref{si_c_2} show the \siiv\ and \civ\ doublets. Notice that the separation between the two members of the \civ\ doublet is $\sim500$ \kms, and so in the red side of the thick trace (or the blue side of the thin trace) one sees the beginning of the other member. Figures \ref{si_c_1_sub} and \ref{si_c_2_sub} show the same doublets after subtracting the \htwo\ emission.  

The blue member of the \siiv\ doublet is affected by R(0)0-5 (at 1393.719 \AA) and R(1)0-5 (at 1393.961 \AA). The red member is affected by P(3)0-5 (at 1401.648 \AA). These \htwo\ lines are responsible for the redshifted peak of the blue line and the blueshifted peak of the red line. The red member of the \civ\ line is affected by R(3)1-8, which can be seen most clearly in the blue shoulder of the blue line of \civ\ in BP Tau.  We have searched the \citet{1993A&AS..101..273A} database for other lines affecting the doublets and we conclude that in our spectra these are the most important lines. Using the fact that these \htwo\ lines are optically thin, we can use other lines from the same upper level to obtain estimates of their fluxes. We start with an analysis of the \siiv\ doublet, followed by the \civ\ doublet. 

\begin{figure}
\plotone{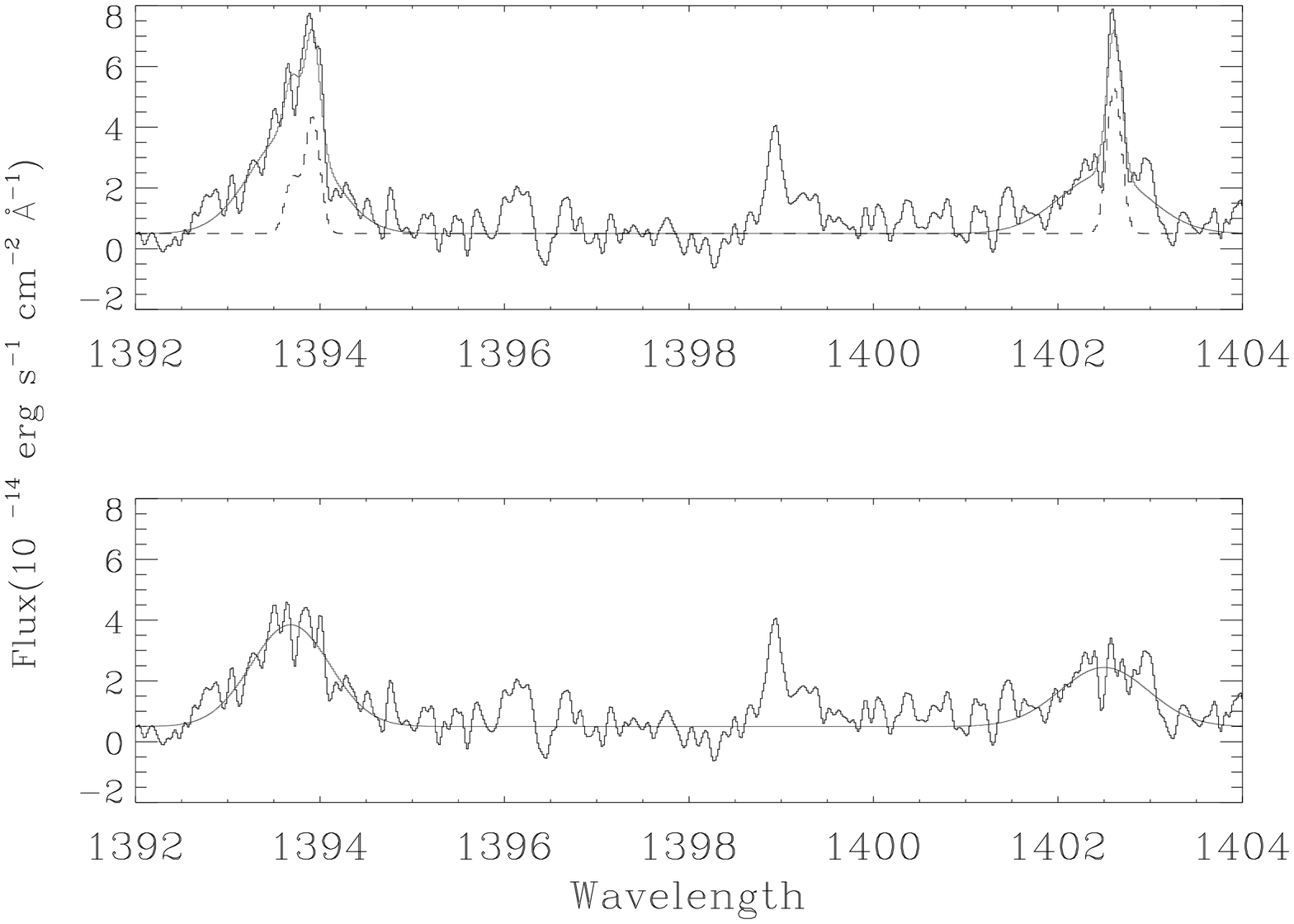}
\caption{\label{decomp}Example of deblending procedure for T Tau. The Figure shows the wavelength range that contains \siiv. The data are the jagged lines. In the top plot is the observed spectrum. The dashed line is the contribution from \htwo\ (Lines R(0)0-5, R(1)0-5, and P(3)0-5). We have obtained the strength of these lines and widths for these lines from other lines in the same fluorescent route (see Table \ref{h2_lines}), assuming that all are optically thin. The centroids are obtained by averaging the centroids of other lines in the same route. The bottom plot shows the data after subtracting the \htwo\ contribution. In both plots, the smooth solid line is the fit to the spectra, before (top) and after (bottom) subtraction. }
\end{figure}

R(0)0-5 and P(2)0-5 come from the same upper level and the flux from P(2)0-5 is expected to be twice as large as that from R(0)0-5. We can measure P(2)0-5 with confidence given that it is isolated between the two members of the \siiv\ doublet.  Given the size of the P(2)0-5, contribution of R(0)0-5 to the blue member of the \siiv\ profile is negligible in most cases. R(1)0-5 and P(3)0-5 belong to the same fluorescent route, and the later is expected to be 1.4 times more intense than the former. For T Tau, DF Tau and DR Tau, we have additional spectral ranges that allow us to determine the contribution of \htwo\ to the doublet from other \htwo\ lines  present in the spectra from the same fluorescent route. In Figure \ref{decomp} we indicate how is the decomposition performed. For BP Tau, RW Aur, RY Tau, RU Lup, and DG Tau, these are the only strong lines from this fluorescent route present in the spectra. Therefore, to obtain estimates of their influence in the \siiv\ profiles, we need to resort to indirect methods. As we argue below, we make positive detections of \siiv\ only on BP Tau, RU Lup, RW Aur, and T Tau. In the case of HBC 388 there is no indication that the \siiv\ lines are affected by \htwo\ emission. Such emission, if present, is buried under the noise for all \htwo\ lines identified reliably in other stars. We now indicate how to carry out the deblending procedure for some individual stars.

For those stars without additional members from the same route as R(1)0-5 and P(3)0-5 we estimate their intensities by the following reasoning. In STIS spectrum of TW Hya \citep{TWspec}, the ratio between the peak values of P(2)0-5 and R(1)0-5 is  $\sim0.7$. If one assumes this ratio is the same for all CTTS, from the measured flux of P(2)0-5 we obtain R(0)0-5 and R(1)0-5, and then P(3)0-5. The comparison with TW Hya is useful because, if \htwo\ is due to \lya\ fluorescence, both P(2)0-5 and R(1)0-5 are excited from $v''=2$ and their excitation wavelengths are $\sim100$ \kms\ apart, so the \htwo\ populations feels a similar \lya. If the FWHM of the \lya\ emission for a star in our sample is larger than that of TW Hya, the ratio of the two lines would be closer to 1. This is the case for DG Tau, for example. We have ignored this complication and assume the same ratio always. Alternatively, we could use the populations derived in section \ref{molhyd} to obtain estimates of the line ratios for the two different routes. Such procedure produces similar results (within the errors) as the one outlined here.

For BP Tau, the contamination of \siiv\ by \htwo\ is not very important. Given the strength of P(2)0-5, R(0)0-5 should be within the noise level. Using TW Hya we conclude that R(1)0-5 and P(3)0-5 are at noise level. For BP Tau star, the values of \siiv\ in Table \ref{all_lines} are obtained without any correction for \htwo. In the case of RY Tau, a similar argument implies that whatever we see in the neighborhood of the \siiv\ doublet is due to \htwo. For RU Lup, the fact that the blue line is stronger than the red one (unlike what is observed in pure \htwo\ emission) implies that some \siiv\ must be present, which we indicate in Table \ref{all_lines}. In the case of RW Aur, the red \htwo\ emission is clearly visible, superimposed to the broad \siiv\ emission. We use the measured parameters to obtain the blue \htwo\ line.

For DG Tau, sharp emission peaks are present at the position of the \siiv\ doublet. Their positions coincide with those of R(0)0-5, R(1)0-5 and P(3)0-5. Their width is similar to other \htwo\ lines in the spectrum, like P(5)1-8. The ratio of the peak emission in the red \siiv\ line to the peak emission in the blue \siiv\ is $\sim1.5$ as expected if they are optically thin \htwo\ lines. The same is true for P(2)0-5 and the blue peak of the blue \siiv\ line. Note that the P(2)0-5 and R(1)0-5 lines have approximately the same flux, which implies the the \lya\ emission is wider for this star that for TW Hya. We conclude that the emission in the \siiv\ wavelengths is due to \htwo.

DF Tau presents a difficult case. As mentioned in section \ref{molhyd} if one identifies the lines near \siiv\ wavelengths as being just R(0)0-5 and P(3)0-5 (i.e., assuming no \siiv), the resulting lines are very wide and do not have the expected intensity of optically thin lines. On the other hand, using the two other available lines in the same fluorescent route to predict their intensity, absorbs most of the lines in the 1400 \AA\ region. This indicates that \siiv\ emission is present, but at the noise level. The upper-limits for \siiv\ quoted in Table \ref{all_lines} will contain all the flux.

The \civ\ doublet is contaminated mainly by R(3)1-8, which should have similar strength to P(5)1-8, assuming both are optically thin. No other \htwo\ lines are strong enough to affect the \civ\ doublet. Furthermore, the effect of other atomic lines in the \civ\ doublet members is negligible. \citet{2000A&A...357..951E} have argued that the shapes of the profiles in RW Aur are due to absorption by \feii\ in the wind (see next section). If this is true for other stars, we should detect a similar pattern of absorptions, and we do not. Furthermore, \citet{1986ApJS...61..801S} show that in the Sun one observes strong \sii\ lines in this region. As the signal-to-noise in this region of our spectra is not very high, it is possible that some \sii\ lines are present. We searched for these lines and obtained estimates of their strengths, using other lines of the same multiplets present in the spectra. We conclude that the contribution of \sii\ to the \civ\ profiles is negligible. We detect \civ\ in all the stars. 

The spectra from DG Tau show strong blueshifted features in both lines. The subtraction of R(3)1-8 does not alter the appearance of the doublet very much. The signal-to-noise of the features to the red of the red member (Figure \ref{si_c_2}) is close to 1, so they are not significant. The fact that both members show emission at similar velocities lead us to conclude that the emission is indeed due to \civ. The signal-to-noise in the region is close to two, so the double peaked structure is not significant.

\subsection{The \siiv\ and \civ\ Profiles with no \htwo\ \label{si_c_no_h2}}
\subsubsection{Morphology}
\begin{figure}
\vskip -0.2in
\hskip 0.3in
\centering \includegraphics[angle=90,width=5in]{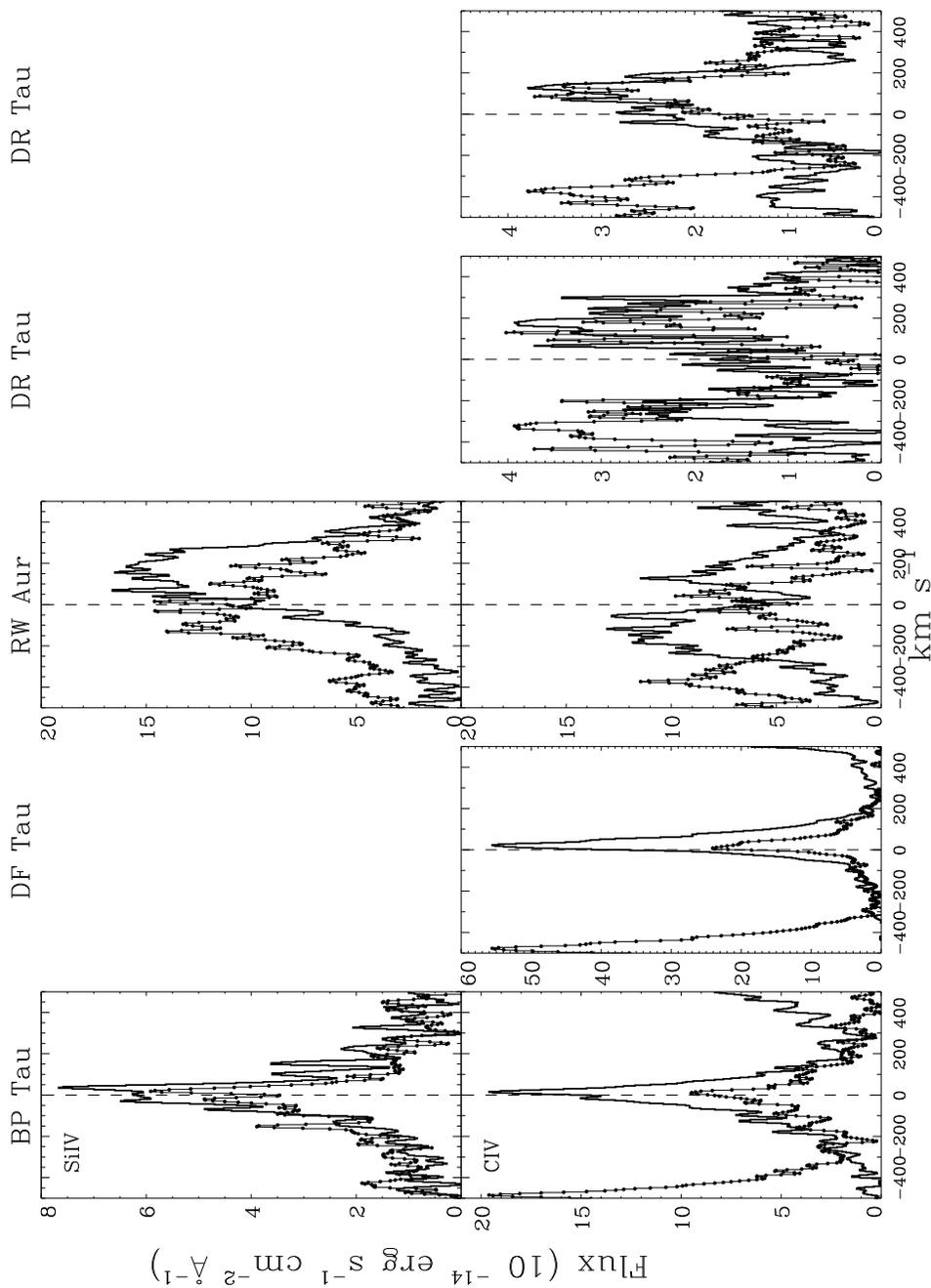}
\vskip 0.1in
\caption{\label{si_c_1_sub}Doublet plots (\siiv\ and \civ), with \htwo\ subtracted. The conventions are the same as in figures \ref{si_c_1}. Note that the vertical scale has changed. Blank spaces indicate stars for which the subtraction of the \htwo\ lines produces profiles consistent with noise (i.e., profiles in which no \siiv\ is detected).}
\end{figure}

\begin{figure}
\vskip -0.2in
\hskip 0.3in
\centering \includegraphics[angle=90,width=5in]{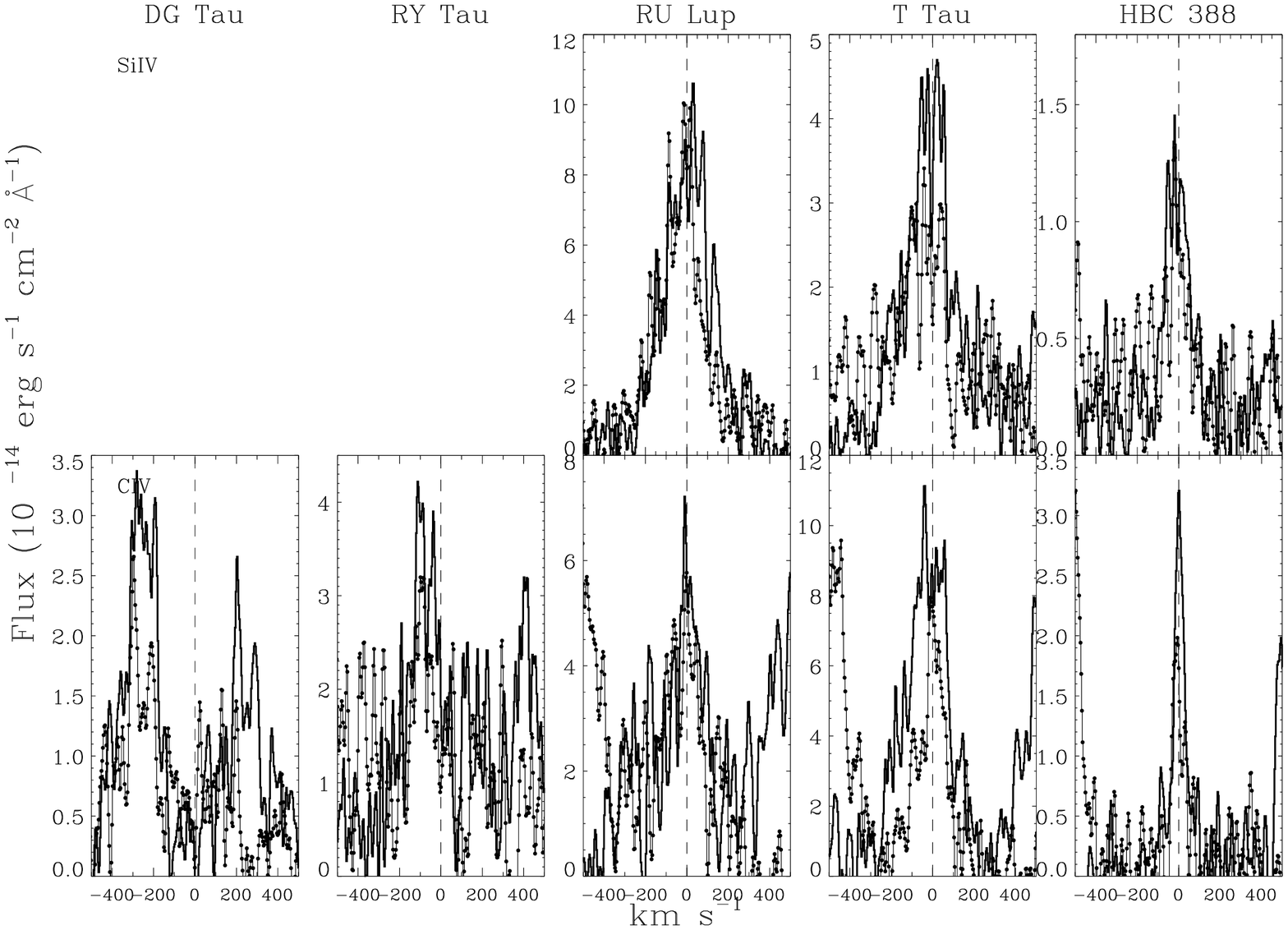}
\vskip 0.1in
\caption{\label{si_c_2_sub}Doublet plots (\siiv\ and \civ) without \htwo, cont.}
\end{figure}

In this section, we study the profiles resulting from subtracting the \htwo\ contribution (Figures \ref{si_c_1_sub} and \ref{si_c_2_sub}). In particular we are interested in the process responsible for the shape (width and components) and the origin of the transition region lines. The \civ\ line profiles show a wide array of shapes, from the centrally peaked emissions of DF Tau to the very asymmetric ones of DG Tau and DR Tau.


\begin{figure}
\plotone{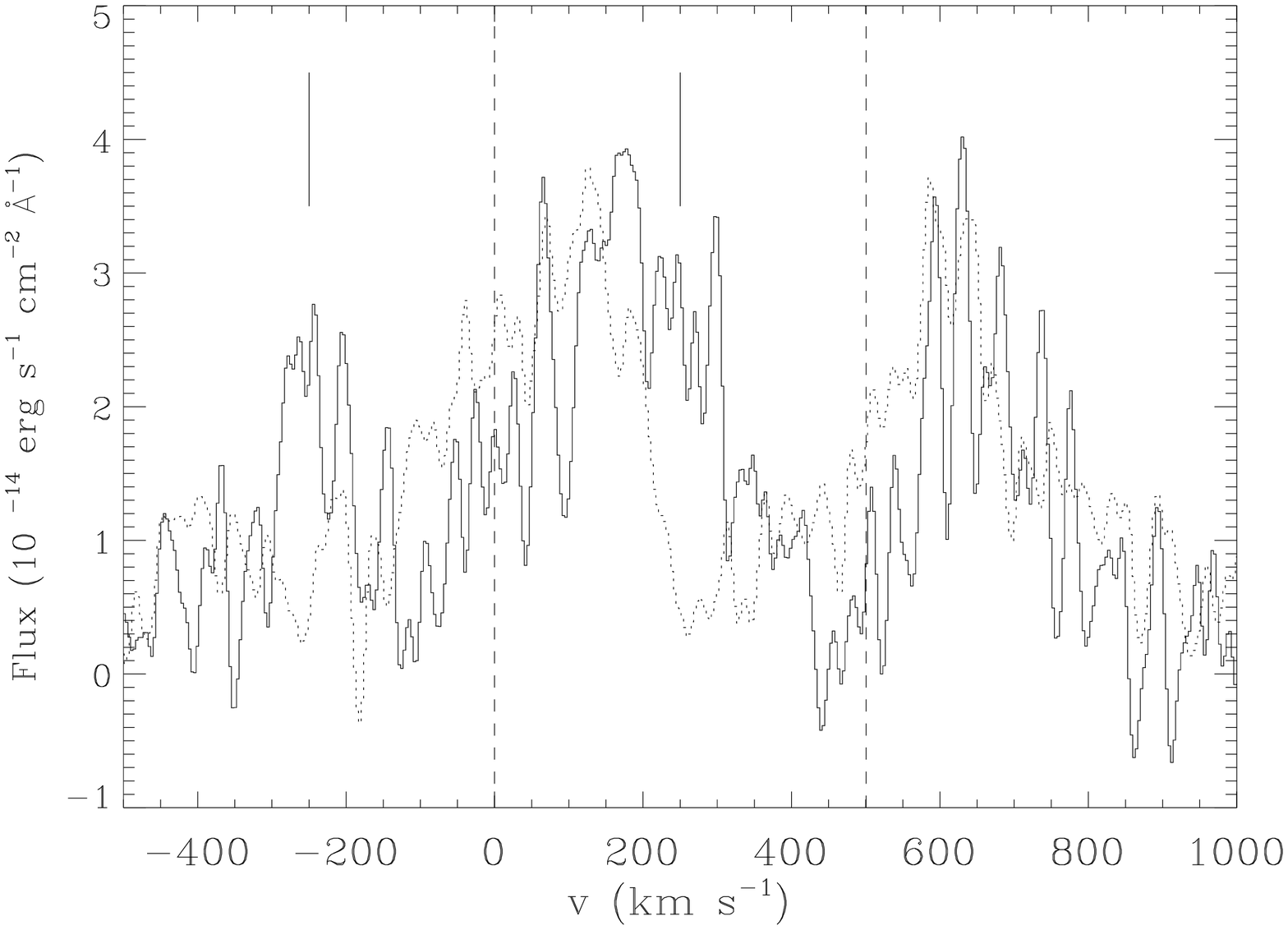}
\caption{\civ\ profiles for DR Tau. We have subtracted the contribution from R(3)1-8. The solid (dotted) lines is the first (second) epoch of observations. The velocity scale is centered on the blue member of the doublet. The vertical dashed lines indicate the position of the doublet members. The vertical solid lines indicate the positions of the transient emissions in the first epoch profile.\label{dr_civ}}
\end{figure}

DR Tau shows a remarkable variation in \civ\ from the first to the second epoch of observations. The first epoch shows a blueshifted emission at $\sim-250$ \kms\ separated from the main blue peak (Figure \ref{dr_civ}). Furthermore, for the first epoch, the blue peak shows a red excess when compared to the red peak. Such excess is absent from the second epoch, as is the blueshifted emission at $\sim-250$ \kms. We interpret this morphology as a transient event in the first epoch. This event produced \civ\ emission 250 \kms\ to the blue of both \civ\ lines. The apparent excess to the red of the blue line actually belongs to this event in the red line. Without this extra emission the red and blue lines in both epochs look remarkably similar. In Table \ref{all_lines} we have decomposed the \civ\ lines from the first epoch in a blue and a redshifted component (see section \ref{origin}).

Table \ref{comparison} summarizes the results of our measurements and compares them with those obtained by \citet{2000ApJS..129..399V}, who use average low resolution IUE data. IUE data can barely resolve the two members of the \siiv\ doublet, but not the \civ\ doublet. Also, it cannot be used to subtract the contribution of \htwo\ to these lines. As Table \ref{all_lines} shows, this correction is small, $\sim10\%$ or less in all cases. A similar estimate was obtained by \citet{2000ApJ...539..815J}, using the much larger IUE aperture.

%\documentclass{aastex}
%\begin{document}
%\newcommand \siiv{\ion{Si}{4}~}
%\newcommand \civ{\ion{C}{4}~}

\begin{deluxetable}{cccccc} 
\tabletypesize{\small}
\tablecolumns{6} 
\tablewidth{0pc} 
\tablecaption{\label{comparison}Fluxes of Transition Region Lines} 
\tablehead{ 
\colhead{}    &  \multicolumn{2}{c}{This Work} &   \colhead{}   & 
\multicolumn{2}{c}{Valenti et al. 2000} \\ 
\cline{2-3} \cline{5-6} \\ 
\colhead{Name} & \colhead{\siiv}   & \colhead{\civ}     & 
\colhead{}    & \colhead{\siiv}   & \colhead{\civ}   }
\startdata 
BP Tau & $6\pm1$, $4\pm1$ & $24.0\pm0.8$ && 
$4.9\pm0.2$, $3.9\pm0.3$ & $25.2\pm0.4$ \\
T Tau & $3.6\pm.4$, $2.2\pm0.4$ & $17\pm1$ &&
$7.3\pm0.6$, $5.9\pm0.6$ & $44.5\pm0.9$ \\
DF Tau & $<4$, $<4$ & $45.7\pm3$ &&
$4.4\pm0.1$, $3.3\pm.1$ & $25.2\pm0.4$\\
RW Aur & $19\pm2$, $25\pm2$ & $23\pm6$ &&
$12.9\pm0.5$, $10.7\pm0.5$ & $21\pm1$\\
DG Tau & $<1.3$, $<1.3$ & $4.3\pm0.4$ &&
$1.4\pm.2$, $1.1\pm.2$ & $ 5.0\pm0.6$\\
DR Tau & $<1.5$, $<1.5$ & $6\pm1$, $5.6\pm0.4$ &&
$2.5\pm.1$, $2.7\pm0.1$ & $17.3\pm0.6$\\
RY Tau & $<2$, $<2$ & $4.2\pm0.5$ &&
$1.64\pm0.07$, $2.90\pm0.07$ & $8.7\pm0.3$\\
RU Lup & $10.8\pm0.7$, $8.5\pm0.6$ & $14\pm1$ &&
$27.2\pm0.7$, $19.8\pm0.7$ & $76\pm1 $\\
HBC 388 &$0.7\pm0.1$, $0.4\pm0.1$ &$1.4\pm0.2$ &&
\nodata & \nodata \\ 
\enddata
\tablecomments{All the fluxes are in units of $10^{-14}\rm{ergs\ sec^{-1}\ cm^{-2}}$. The fluxes are not corrected for reddening. In the \siiv\ columns, both members are quoted separately. In the \civ\ columns relating to this work, the fluxes in each member of the doublet have been added. We have two different epochs for DR Tau \civ}
\end{deluxetable} 

%\end{document}

In the Table we quote upper limits for the \siiv\ that we cannot measure. These upper limits are twice the value obtained by integrating a smoothed noise vector over the same width as the \civ\ line. As such, they are the flux of a line with this FWHM and correspond to a 2-$\sigma$ detection. From Table \ref{comparison} it is clear that the intrinsic variability of TTS dominates over any \htwo\ correction. For example, the \civ\ flux in DF Tau increases by a factor of 2, while the flux in RU Lup decreases by a factor of 5, with respect to the measurements of \citet{2000ApJS..129..399V}. 

\begin{figure}
\plotone{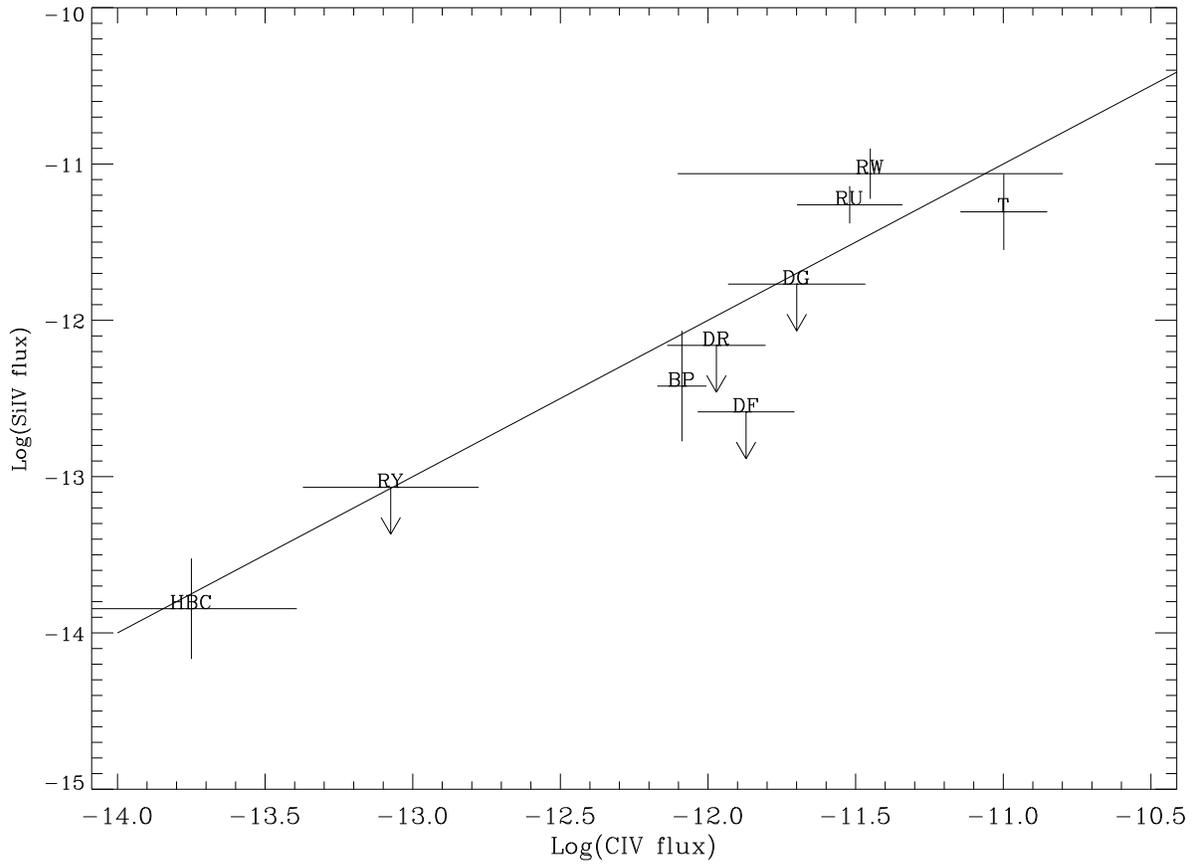}
\caption{\label{civ_vs_siv} \siiv\ vs. \civ. The bars indicate errors. Upper limits are indicated by arrows. The solid line is the identity relationship.}
\end{figure}

For stars in which we have both \civ\ and \siiv\ profiles, we observe two different behaviors in the ratio of their fluxes. For BP Tau, T Tau and HBC 388, the \civ\ blue line is stronger than the \siiv\ blue line. For RW Aur and RU Lup, the situation is reversed. As comparison with the SUMER Atlas of Solar-Disk features shows\footnote{http://www.uio.no/$\sim$paalb/sumer\_atlas.html.}, the flux ratio of  \civ\ to \siiv\ can change from $\sim0.7$ to $\sim4$ as one goes from the network to the quiet Sun \citep{2001astro.ph}. In our target stars the ratio from \civ\ to \siiv\ varies between 0.4 (RW Aur) to more than 5 (DF Tau). The number of stars for which we actually measure both lines is small (only four CTTS, for which we get that the average ratio of \civ\ to \siiv\ fluxes is 1.3), and so it is difficult to explain the cause of these differences. The relationship between the \civ\ and \siiv\ emission is clarified by Figure \ref{civ_vs_siv}, which shows an almost linear correlation between the two. Therefore the flux in \civ\ predicts the flux in \siiv.  \civ\ and \siiv\ fluxes are also correlated in stellar chromospheres \citep*[e.g.][]{1981ApJ...247..545A}.

\begin{figure}
\plotone{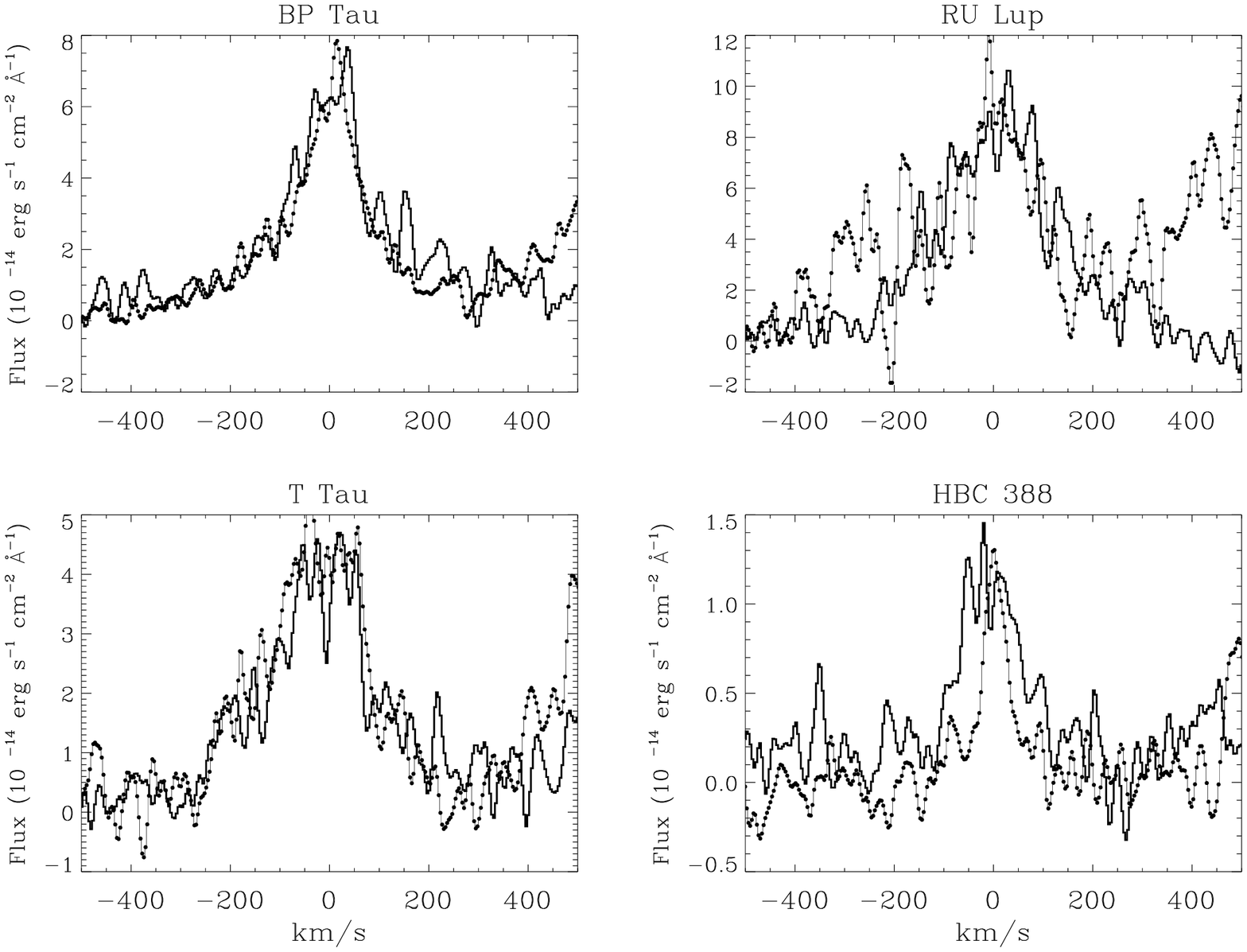}
\caption{\label{si_c_comp}Comparison between the red members of \siiv\ (thick) and \civ\ (thin dotted). The \civ\ lines have been scaled to match the peak of the \siiv\ in each star.}
\end{figure}

Figure \ref{si_c_comp} compares the \siiv\ and \civ\ red members for those stars for which we have both, not including RW Aur (we discuss the RW Aur case below). The \siiv\ and \civ\ profiles look very similar to each other in all stars, which indicates that they are formed in the same region. For HBC 388, the \siiv\ line is broader than the \civ\ line. This is unlike SUMER observations of the Sun, for which both lines are very similar. The \siiv\ and \civ\ observations are not simultaneous, so it is not impossible that a flare event is affecting the \siiv\ observations of HBC 388. 

The Gaussian fits to the \siiv\ and \civ\ profiles, indicated in Table \ref{all_lines}, show that the widths vary from  $\sim200$ to $\sim300$ \kms. For BP Tau and DF Tau the signal to noise is high enough to perform a decomposition in multiple Gaussian elements. In the case of BP Tau we perform a two element Gaussian decomposition to each line, which results in a NC and a BC. For DF Tau, we perform a 3 element Gaussian decomposition of the \civ\ profile. In general, we find a BC of width $\sim300$\kms and a NC of width $\sim100$ \kms. For DF Tau we also find a narrower component with width $\sim50$ \kms. Multiple components may indicate that the emission comes from more than one region: for example, the NC may come from the post-shock while the BC comes from the pre-shock. On the other hand, the shape of the line may be the result of the area of the accretion ring oriented towards the observer (Section \ref{origin}) and so a Gaussian decomposition is not really valid. 

In these analyses RW Aur stands in a class by itself. Not only are the \civ\ and \siiv\ lines very different to each other, but the individual members of each doublet are very different among themselves. \citet{2000A&A...357..951E} have analyzed the \civ\ doublet for RW Aur and they conclude that its appearance can be explained by \feii\ absorption from a $10000$K stellar wind. For this explanation they invoke FeII] absorption, together with permitted FeII lines. Without more careful modeling of the wind environment of RW Aur is not clear if this is the correct explanation, especially since nearby FeII] lines are absent. The red \siiv\ line may be affected by \oiv, and both members are affected by \feii. We do not observe any other example of these \feii\ absorptions (or emissions) in the other stars in our sample (although \citealt{2000AstL...26..225L} reports strong \feii\ features in RU Lup, in the 2325 \AA\ range), which may indicate that RW Aur has the coolest and/or densest wind of all our stars. Without higher resolution and higher signal-to-noise spectra to disentangle the contribution of \feii\  lines to the RW Aur spectrum, it is premature to advance any definite conclusion.

\subsubsection{\label{rel_acc}Relationship to Accretion}

We have seen that the \siiv\ flux is correlated with the \civ\ line flux. We know show that the \civ\ line flux is also correlated with the intensity of \htwo\ and therefore can be used to predict the overall appearance of the spectra.
 
\begin{figure}
\plotone{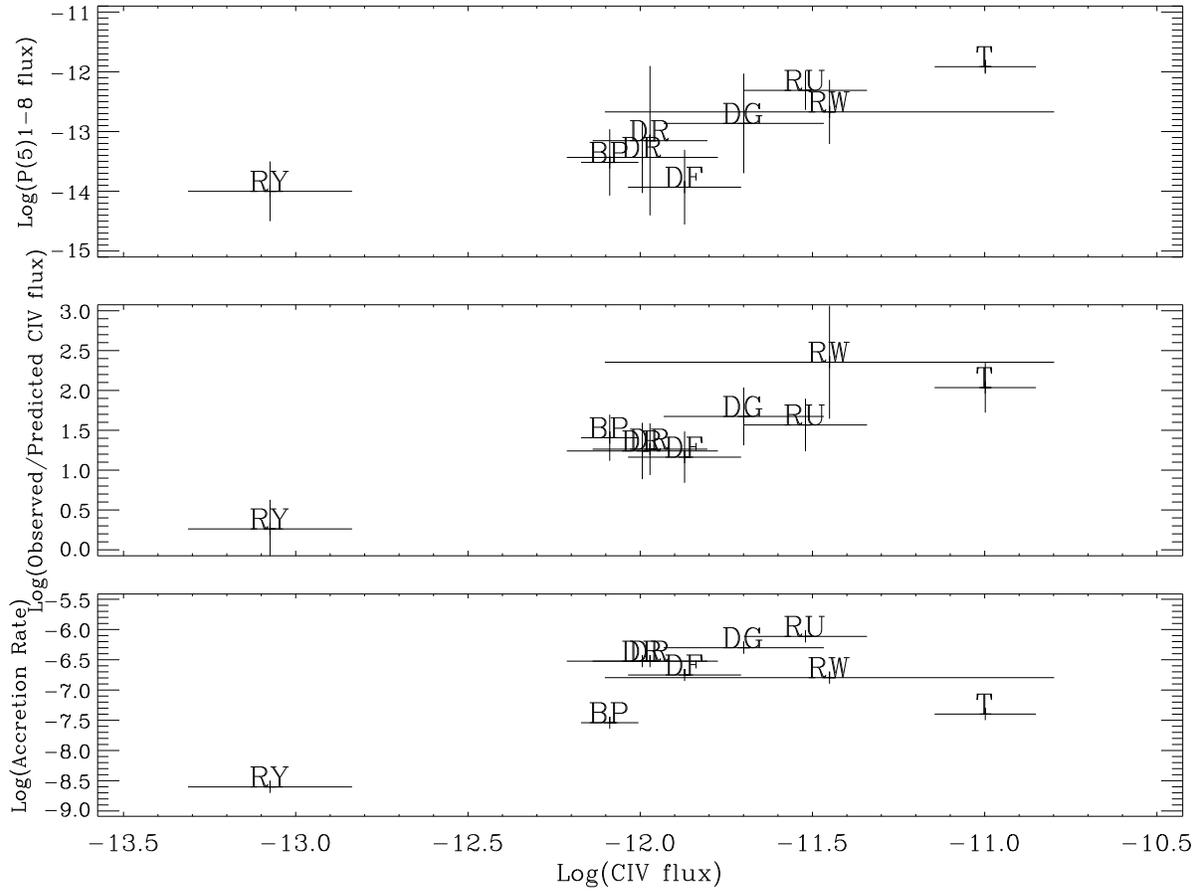}
\caption{\label{comp_civ}Relationship between \civ\ doublet flux (corrected for \htwo and extinction) and the flux of the \htwo\ line P(5)1-8 (top), the ratio of observed to predicted \civ\ fluxes (middle), and the accretion rate (bottom).}
\end{figure}

We observe a correlation between \civ\ and \htwo\ (Figure \ref{comp_civ}, top plot). In the Figure we use the  \htwo\ line P(5)1-8 at 1562.39 \AA. While this line is weak in most of our spectra, it is present in all of them. For \civ\ we sum the flux in both elements of the doublet (once they have been corrected for \htwo\ emission). For an individual star, a correlation between multiple observations of \civ\ and P(5)1-8 is expected, because even with no accretion \lya\ and \civ\ are both indicators of accretion activity. It is surprising, however that we observe the correlation for multiple stars. This indicates that what limits the emission is pump photons, not the state of \htwo\ or the stellar inclination.          

In the CTTSs, the fluxes from \siiv, \civ, and \htwo are most likely driven by the accretion rate. To understand what fraction of emission in the transition lines of CTTS is due to accretion processes we can look at HBC 388. This is a WTTS and so the emission in \civ\ and \siiv\ is not related to accretion processes. It is a fairly massive TTS, with one of the smallest radii in our sample. The emission is very weak. If all the stars shared the same intrinsic line surface flux, the observed unreddened fluxes would scale with their respective surface areas. In Figure \ref{comp_civ} (middle) we show the ratio between the observed \civ\ flux and the predicted one, assuming that all the stars have the \civ\ surface flux of HBC 388. For RY Tau and T Tau, which are stars of similar spectral type as HBC 388, this assumption underpredicts the observed \civ\ flux by a factor of 2 and a hundred, respectively. On the basis of the Figure we conclude that the \civ\ flux is dominated by accretion-driven emission processes, as opossed to solar-like stellar chromospheric activity, even though at low accretion rates the chromospheric contribution becomes important. The accretion process has enough energy to drive the emission, as the luminosity in \civ\ is only a small fraction ($<$0.1\%) of the accretion luminosity \citep{1996csss....9..419C}. It should be kept in mind that the internal structure of HBC 388 (a K1 star) may be different from that of lower mass stars in our sample (which are K7 or later), and so it is not clear how representative the chromospheric emission of this WTTS is.

Figure \ref{comp_civ} shows the relationship between \civ\ and accretion rate.  It should be noted that a relationship similar to this one is obtained by \citet{2000ApJ...539..815J}. On the basis of such relationship, they conclude that the accretion rate for T Tau is $30\ 10^{-8}\msun$/yr. Using this accretion rate the point for T Tau in Figure \ref{comp_civ} moves upward to the position of RW Aur. This Figure certainly suggests that accretion is the single engine behind the strong hot lines.

\subsubsection{The origin of the hot line emission\label{origin}}

In this section we study what the line shapes can tell us about the emission line region. 

As \citet*{1998AJ....116..455M} argue on the basis of the appearance of the \nad\ and Balmer lines, the temperature in the funnel flow is around 10000 K. Therefore, the \civ\ and \siiv\ lines cannot originate in the funnel. The shock region itself and the stellar atmosphere are prime candidates for being the source of the emission. 
In the shock at the end of the accretion funnel there are two different regions that may contribute to the emission: the pre-shock or radiative precursor, and the post-shock. The former has velocities of $\sim$300 \kms\ and temperatures $\sim10^4$K, while immediately after the shock the velocities are  $\sim$70 \kms\ and the temperatures are $10^5$K \citep{1998ApJ...509..802C}. Both regions may emit hot transition region lines. The shape of this shock on the surface of the star is unknown: assuming a dipolar magnetic field produces a ring on the stellar surface.

The ratio between the emission from the radiative precursor and that from the post-shock depends on the velocity, according to \citet{1998ARep...42..322L}, in such a way that the post-shock dominates for infalling velocities larger than 300 \kms\ (at infalling densities of $10^{12}$ cm$^{-3}$). The post-shock emission is produced at very low velocities. In general, one should therefore observe a broad redshifted component (the pre-shock emission) and a narrow centered component. For DF Tau and BP Tau, the only stars for which a decomposition in multiple components is possible this is not observed. More generally, as infalling velocities are almost certainly less than 300 \kms, pre-shock emission which preferentially produces redshifted profiles, should dominate. This is an attractive possibility because the emission should be broad. However, from Figures 13 and 14, four stars in our sample have centered profiles, which suggests that a different region is responsible for the emission.

\begin{figure}
\plotone{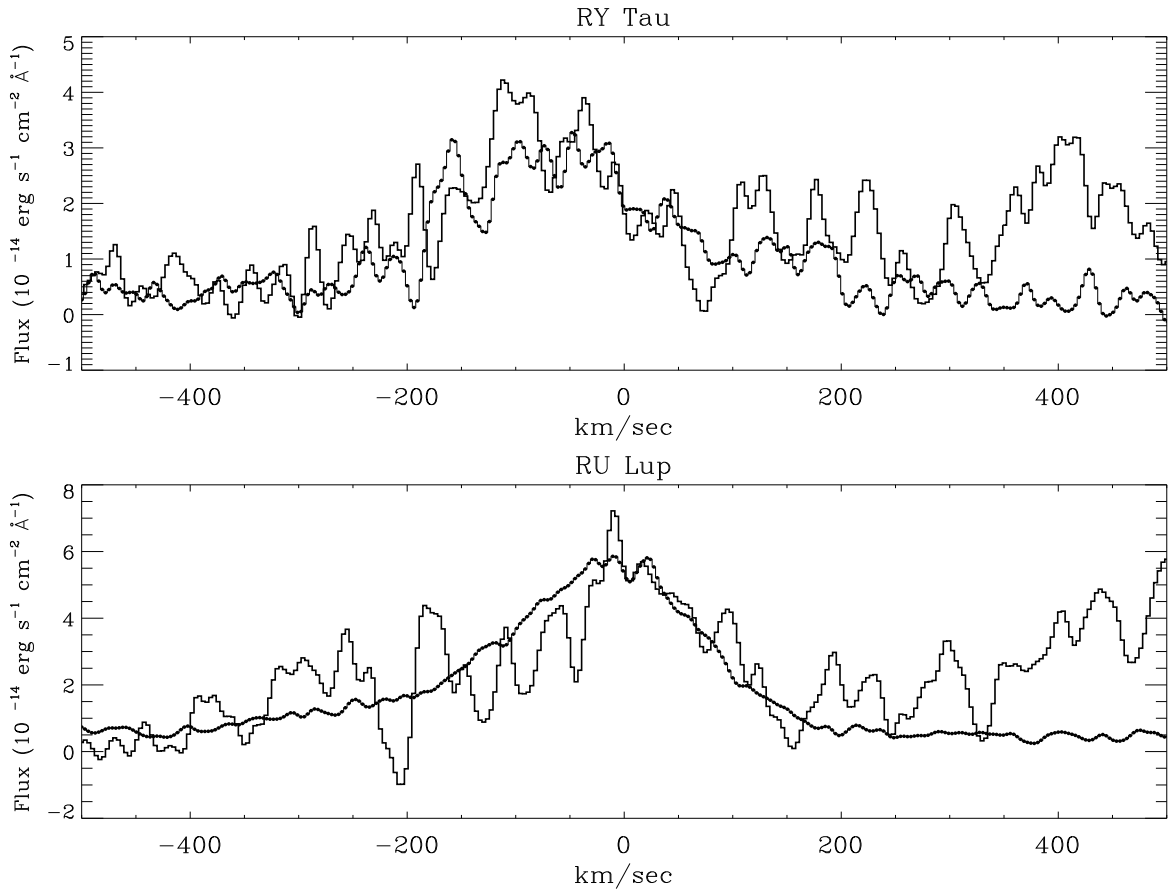}
\caption{\label{lamzin_comp}Comparison between the red members of \civ\ (thick line) with the \siiii\ (1892 \AA) line (thin dotted line). For RY Tau, we have scaled the observed \siiii\ profile by 0.25. \citet{2000ARep...44..323L} argue that the width of \siiii\ is due only to kinematic broadening.}
\end{figure}

\citet{2000ARep...44..323L} modeled the effect of the ring geometry in the \siiii\ line (1892 \AA. When collisionally excited in a low density medium, the ion is formed at 30000 K, see \citealt{1992ApJ...398..394A}) for RY Tau and RU Lup, using the spectra shown in Figure \ref{siii_spec}. They assume that the \siiii\ line emission comes from the pre-shock, in a gas that moves at $\sim300$ \kms. The fast moving gas produces the extended wings in the lines. They argue that a tilted dipole can reproduce the shape of the line, assuming that it is the result of the orientation of the accretion region with respect to the observer. In Figure \ref{lamzin_comp} we compare the red lines of \siiv\ and \civ\ (without \htwo) with the \siiii\ line for these two stars. As can be seen, the centroids and widths of both lines are similar, which supports the hypothesis that the main contribution to the \siiv\ and \civ\ lines comes from the same region. 

\citet{2000ARep...44..323L} argue that to reproduce the \siiii\ line for RU Lup one needs an inclined rotation axis for the star but symmetric accretion zones aligned with the disk axis. As they recognize, it is not clear what the physical plausibility of this model is. To reproduce the observed profiles, their models assume that the inclination of the observer is 10 degrees for RU Lup and 12 degrees for RY Tau. Observationally, however, the inclination of RY Tau is likely to be much larger, close to 90 degrees, given that $v \sin i=52$ \kms\ and $P_{rot.}=5.6$ days. 

While these geometrical models that assume that the lines are formed in the pre-shock can produce symmetric, redshifted, and slightly blueshifted profiles, the strong emission at $\sim-300$ \kms\ in DG Tau or the episodic emission at the red of DR Tau must be produced by a different mechanism. DG Tau is almost pole-on and it is known to have a strong jet, in which velocities as large as 300 \kms\ have been observed \citep{2000ApJ...537L..49B}. However, wind models predict \citep{1994ApJ...429..781S} that its temperature is at most 10000 K close to the launching point, which rules out the base of the wind as the source of the \civ\ emission. Models by \citet*{1999ApJ...524..142G} show that fast episodic emission of very hot plasma in the magnetosphere of a CTTS is likely to occur, due to entangling and reconnection of the stellar magnetic field. It is possible that this is what we are observing in the case of DG Tau and DR Tau.

A direct measurement of the physical conditions in the emission line region is provided by the line ratios. Given that the f-value of the blue doublet member is twice that of the red doublet member (for both \civ\ and \siiv), optically thin (or effectively thin) conditions imply that the blue member should be twice as strong as the red member. Table \ref{ratios} shows the ratios of each doublet member for \siiv\ and \civ. T Tau and HBC 388 are the only stars for which \siiv\ may indicate optically thin emission. For all stars, except RU Lup and DR Tau (9/7/95) (and perhaps DG Tau), the \civ\ ratios indicate optically thin emission. For comparison, the SUMER Atlas for the Sun shows an almost perfect 2-to-1 ratio between the blue and red members of the \civ\ and \siiv\ doublets.

%\documentclass{aastex}
%\begin{document}
%\newcommand \siiv{\ion{Si}{4}~}
%\newcommand \civ{\ion{C}{4}~}

\begin{deluxetable}{ccc} 
\tablecolumns{3} 
\tablewidth{0pc} 
\tablecaption{\label{ratios}Ratios of Transition Line Doublets} 
\tablehead{ 
\colhead{Name} & \colhead{\siiv}   & \colhead{\civ} }
\startdata 
BP Tau & $1.4\pm0.5$& $2.0\pm0.2$ \\
T Tau & $1.7\pm0.3$ & $2.4\pm0.4$ \\
DF Tau &\nodata & $2.13\pm0.3$ \\
RW Aur & $0.8\pm0.1$ & $1.6\pm0.9$ \\
DG Tau &\nodata& $1.6\pm0.3$   \\
DR Tau &\nodata & $1.0\pm0.4$, $1.4\pm0.2$ \\
RY Tau &\nodata & $1.8\pm0.5$ \\
RU Lup & $1.3\pm0.1$ & $1.2\pm0.2$ \\
HBC 388 & $1.8\pm0.7$ & $1.9\pm0.5 $ \\
\enddata
\tablecomments{All the ratios have been calculated once the contribution of \htwo\ has been subtracted}
\end{deluxetable} 

%\end{document}

Almost certainly, the radiative precursor is effectively thin in \civ\ and \siiv\ \citep{1998ARep...42..322L}. Therefore, \siiv\ cannot be formed here. Due to the rapid motion of the gas flow in the post-shock, it is possible that this hot lines are formed close to the star, at temperatures of the order of $\sim10^4$K. At such temperatures, He and H are not fully ionized and may provide enough opacity to significantly absorb \siiv. In this case, however one is left with the problem of explaining the widths of the lines.

In summary, while the relative ratios of the resonance doublets are consistent with the lines being formed after the accretion shock, their  widths are consistent with formation in the radiative precursor. The centroids are difficult to understand assuming formation in one region or the other only. Models of formation before the shock have to explain the nearly symmetric appearance of the doublets in the face of multiple geometric configurations. Models of formation after the shock have to explain the line width.

\section{Conclusions}

We have analyzed GHRS data of eight CTTS and one WTTS. The GHRS data consists of spectral ranges 40 \AA\ wide centered on 1345, 1400, 1497, 1550, and 1900 \AA. These UV spectra show strong \siiv, and \civ\ emission, and large quantities of sharp ($\sim40$ \kms in average) \htwo\ lines. In a companion paper we analyze the \mgii\ resonance doublet centered on 2800 \AA.

All the \htwo\ lines belong to the Lyman band (transitions from $B^1\Sigma^+_u$ to $X^1\Sigma^+_g$). We identify seven different routes, where the lines in the same route all come from the same upper level. All the observed lines are single peaked and we verify that they are all optically thin. 

We do not observe any correlation between the intensity of the measured \htwo\ lines, their center, and their FWHM with each other or with the inclination of the systems. The lack of correlation between line flux and inclination suggests that the gas responsible for the \htwo\ emission is not limited to the disk. We observe that the average of the centroids for all the lines measured in each star is blueshifted. Centering errors in the spectrograph may produce shifts from star to star of up to 20 \kms. Also, an asymmetric, non-uniform, spatially-extended distribution would produce velocity shifts. However, it would be surprising if all the shifts were to be in the same direction. Line widths vary from about 20 to 60 \kms. Turbulence, or inhomogeneities in the extended spatial distribution of \htwo\ in TTSs (as has been observed in T Tau) may be responsible for these widths. 

Regarding the origin of the emission, if \htwo\ is formed in the surface layers of the disk, the observed profiles should be double peaked. The GHRS resolution is such that we should be able to observe double peaked profiles from within  $\sim10(M/0.5\msun)\sin^2 i$ AU of the central star, where $M$ is the mass of the star and $i$ is the inclination. All the observed profiles are single peaked, and we do not observe any correlation of width and inclination. Together with the fact that the averages of the lines for each star are blueshifted, this suggests that \htwo\ is formed in a material moving towards the observer, i.e. an outflow. On the other hand, it is not difficult to imagine circumstances in which this is not true: the \htwo\ emission comes from a large disk region, the blueshift of the lines is accidental, due to either centering errors (which would have to be in the same direction in all the stars!) or asymmetric extended \htwo\ emission.  

We interpret the emission in \htwo\ as being due to fluorescence mostly by \lya: this line is one order of magnitude stronger than other atomic lines in the spectra. Furthermore, we can track every observed \htwo\ line to an excitation line within 1000 \kms\ of \lya. Fluorescent \htwo\ lines in CTTSs have already been reported widely in the literature \citep[e.g.][]{1981Natur.290...34B,1984MNRAS.207..831B,2000ApJS..129..399V}. The observed fluorescent routes are excited from levels N(v'',J'')=N(2,0), N(2,1), N(2,2), N(2,5), N(2,6), and N(1,13). None of the observed routes is excited from the blue wing of \lya, even though there are strong transitions present there that would produce fluorescent cascades strong enough to be seen in our spectra. Higher signal-to-noise data would be necessary before we can say whether this is due to the exotic (i.e. non-thermal) level populations or to the fact that absorption of \lya\ photons by \htwo\ occurs relatively far from the star, where the blue wing of \lya\ has already been absorbed by the wind.

We use previous measurements of the \lya\ flux by IUE to obtain estimates of the column densities and optical depths of the states from which the excitation occurs. This is an uncertain process given the low signal-to-noise of the IUE spectra in this region and the low number of them. We assume that \lya\ can be modeled by a Gaussian, that the strength is 30 times larger than the blue member of the \civ\ doublet (valid for RU Lup), and that the Gaussian has $\sigma=1.2$ \AA\ (valid for TW Hya). From these assumptions we obtain optical depths of the exciting transitions ($\sim1$ or less) and column densities of the levels from which the transitions originate ($10^{12}$ to $10^{15}\rm{cm^{-2}}$). Errors in the intensity of the \lya\ flux will change the absolute column densities but not their relative values. We conclude that the populations are far from being in thermal equilibrium, an important constraint in modeling efforts. This conclusion is independent of the exact line shape (Gaussian or Voigt) as long as the \lya\ is centered at rest wavelengths and its intensity decreases to the red. A comparison with ISO observations by \citet{1999A&A...348..877V} shows that our results are within the parameter region predicted by a thermal gas at 1500 K with a total column density of $2\times10^{18}\rm{cm^{-2}}$ (derived from far IR lines). Obviously, this coincidence is not incompatible with the lack of a thermal distribution for upper levels of \htwo: a mechanism that produces a thermal distribution in the lower levels does not necessarily have an effect in the upper levels. 

The main weakness of this approach is that the IUE observations show that \lya\ is very redshifted, perhaps because its blue wing has been absorbed by an outflow or by the ISM (which means that the flux absorbed by \htwo\ is probably 3-4 times larger as the one we use). What the exact shape of the line producing the excitation of \htwo\ is, we cannot tell from these data. The fact that all the observed routes are excited from transitions to the red of the rest wavelength of \lya\ suggests that at least some absorption has already taken place when the line radiation arrives to the location of \htwo. What we have shown is that if \lya\ can be approximated by a Gaussian centered at 1215.67 \AA, the populations of the \htwo\ states that absorb \lya\ are not thermal. If one accepts these assumptions, our results firmly suggest that better models of excitation of \htwo\ by radiation are needed, given that even at v''=2 the levels do not have a thermal population.

If the ions are collisionally formed, the \civ\ and \siiv\ lines arise in hot ($10^5$ K) regions. In the past, analyses of these lines have been carried out using the low resolution spectra provided by IUE. In principle, these kind of analyses have the problem that there are strong \htwo\ lines whose positions coincide with those of the transition lines. In this work we correct for \htwo\ emission by using other lines from the same fluorescent route to calculate the contribution to the transition lines (as we have confirmed that the \htwo\ lines are optically thin). In this way, we obtain unblended profiles of the \civ\ and \siiv\ lines, and conclude that the contribution of the \htwo\ lines to the profiles is less than 10\% and that the variability of the fluxes is dominated by variability in \civ\ and \siiv.  

After subtracting the \htwo\ contribution, we show that the shape of the \civ\ lines is the same as that of the \siiv\ lines (with the exception of RW Aur), and that their fluxes are linearly correlated in log space. In spite of this correlation, a wide variation is observed in the ratio of \civ\ to \siiv\ flux. It is not clear what is the origin of such variation. For RW Aur, \citet{2000A&A...357..951E} argues that strong \feii\ absorptions are present in the \civ\ and \siiv\ profiles. We do not see any example of these absorptions in any other star. 

The profiles have full-widths at half-maxima that vary from 200 to 300 \kms, and a wide array of centroids (blueshifted, centered, redshifted). In particular, blueshifted emission at $\sim-300$\kms\ is observed in the \civ\ lines of DR Tau and DG Tau. In the first star, the emission is episodic, as it is shown by the two separate epochs of observations in our data. These blueshifted emissions are perhaps evidence of reconnection processes in the magnetosphere that behave as a coronal wind, given that the standard magnetocentrifugally driven T Tauri wind is expected to be too cold to produce this kind of emission. According to  \citet{1994ApJ...429..781S} such reconnection events occur as a consequence of mismatches between the rotation rate of the star and the point at which the disk is truncated, and they may be the cause of much of the variability observed in TTS spectra. They have been modeled in detail by \citet{1999ApJ...524..142G}. Evidence from hot episodic emission has been observed in \hal\  \citep{AlencaronDR} and in \hei\ \citep*{2001ApJ...551.1037B}.

We observe a correlation between the \civ\ line and the \htwo\ emission. Such correlation has also been observed by \citet{2000ApJ...539..815J} for a large sample of CTTSs. It is surprising given that the physical conditions of \htwo\ are different for each star and indicates that the \htwo\ emission is limited by the excitation mechanism and not by the \htwo\ population. This is a useful fact that any modeling attempt should take into account. For the stars of this paper, the observed \civ\ and \siiv\ lines are most likely the result of accretion-related (as opposed to chromospheric) processes. If we assume that the transition region emission of our CTTSs is like that of HBC 388 (a WTTS), we conclude that the dominant contribution to the \civ\ line flux comes from accretion related processes. Our results also suggest that for accretion rates of the order of $10^{-9} \msun$/yr most of the line flux in \civ\ should come from the stellar chromosphere. A correlation is also observed (albeit not a very strong one) between the \civ\ line flux and the accretion rate. The lack of a strong correlation is not a surprise: in addition the fact that our sample is very small, accretion rates are measured at one point in time and are known to change \citep*{2000ApJ...539..834A}, as are the measured fluxes. Without a theory of the behavior of the transition region lines as a function of accretion rate is not clear what the correlation presented here (or the more extensive one obtained by \citealt{2000ApJ...539..815J}) is really indicating.

While it seems clear that the \siiv\ and \civ\ lines are related to the accretion shock formed when the gas flow impacts the stellar surface, the exact region of formation remains a mystery. The shape and centroid of \civ\ is similar to that of the \siiii\ line, believed to be emitted from the radiative precursor of the shock \citep{2000ARep...44..323L}. Formation in the pre-shock may explain the widths of the lines as being due to the projection of the accretion ring in the stellar surface with respect to the observer. Emission from this region would preferentially produce redshifted profiles, as the gas flow is still moving at almost free-fall velocities. However, a sizable fraction (4 out of 8) of the observed profiles have zero velocity centroids, a remarkable statistical coincidence if the position of the centroids is due to the apparent position of the emitting region in the stellar surface. Furthermore, while \civ\ lines tend to be optically thin, the \siiv\ profiles are not (the exceptions are HBC 388 and T Tau, for which the flux ratios indicate that the \siiv\ profiles are optically thin). Models of the radiative precursor indicate that this should be effectively thin in both lines.

If most of the emission is originated in the post-shock, line formation will occur close to the star, where the gas has temperatures of $\sim10^4$ K. At this point, the column is starting to become optically thick to its own radiation and so emission from the region is consistent with the observed flux ratios. On the other hand, such region will produce centered, narrow profiles, unlike the observed ones.

Multiple emission regions are possible. The \civ\ and \siiv\ lines of BP Tau can be decomposed in multiple Gaussian components, with a broad ($\sim300$ \kms) and a narrow component ($\sim100$ \kms). For DF Tau, the only star in which the signal-to-noise is large enough, we observe a third component, with $\sim50$ \kms width. This suggests that the emission in the hot lines may come from more than one region. However, all components have velocities near zero. As suggested above, simultaneous formation in the pre- and post-shock should produce widely different velocity centroids. 

The study of these ultraviolet ranges provide us with yet another window into which to look at T-Tauri phenomena. What these analyses show is that there is still a lot of work to be done in the theoretical front before we can understand the innermost regions of TTS. Observationally too, it is clear that high resolution spatial and spectral observations with good time coverage are necessary before a more careful model of the region can be attempted.

\acknowledgments

This paper is based on observations made with the NASA/ESA
{\it Hubble Space Telescope}, obtained at the Space Telescope Science Institute, which is operated by the Association of Universities for Research in
Astronomy, Inc., under NASA contract NAS 5-26555. DRA and GB acknowledge support from the National Science Foundation (grant ASI952872) and NASA (grant NAG5-3471). FMW acknowledges support through NASA grants NAG5-1862, STScI-GO-5317-0193A, STScI-GO-058750294A, and STScI-GO-3845.02-91A to SUNY Stony Brook. CMJK acknowledges support through NASA grant NAG5-8209.

We thank G. Herczeg and J. Linsky for supplying us with their STIS TW Hya spectrum in advance of publication. We also thank D. Hollenbach, M. McMurry, J. Black,  and C. Jordan for fruitful discussions on molecular hydrogen.

\bibliographystyle{apj}
%\bibliography{/garavito/ardila/tex/bibtex/ardila}

\begin{thebibliography}{65}
\expandafter\ifx\csname natexlab\endcsname\relax\def\natexlab#1{#1}\fi

\bibitem[{{Abgrall} {et~al.}(1993){Abgrall}, {Roueff}, {Launay}, {Roncin}, \&
  {Subtil}}]{1993A&AS..101..273A}
{Abgrall}, H., {Roueff}, E., {Launay}, F., {Roncin}, J.~Y., \& {Subtil}, J.~L.
  1993, \aaps, 101, 273

\bibitem[{{Alencar} \& {Basri}(2000)}]{2000AJ....119.1881A}
{Alencar}, S. H.~P. \& {Basri}, G. 2000, \aj, 119, 1881

\bibitem[{{Alencar} {et~al.}(2001){Alencar}, {Johns-Krull}, \&
  {Basri}}]{AlencaronDR}
{Alencar}, S. H.~P., {Johns-Krull}, C.~M., \& {Basri}, G. 2001, submitted to \apj

\bibitem[{{Ardila} \& {Basri}(2000)}]{2000ApJ...539..834A}
{Ardila}, D.\ R. and {Basri}, G. 2000, \apj, 539, 834

\bibitem[{{Arnaud} \& {Raymond}(1992)}]{1992ApJ...398..394A}
{Arnaud}, M. \& {Raymond}, J. 1992, \apj, 398, 394

\bibitem[{{Ayres} {et~al.}(1981){Ayres}, {Marstad}, \&
  {Linsky}}]{1981ApJ...247..545A}
{Ayres}, T.~R., {Marstad}, N.~C., \& {Linsky}, J.~L. 1981, \apj, 247, 545

\bibitem[{{Bacciotti} {et~al.}(2000){Bacciotti}, {Mundt}, {Ray},
  {Eisl{\"o}ffel}, {Solf}, \& {Camezind}}]{2000ApJ...537L..49B}
{Bacciotti}, F., {Mundt}, R., {Ray}, T.~P., {Eisl{\"o}ffel}, J., {Solf}, J., \&
  {Camezind}, M. 2000, \apjl, 537, L49

\bibitem[{{Bartoe} {et~al.}(1979){Bartoe}, {Brueckner}, {Nicolas}, {Sandlin},
  {Vanhoosier}, \& {Jordan}}]{1979MNRAS.187..463B}
{Bartoe}, J. .~F., {Brueckner}, G.~E., {Nicolas}, K.~R., {Sandlin}, G.~D.,
  {Vanhoosier}, M.~E., \& {Jordan}, C. 1979, \mnras, 187, 463

\bibitem[{{Beristain} {et~al.}(2001){Beristain}, {Edwards}, \&
  {Kwan}}]{2001ApJ...551.1037B}
{Beristain}, G., {Edwards}, S., \& {Kwan}, J. 2001, \apj, 551, 1037

\bibitem[{{Black} \& {van Dishoeck}(1987)}]{1987ApJ...322..412B}
{Black}, J.~H. \& {van Dishoeck}, E.~F. 1987, \apj, 322, 412

\bibitem[{{Blackwell} {et~al.}(1993){Blackwell}, {Shore}, {Robinson},
  {Feggans}, {Lindler}, {Malumuth}, {Sandoval}, \& {Ake}}]{blackwell1993}
{Blackwell}, J., {Shore}, S., {Robinson}, R., {Feggans}, K., {Lindler}, D.,
  {Malumuth}, E., {Sandoval}, J., \& {Ake}, T. 1993, A User's Guide to the GHRS
  Software (Baltimore:GSFC)

\bibitem[{{Blondel} {et~al.}(1993){Blondel}, {Talavera}, \&
  {Djie}}]{1993A&A...268..624B}
{Blondel}, P. F.~C., {Talavera}, A., \& {Djie}, H. R. E. T.~A. 1993, \aap, 268,
  624

\bibitem[{{Brandt} {et~al.}(1994){Brandt}, {Heap}, {Beaver}, {Boggess},
  {Carpenter}, {Ebbets}, {Hutchings}, {Jura}, {Leckrone}, {Linsky}, {Maran},
  {Savage}, {Smith}, {Trafton}, {Walter}, {Weymann}, {Ake}, {Bruhweiler},
  {Cardelli}, {Lindler}, {Malumuth}, {Randall}, {Robinson}, {Shore}, \&
  {Wahlgren}}]{1994PASP..106..890B}
{Brandt}, J.~C., {et~al.} 1994, \pasp, 106, 890

\bibitem[{{Brandt} {et~al.}(1993){Brandt}, {Heap}, {Beaver}, {Boggess},
  {Carpenter}, {Ebbets}, {Hutchings}, {Jura}, {Leckrone}, {Linsky}, {Maran},
  {Savage}, {Smith}, {Trafton}, {Walter}, {Weymann}, {Snow}, {Randall},
  {Lindler}, {Shore}, {Morris}, {Gilliland}, {Lu}, \&
  {Robinson}}]{1993AJ....105..831B}
{Brandt}, J.~C., {et~al.} 1993, \aj, 105, 831

\bibitem[{{Brown} {et~al.}(1984){Brown}, {de M.\ Ferraz}, \&
  {Jordan}}]{1984MNRAS.207..831B}
{Brown}, A., {de M.\ Ferraz}, M.~C., \& {Jordan}, C. 1984, \mnras, 207, 831

\bibitem[{{Brown} {et~al.}(1981){Brown}, {Jordan}, {Millar}, {Gondhalekar}, \&
  {Wilson}}]{1981Natur.290...34B}
{Brown}, A., {Jordan}, C., {Millar}, T.~J., {Gondhalekar}, P., \& {Wilson}, R.
  1981, \nat, 290, 34

\bibitem[{{Calvet} \& {Gullbring}(1998)}]{1998ApJ...509..802C}
{Calvet}, N. \& {Gullbring}, E. 1998, \apj, 509, 802

\bibitem[{{Calvet} {et~al.}(1996){Calvet}, {Hartmann}, {Hewett}, {Valenti},
  {Basri}, \& {Walter}}]{1996csss....9..419C}
{Calvet}, N., {Hartmann}, L., {Hewett}, R., {Valenti}, J.~A., {Basri}, G., \&
  {Walter}, F. 1996, in ASP Conf. Ser. 109: Cool Stars, Stellar Systems, and
  the Sun, ed. R. Pallavicini \& A. K. Dupree (San Francisco: ASP), p.419


\bibitem[{{Curdt} {et~al.}(2000){Curdt}, {Brekke}, {Feldman}, {Dwivedi},
  {Sch\"ule}, \& {Lemaire}}]{2001astro.ph}
{Curdt}, W., {Brekke}, P., {Feldman}, U., {Dwivedi}, B.~N., {Sch\"ule}, U., \&
  {Lemaire}, P. 2000, in Submitted to A\&A Supplement.

\bibitem[{{Errico} {et~al.}(2000){Errico}, {Lamzin}, \&
  {Vittone}}]{2000A&A...357..951E}
{Errico}, L., {Lamzin}, S.~A., \& {Vittone}, A.~A. 2000, \aap, 357, 951

\bibitem[{{Gahm} {et~al.}(1999){Gahm}, {Petrov}, {Duemmler}, {Gameiro}, \&
  {Lago}}]{1999A&A...352L..95G}
{Gahm}, G.~F., {Petrov}, P.~P., {Duemmler}, R., {Gameiro}, J.~F., \& {Lago}, M.
  T. V.~T. 1999, \aap, 352, L95

\bibitem[{{Ghez} {et~al.}(1993){Ghez}, {Neugebauer}, \&
  {Matthews}}]{1993AJ....106.2005G}
{Ghez}, A.~M., {Neugebauer}, G., \& {Matthews}, K. 1993, \aj, 106, 2005

\bibitem[{{Ghez} {et~al.}(1997){Ghez}, {White}, \&
  {Simon}}]{1997ApJ...490..353G}
{Ghez}, A.~M., {White}, R.~J., \& {Simon}, M. 1997, \apj, 490, 353

\bibitem[{{Gomez de Castro} \& {Lamzin}(1999)}]{1999MNRAS.304L..41G}
{Gomez de Castro}, A.~I. \& {Lamzin}, S.~A. 1999, \mnras, 304, L41

\bibitem[{{Gomez de Castro} {et~al.}(1994){Gomez de Castro}, {Lamzin}, \&
  {Shatskii}}]{1994ARep...38..540G}
{Gomez de Castro}, A.~I., {Lamzin}, S.~A., \& {Shatskii}, N.~I. 1994, Astronomy
  Reports, 38, 540

\bibitem[{{Goodson} {et~al.}(1999){Goodson}, {B{\"o}hm}, \&
  {Winglee}}]{1999ApJ...524..142G}
{Goodson}, A.~P., {B{\"o}hm}, K., \& {Winglee}, R.~M. 1999, \apj, 524, 142

\bibitem[{{Gullbring} {et~al.}(2000){Gullbring}, {Calvet}, {Muzerolle}, \&
  {Hartmann}}]{2000ApJ...544..927G}
{Gullbring}, E., {Calvet}, N., {Muzerolle}, J., \& {Hartmann}, L. 2000, \apj,
  544, 927

\bibitem[{{Gullbring} {et~al.}(1998){Gullbring}, {Hartmann}, {Briceno}, \&
  {Calvet}}]{1998ApJ...492..323G}
{Gullbring}, E., {Hartmann}, L., {Briceno}, C., \& {Calvet}, N. 1998, \apj,
  492, 323

\bibitem[{{Hartigan} {et~al.}(1995){Hartigan}, {Edwards}, \&
  {Ghandour}}]{1995ApJ...452..736H}
{Hartigan}, P., {Edwards}, S., \& {Ghandour}, L. 1995, \apj, 452, 736

\bibitem[{{Hartmann} {et~al.}(1991){Hartmann}, {Stauffer}, {Kenyon}, \&
  {Jones}}]{1991AJ....101.1050H}
{Hartmann}, L., {Stauffer}, J.~R., {Kenyon}, S.~J., \& {Jones}, B.~F. 1991,
  \aj, 101, 1050

\bibitem[{{Heap} {et~al.}(1995){Heap}, {Brandt}, {Randall}, {Carpenter},
  {Leckrone}, {Maran}, {Smith}, {Beaver}, {Boggess}, {Ebbets}, {Garner},
  {Hutchings}, {Jura}, {Linsky}, {Savage}, {Cardelli}, {Trafton}, {Walter},
  {Weymann}, {Ake}, {Crenshaw}, {Malamuth}, {Robinson}, {Sandoval}, {Shore},
  {Wahlgren}, {Bruhweiler}, {Lindler}, {Gilliland}, {Hulbert}, \&
  {Soderblom}}]{1995PASP..107..871H}
{Heap}, S.~R. {et~al.} 1995, \pasp, 107, 871

\bibitem[{{Herbig} \& {Bell}(1988)}]{1988cels.book.....H}
{Herbig}, G.~H. \& {Bell}, K.~R. 1988, Catalog of emission line stars of the
  orion population (Santa Cruz: Lick Observatory)

\bibitem[{{Herbst} {et~al.}(1997){Herbst}, {Robberto}, \&
  {Beckwith}}]{1997AJ....114..744H}
{Herbst}, T.~M., {Robberto}, M., \& {Beckwith}, S. V.~W. 1997, \aj, 114, 744

\bibitem[{{Herczeg} {et~al.}(2001){Herczeg}, {Linsky}, {Valenti}, \&
  {Johns-Krull}}]{TWspec}
{Herczeg}, G.~H., {Linsky}, J.~L., {Valenti}, J.~A., \& {Johns-Krull}, C.~M.
  2001, in Submitted to ApJ

\bibitem[{{Hughes} {et~al.}(1994){Hughes}, {Hartigan}, {Krautter}, \&
  {Kelemen}}]{1994AJ....108.1071H}
{Hughes}, J., {Hartigan}, P., {Krautter}, J., \& {Kelemen}, J. 1994, \aj, 108,
  1071

\bibitem[{{Johns-Krull} {et~al.}(1999){Johns-Krull}, {Valenti}, {Hatzes}, \&
  {Kanaan}}]{1999ApJ...510L..41J}
{Johns-Krull}, C.~M., {Valenti}, J.~A., {Hatzes}, A.~P., \& {Kanaan}, A. 1999,
  \apjl, 510, L41

\bibitem[{{Johns-Krull} {et~al.}(2000){Johns-Krull}, {Valenti}, \&
  {Linsky}}]{2000ApJ...539..815J}
{Johns-Krull}, C.~M., {Valenti}, J.~A., \& {Linsky}, J.~L. 2000, \apj, 539, 815

\bibitem[{{Jordan} {et~al.}(1978){Jordan}, {Brueckner}, {Bartoe}, {Sandlin}, \&
  {Vanhoosier}}]{1978ApJ...226..687J}
{Jordan}, C., {Brueckner}, G.~E., {Bartoe}, J. .~F., {Sandlin}, G.~D., \&
  {Vanhoosier}, M.~E. 1978, \apj, 226, 687

\bibitem[{{Kurt} \& {Lamzin}(1995)}]{1995ARep...39..322K}
{Kurt}, V.~G. \& {Lamzin}, S.~A. 1995, Astronomy Reports, 39, 322

\bibitem[{{Lamzin}(1998)}]{1998ARep...42..322L}
{Lamzin}, S.~A. 1998, Astronomy Reports, 42, 322

\bibitem[{{Lamzin}(2000{\natexlab{a}})}]{2000AstL...26..225L}
---. 2000{\natexlab{a}}, Astronomy Letters, 26, 225

\bibitem[{{Lamzin}(2000{\natexlab{b}})}]{2000AstL...26..589L}
---. 2000{\natexlab{b}}, Astronomy Letters, 26, 589

\bibitem[{{Lamzin}(2000{\natexlab{c}})}]{2000ARep...44..323L}
---. 2000{\natexlab{c}}, Astronomy Reports, 44, 323

\bibitem[{{Lamzin} {et~al.}(1996){Lamzin}, {Bisnovatyi-Kogan}, {Errico},
  {Giovannelli}, {Katysheva}, {Rossi}, \& {Vittone}}]{1996A&A...306..877L}
{Lamzin}, S.~A., {Bisnovatyi-Kogan}, G.~S., {Errico}, L., {Giovannelli}, F.,
  {Katysheva}, N.~A., {Rossi}, C., \& {Vittone}, A.~A. 1996, \aap, 306, 877

\bibitem[{{Lamzin} {et~al.}(2001){Lamzin}, {Vittone}, \& {Errico}}]{LamzinonDF}
{Lamzin}, S.~A., {Vittone}, A.~A., \& {Errico}, L. 2001, Astronomy Letters, 27,
  313

\bibitem[{{Leinert} {et~al.}(1993){Leinert}, {Zinnecker}, {Weitzel},
  {Christou}, {Ridgway}, {Jameson}, {Haas}, \& {Lenzen}}]{1993A&A...278..129L}
{Leinert}, C., {Zinnecker}, H., {Weitzel}, N., {Christou}, J., {Ridgway},
  S.~T., {Jameson}, R., {Haas}, M., \& {Lenzen}, R. 1993, \aap, 278, 129

\bibitem[{{Martin} \& {Mandy}(1995)}]{1995ApJ...455L..89M}
{Martin}, P.~G. \& {Mandy}, M.~E. 1995, \apjl, 455, L89

\bibitem[{{McMurry} {et~al.}(1999){McMurry}, {Jordan}, \&
  {Carpenter}}]{1999MNRAS.302...48M}
{McMurry}, A.~D., {Jordan}, C., \& {Carpenter}, K.~G. 1999, \mnras, 302, 48

\bibitem[{{Muzerolle} {et~al.}(1998){Muzerolle}, {Hartmann}, \&
  {Calvet}}]{1998AJ....116..455M}
{Muzerolle}, J., {Hartmann}, L., \& {Calvet}, N. 1998, \aj, 116, 455

\bibitem[{{Petrov} {et~al.}(2001){Petrov}, {Gahm}, {Gameiro}, {Duemmler},
  {Ilyin}, {Laakkonen}, {Lago}, \& {Tuominen}}]{2001A&A...369..993P}
{Petrov}, P.~P., {Gahm}, G.~F., {Gameiro}, J.~F., {Duemmler}, R., {Ilyin},
  I.~V., {Laakkonen}, T., {Lago}, M. T. V.~T., \& {Tuominen}, I. 2001, \aap,
  369, 993

\bibitem[{{Preibisch} \& {Smith}(1997)}]{1997A&A...322..825P}
{Preibisch}, T. \& {Smith}, M.~D. 1997, \aap, 322, 825

\bibitem[{{Sandlin} {et~al.}(1986){Sandlin}, {Bartoe}, {Brueckner}, {Tousey},
  \& {Vanhoosier}}]{1986ApJS...61..801S}
{Sandlin}, G.~D., {Bartoe}, J. .~F., {Brueckner}, G.~E., {Tousey}, R., \&
  {Vanhoosier}, M.~E. 1986, \apjs, 61, 801

\bibitem[{{Sartoretti} {et~al.}(1998){Sartoretti}, {Brown}, {Latham}, \&
  {Torres}}]{1998A&A...334..592S}
{Sartoretti}, P., {Brown}, R.~A., {Latham}, D.~W., \& {Torres}, G. 1998, \aap,
  334, 592

\bibitem[{{Shu} {et~al.}(1994){Shu}, {Najita}, {Ostriker}, {Wilkin}, {Ruden},
  \& {Lizano}}]{1994ApJ...429..781S}
{Shu}, F., {Najita}, J., {Ostriker}, E., {Wilkin}, F., {Ruden}, S., \&
  {Lizano}, S. 1994, \apj, 429, 781

\bibitem[{{Stapelfeldt} {et~al.}(1998){Stapelfeldt}, {Burrows}, {Krist},
  {Watson}, {Ballester}, {Clarke}, {Crisp}, {Evans}, {Gallagher}, {Griffiths},
  {Hester}, {Hoessel}, {Holtzman}, {Mould}, {Scowen}, {Trauger}, \&
  {Westphal}}]{1998ApJ...508..736S}
{Stapelfeldt}, K.~R. {et~al.} 1998, \apj, 508, 736

\bibitem[{{Thi} {et~al.}(1999){Thi}, {van Dishoeck}, {Blake}, {van Zadelhoff},
  \& {Hogerheijde}}]{1999ApJ...521L..63T}
{Thi}, W., {van Dishoeck}, E.~F., {Blake}, G.~A., {van Zadelhoff}, G., \&
  {Hogerheijde}, M.~R. 1999, \apjl, 521, L63

\bibitem[{{Thiebaut} {et~al.}(1995){Thiebaut}, {Balega}, {Balega}, {Belkine},
  {Bouvier}, {Foy}, {Blazit}, \& {Bonneau}}]{1995A&A...304L..17T}
{Thiebaut}, E., {Balega}, Y., {Balega}, I., {Belkine}, I., {Bouvier}, J.,
  {Foy}, R., {Blazit}, A., \& {Bonneau}, D. 1995, \aap, 304, L17

\bibitem[{{Valenti} {et~al.}(1993){Valenti}, {Basri}, {Walter}, {Hartmann}, \&
  {Calvet}}]{1993AAS...183.4007V}
{Valenti}, J.~A., {Basri}, G., {Walter}, F., {Hartmann}, L., \& {Calvet}, N.
  1993, BAAS, 183, 4007

\bibitem[{{Valenti} {et~al.}(2000){Valenti}, {Johns-Krull}, \&
  {Linsky}}]{2000ApJS..129..399V}
{Valenti}, J.~A., {Johns-Krull}, C.~M., \& {Linsky}, J.~L. 2000, \apjs, 129,
  399

\bibitem[{{van den Ancker} {et~al.}(1999){van den Ancker}, {Wesselius},
  {Tielens}, {van Dishoeck}, \& {Spinoglio}}]{1999A&A...348..877V}
{van den Ancker}, M.~E., {Wesselius}, P.~R., {Tielens}, A. G. G.~M., {van
  Dishoeck}, E.~F., \& {Spinoglio}, L. 1999, \aap, 348, 877

\bibitem[{{van Dishoeck} {et~al.}(1997){van Dishoeck}, {Thi}, {Blake},
  {Mannings}, {Sargent}, {Koerner}, \& {Mundy}}]{1997Ap&SS.255...77V}
{van Dishoeck}, E.~F., {Thi}, W.~F., {Blake}, G.~A., {Mannings}, V., {Sargent},
  A.~I., {Koerner}, D., \& {Mundy}, L.~G. 1997, \apss, 255, 77

\bibitem[{{Walter} {et~al.}(1988){Walter}, {Brown}, {Mathieu}, {Myers}, \&
  {Vrba}}]{1988AJ.....96..297W}
{Walter}, F.~M., {Brown}, A., {Mathieu}, R.~D., {Myers}, P.~C., \& {Vrba},
  F.~J. 1988, \aj, 96, 297

\bibitem[{{Walter} \& {Liu}(1998)}]{1998sigh.conf..360W}
{Walter}, M. \& {Liu}, Y. 1998, in ASP Conf. Ser. 143: The Scientific Impact of
  the Goddard High Resolution Spectrograph, ed. J. C. Brandt, T. B. Ake, \& C. C. Petersen, 360

\bibitem[{{Weintraub} {et~al.}(2000){Weintraub}, {Kastner}, \&
  {Bary}}]{2000ApJ...541..767W}
{Weintraub}, D.~A., {Kastner}, J.~H., \& {Bary}, J.~S. 2000, \apj, 541, 767

\bibitem[{{Wolk} \& {Walter}(1996)}]{1996AJ....111.2066W}
{Wolk}, S.~J. \& {Walter}, F.~M. 1996, \aj, 111, 2066

\end{thebibliography}

\end{document}